\pdfoutput=1
\documentclass[10pt, logo, copyright, nonumbering]{nvidiatechreport}

\usepackage[utf8]{inputenc} % allow utf-8 input
\usepackage[T1]{fontenc}    % use 8-bit T1 fonts
\usepackage{hyperref}       % hyperlinks
\usepackage{url}            % simple URL typesetting
\usepackage{booktabs}       % professional-quality tables
\usepackage{amsfonts}       % blackboard math symbols
\usepackage{nicefrac}       % compact symbols for 1/2, etc.
\usepackage{microtype}      % microtypography
\usepackage{xcolor}         % colors
\usepackage{amsmath}
\usepackage{amssymb}
\usepackage{mathtools}
\usepackage{amsthm}
\usepackage{siunitx}
\usepackage{enumitem}
\usepackage{xspace}
\usepackage{xcolor}
\usepackage{wrapfig}
\usepackage{caption}
\usepackage{amssymb}
\usepackage{multirow}
\usepackage{adjustbox}
\usepackage{makecell}
\usepackage{mathtools}
\usepackage{algorithm}
\usepackage{algorithmic}
\usepackage{colortbl}
\usepackage{tcolorbox}
\usepackage{babel}
\usepackage{bbding}
\usepackage{pifont}
\usepackage{nicematrix}
\usepackage{enumitem}
\usepackage{booktabs}
\usepackage[flushleft]{threeparttable}

\usepackage{thmtools}
\usepackage{thm-restate}

\definecolor{myblue}{RGB}{68, 114, 196}
\definecolor{myorange}{RGB}{237, 125, 49}
\definecolor{lightgray}{RGB}{229, 232, 232}
\definecolor{mydarkblue}{rgb}{0,0.08,0.45}

\newcommand{\method}{Prote\'{i}na-Atom\'{i}stica\xspace}
\newcommand{\basemodel}{Prote\'{i}na\xspace}
\newcommand{\laprot}{La-Prote\'{i}na\xspace}
\newcommand{\ourmodel}{Prote\'{i}na-Atom\'{i}stica\xspace}
\newcommand{\ourmodelnoit}{Prote\'{i}na-Atom\'{i}stica\xspace}
\newcommand{\codesmodel}{Prote\'{i}na-Co-design\xspace}
\newcommand{\mask}{\text{M}}
\newcommand{\ca}{\rvx}
\newcommand{\nonca}{\rvz}
\newcommand{\seq}{\rvr}
\newcommand{\geniedata}{$\mathcal{D}_\mathrm{AFDB-clstr}$\xspace}
\newcommand{\ourdata}{$\mathcal{D}_\mathrm{SYN-ours}$\xspace}
\newcommand{\xmark}{\ding{55}}

\usepackage{amsmath,amsfonts,bm}

\def\eqref#1{equation~\ref{#1}}
\def\1{\bm{1}}

\def\eps{{\epsilon}}

\def\rvc{{\mathbf{c}}}

\def\rvu{{\mathbf{i}}}

\def\rvr{{\mathbf{r}}}
\def\rvs{{\mathbf{s}}}
\def\rvt{{\mathbf{t}}}
\def\rvu{{\mathbf{u}}}
\def\rvv{{\mathbf{v}}}

\def\rvx{{\mathbf{x}}}

\def\rvz{{\mathbf{z}}}

\def\mI{{\bm{I}}}

\DeclareMathAlphabet{\mathsfit}{\encodingdefault}{\sfdefault}{m}{sl}
\SetMathAlphabet{\mathsfit}{bold}{\encodingdefault}{\sfdefault}{bx}{n}

\def\gL{{\mathcal{L}}}

\def\gN{{\mathcal{N}}}

\newcommand{\E}{\mathbb{E}}

\newcommand{\R}{\mathbb{R}}

\newcommand\sref{Sec. \ref}
\newcommand\aref{Algorithm~\ref}
\newcommand\eref{Eq.~\ref}
\newcommand\fref{Fig.~\ref}
\newcommand\tref{Table~\ref}
\usepackage{nicefrac}       % compact symbols for 1/2, etc.
\usepackage{multirow}
\usepackage{multicol}
\usepackage{graphicx}
\usepackage{subcaption}
\usepackage[numbers]{natbib}
\usepackage{tabto}
\usepackage{xspace}
\usepackage{adjustbox}
\usepackage{enumitem}
\usepackage{wrapfig}
\usepackage{dblfloatfix}
\usepackage{sidecap}

\usepackage{amsmath}
\usepackage{amssymb}
\usepackage{mathtools}
\usepackage{amsthm}
\usepackage{siunitx}
\usepackage{url}            % simple URL typesetting
\usepackage{booktabs}       % professional-quality tables
\usepackage{amsfonts}       % blackboard math symbols
\usepackage{nicefrac}       % compact symbols for 1/2, etc.
\usepackage{microtype}      % microtypography
\usepackage{xcolor}         % colors
\usepackage{colortbl}
\usepackage{graphicx}
\usepackage{amsmath,amsfonts,amssymb,amsthm}
\usepackage{color}
\usepackage{caption}
\usepackage{array}
\usepackage{tikzducks}
\usepackage{pifont}
\usepackage{wrapfig}
\usepackage{enumitem}
\usepackage{algorithm,algorithmic}

\usepackage{xcolor}         % colors

\usepackage{colortbl}

\definecolor{motifred}{rgb}{1.0, 0.41, 0.38}

\definecolor{figureblue}{rgb}{0.1216, 0.4667, 0.7059}
\definecolor{figurered}{rgb}{0.8392, 0.1529, 0.1569}

\definecolor{niceblue}{rgb}{0.1,0.2,0.6}

\usepackage[toc,page,header]{appendix}
\usepackage{minitoc}

\usepackage{float}
\doparttoc                   % Crucial: Enable part-level minitocs
\mtcsetdepth{parttoc}{3}

\usepackage[capitalize,noabbrev]{cleveref}
\usepackage[capitalize]{cleveref}
\crefname{appendix}{App.}{Apps.}
\crefname{section}{Sec.}{Secs.}
\Crefname{section}{Sec.}{Sections}
\Crefname{table}{Tab.}{Tabs.}
\crefname{table}{Tab.}{Tabs.}
\Crefname{figure}{Fig.}{Figs.}
\Crefname{equation}{Eq.}{Eqs.}

\title{Consistent Synthetic Sequences Unlock Structural Diversity in Fully Atomistic De Novo Protein Design }
\author{

Danny Reidenbach\textsuperscript{1*\dag}\quad
Zhonglin Cao\textsuperscript{1*} \quad
Zuobai Zhang\textsuperscript{1,3,4*} \quad
Kieran Didi\textsuperscript{1,2} \quad
Tomas Geffner\textsuperscript{1} \quad
Guoqing Zhou\textsuperscript{1} \quad
Jian Tang\textsuperscript{3,5,6} \quad
Christian Dallago\textsuperscript{1} \quad
Arash Vahdat\textsuperscript{1} \quad
Emine Kucukbenli\textsuperscript{1} \quad
Karsten Kreis\textsuperscript{1} \quad
}
\correspondingauthor{dreidenbach@nvidia.com}

\begin{abstract}
\textbf{Abstract:} 
High-quality training datasets are crucial for the development of effective protein design models, but existing synthetic datasets often include unfavorable sequence-structure pairs, impairing generative model performance. We leverage ProteinMPNN, whose sequences are experimentally favorable as well as amenable to folding, together with structure prediction models to align high-quality synthetic structures with recoverable synthetic sequences. In that way, we create a new dataset designed specifically for training expressive, fully atomistic protein generators. By retraining \laprot, which models discrete residue type and side chain structure in a continuous latent space, on this dataset, we achieve new state-of-the-art results, with improvements of +54\% in structural diversity and +27\% in co-designability. To validate the broad utility of our approach, we further introduce \textit{\method}, a unified flow-based framework that jointly learns the distribution of protein backbone structure, discrete sequences, and atomistic side chains without latent variables. We again find that training on our new sequence-structure data dramatically boosts benchmark performance, improving \method's structural diversity by +73\% and co-designability by +5\%.
Our work highlights the critical importance of aligned sequence-structure data for training high-performance de novo protein design models.
Our new dataset, the \href{https://catalog.ngc.nvidia.com/orgs/nvidia/teams/clara/resources/proteina-atomistica_data/files?version=release}{Consistency Distilled Synthetic Protein Database}, is made available as an open-source resource.
\end{abstract}

\begin{document}

\maketitle
\doparttoc % Tell to minitoc to generate a toc for the parts
\faketableofcontents % Run a fake tableofcontents command for the partocs

\section{Introduction}

De novo protein design aims to generate functional proteins from scratch, making it a central challenge in molecular biology~\citep{richardson1989protein,huang2016protein,kuhlman2019protein,korendovych2020novo}. 
Recent generative models have made impressive progress to design protein backbones using diffusion and flow-based approaches~\citep{ingraham2022chroma,watson2023rfdiffusion,yim2023framediff,bose2024foldflow,lin2023genie1}.
Several methods have begun to move beyond backbone-only modeling to enable all-atom generation~\citep{geffner2025laproteina, chu2024protpardelle, qu2024pallatom}.
% However, a critical limitation of these models is their focus solely on the backbone, neglecting the essential amino acid sequence and side chains. 
Since the sequence serves as the actual design specification for synthesis, and side chains are pivotal in biochemical interactions, generating complete atomistic structures is crucial for structure-guided protein design. As models must reason about the generated sequence and structure to ensure cross consistency, fully atomistic training data plays a crucial role in fully atomistic de novo design.\looseness=-1

\begin{figure*}[t!]
  % \vspace{-0.7cm}
  \begin{minipage}[c]{0.67\textwidth}
  \vspace{-1mm}
  \centering
    \includegraphics[width=\textwidth]{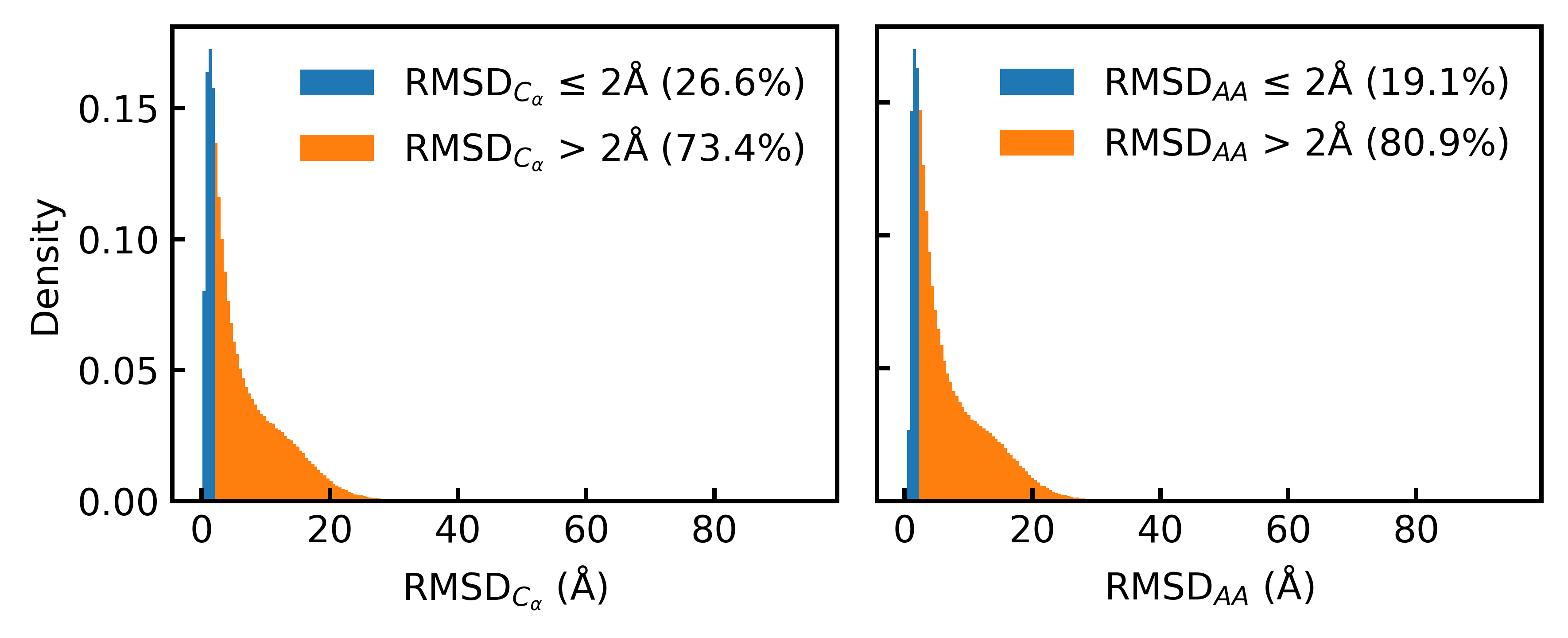}
  \end{minipage}\hfill
  \begin{minipage}[c]{0.31\textwidth}
  \vspace{1mm}
    \caption{\textbf{Co-designability of \geniedata}. Histograms of $C_\alpha$ and all-atom RMSD between AFDB and ESMFold structures show only 26.6\% of protein backbones and 19.1\% of the all-atom structures are considered designable. As a result, most AFDB synthetic structures are not recoverable.\looseness=-1}
    \label{fig:afdb_data_dist}
  \end{minipage}
  % \vspace{-5ex}
\end{figure*}

We identify a critical limitation in commonly used training datasets derived from the AlphaFold Database (AFDB)~\citep{varadi2021afdb}. Specifically, the (real sequence, synthetic structure) pairs in the AFDB are largely not co-designable by ESMFold~\citep{lin2023esm2} (see ~\fref{fig:afdb_data_dist}), AlphaFold2~\citep{jumper2019alphafold2}, or Boltz-1~\citep{wohlwend2024boltz1},\footnote{We used single-sequence mode as well as multiple sequence alignments (MSAs) with different databases, but we were not able to reliably reproduce the AFDB structures.} meaning the sequences do not likely fold into their given structures to the best of available in silico approximations. This is surprising, given that the AFDB was created through computational structure prediction.
% , but we were not able to reproduce the predicted structures via common folding models.
Hence, this data is not well-suited for training joint sequence-structure models, as the data pairs are not consistently reproducible via common folding models.
This motivated us to construct a high-quality dataset from scratch: We leverage ProteinMPNN~\citep{dauparas2022robust}, which is known for strong in silico success and wetlab validation~\citep{watson2023rfdiffusion,pacesa2024bindcraft}, and generate several sequences for each Foldseek AFDB cluster representative structure~\citep{lin2024genie2,barrio2023clustering}. We then re-folded all new synthetic sequences to obtain corresponding fully atomistic structures for the new synthetic sequences. By generating fully atomistic sequence-structure pairs in this manner, we construct a more aligned dataset ideally suited for the training of expressive atomistic protein generators. We will publicly release the dataset.

% Building on this new dataset, we investigate the primary challenge in atomistic protein modeling arises from the dependency of side-chain structures on amino acid identity, \emph{i.e.}, accurate side-chain coordinates cannot be determined without either explicitly or implicitly knowing the sequence. Consequently, many methods rely on a multistage process: backbone generation, sequence prediction, and rotamer-based side-chain packing~\citep{huang2011rosettaremodel,anishchenko2021novo,huang2022backbone, chen2025allatom}.
% Recent efforts take steps toward full-atom co-design by incorporating all-atom representations during structure generation~\cite{chu2024protpardelle,qu2024pallatom,lu2025all,chen2025allatom}. However, none of these methods explicitly model the joint distribution of sequences and atomistic structures in a single, unified model. \laprot~\cite{geffner2025laproteina} was the first to excel in jointly learning sequences and side-chain structure via a continuous latent space, achieving strong performance in de novo design and atomistic motif scaffolding. Continuing along this direction of joint sequence and structure modeling we explore a multi-modal framework that operates in the explicit observable space in contrast to \laprot's latent space.

Next, we used our new dataset to train fully atomistic protein generative models
% Building on our new dataset, we investigate the primary challenge in atomistic protein modeling:
that need to capture the intricate relationship between atomistic structures and amino acid identity. Side-chain coordinates cannot be determined without knowledge of the sequence, either explicitly or implicitly, and co-generating diverse and consistent sequences and atomistic structures is challenging. Therefore, many methods rely on a multistage process: generating the backbone, predicting the sequence, and optionally packing side-chains using rotamers~\citep{huang2011rosettaremodel,anishchenko2021novo,huang2022backbone, chen2025allatom}.
Recent efforts have made progress toward full-atom co-design by incorporating all-atom representations during structure generation~\cite{chu2024protpardelle,qu2024pallatom,lu2025all,chen2025allatom}. However, these methods still do not explicitly model the joint distribution of sequences and atomistic structures in a unified framework. Recently, \laprot~\cite{geffner2025laproteina} jointly learned sequences and side-chain structures via a continuous latent space, achieving strong performance in de novo design and atomistic motif scaffolding. Training \laprot on our new data significantly improves the model samples' structural diversity (+54\%) and co-designability (+27\%), highlighting the importance of well-aligned training data to accurately model the complex sequence-structure relationship.%\looseness=-1

\begin{figure*}[t!]
  % \vspace{-0.7cm}
  \begin{minipage}[c]{0.68\textwidth}
  \vspace{-1mm}
  \centering
    \includegraphics[width=\textwidth]{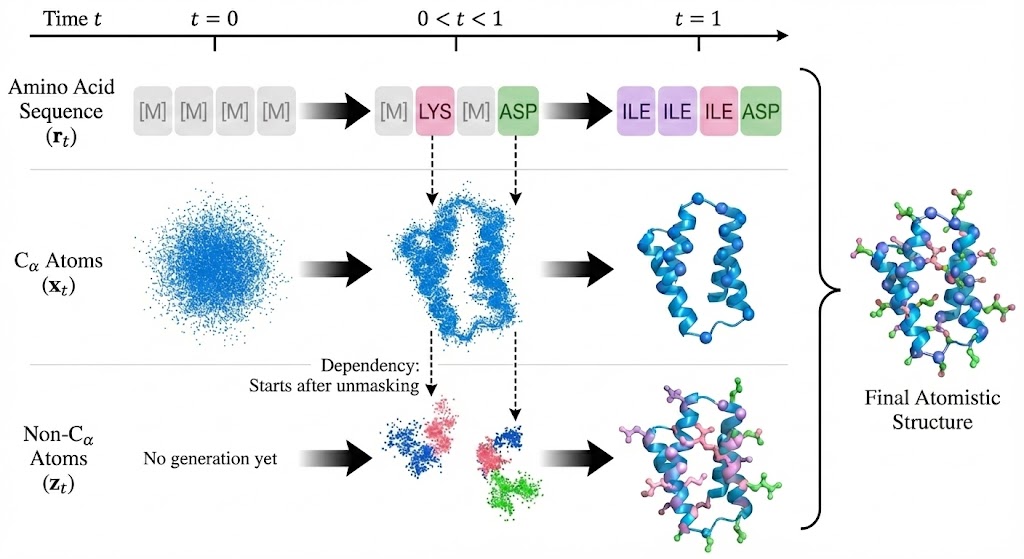} %{figures/pipeline.pdf}
  \end{minipage}\hfill
  \begin{minipage}[c]{0.3\textwidth}
  \vspace{1mm}
    \caption{\small \textbf{\ourmodel.} We use a multimodal flow matching framework to learn a mapping from noise distributions of $C_\alpha$ atoms ($\rvx_t$), amino acid sequences ($\rvr_t$), and non-$C_\alpha$ atoms ($\rvz_t$) to realistic atomistic structures. We prevent leakage by initiating the generation of non-$C_\alpha$ atoms $\rvz_t$ only after their corresponding residues in the sequence $\rvr_t$ are unmasked.}
    \label{fig:overview}
  \end{minipage}
  % \vspace{-5mm}
\end{figure*}
To validate the generality of our approach, we further propose a multi-modal framework that operates in the explicit observable space, providing a complementary approach to \laprot's latent space method.
Specifically, we introduce \textit{\method}, a unified flow-based framework that jointly learns the distribution over fully atomistic protein structure and sequence. We treat this as a joint multi-modal generation task with three co-dependent modalities (\fref{fig:overview}): \textit{(i)} $C_\alpha$ atom positions capture large-scale backbone structure. \textit{(ii)} categorical amino acid identities define the protein sequence. \textit{(iii)} non-$C_\alpha$ backbone and side-chain atoms represent local details.
We again observe that training on our new aligned sequence-structure data dramatically boosts the model's performance---structural diversity by 73\% and co-designability by 5\%. This confirms the broad utility of our newly created, aligned data for training different types of fully atomistic protein generative models.\looseness=-1

% Such explicit modeling strategy enables \ourmodel to reason over different granular levels of protein structure, thus allowing it to capture and expose the intricate explicit relationships between sequence, backbone, and side chain structure.

Our experiments emphasize that consistent synthetic sequences play a significant role in enhancing structural diversity. We also show in ablation studies that simply replacing AFDB structures with those from ESMFold to create a ``100\% designable'' dataset degrades both the ESMFold-based designability and structural diversity of generated proteins. This observation served as a key motivation to leverage ProteinMPNN for predicting new sequences, thereby creating a fundamentally new training dataset that consists of both synthetic structures and synthetic sequences, in contrast to the AFDB. Since our models are directly trained on ProteinMPNN sequences and are the first to surpass ProteinMPNN in co-designability, they remove the need for ProteinMPNN-based re-design at the end of generation—a common step in existing pipelines that requires subsequent side-chain redesign to accommodate changes in sequence space.
% Our studies highlight that the data's structure-sequence consistency significantly enhances model performance for both \ourmodel and \laprot.
% In benchmarks, \ourmodel achieves strong performance on key protein design metrics, including co-designability and structural diversity, while also demonstrating highly accurate side chain conformations.
% Furthermore, we demonstrate the efficacy of our new dataset, which improves the structural diversity of \laprot and \ourmodel by 54\% and 73\%, respectively.

\textbf{Contributions}:
\textit{(i)} We find that AFDB structures are not recoverable with common structure prediction models and argue that the low consistency of AFDB-derived datasets is a critical limiting factor for atomistic structure and sequence co-generation.
\textit{(ii)} To overcome this limitation, we introduce a new high-quality dataset consisting of aligned synthetic sequences and structures, ideally suited for the training of high-performance fully atomistic protein generators.
\textit{(iii)} We introduce \method, a novel unified multi-modal flow-based generative framework that jointly and explicitly models the distribution over fully atomistic protein structures and sequences. %, combining the strengths of discrete and continuous flow-matching methods. 
\textit{(iv)} We show that when trained on our new data \ourmodel outperforms all prior non-unified methods and \laprot achieves new state-of-the-art performance in fully atomistic protein generation.\looseness=-1

\section{Related Work}
% \begin{wrapfigure}{r}{6.2cm}
%   \vspace{-4mm}
%     \includegraphics[width=\linewidth]{figures/afdb_data_codes_dist.png}
%   \vspace{-6mm}
%     \caption{\small \textbf{Co-designability of $\boldsymbol{\mathcal{D}_\mathrm{Genie2}}$}. Histogram of the $C_\alpha$ and all-atom RMSD between AFDB and ESMFold structures.}
%     \label{fig:afdb_data_dist}
%   % \vspace{-4mm}
%   \vspace{-3mm}
% \end{wrapfigure}%

% \begin{figure}[t] % b
% \centering
% \includegraphics[width=\linewidth]{figures/afdb_data_codes_dist.png}
% \caption{\textbf{Co-designability of $\boldsymbol{\mathcal{D}_\mathrm{Genie2}}$}. Histograms of $C_\alpha$ and all-atom RMSD between AFDB and ESMFold structures show only 26.6\% of protein backbones and 19.1\% of the all-atom structures are considered designable. As a result, the majority of the AFDB synthetic structures are not recoverable.\looseness=-1}
% \label{fig:afdb_data_dist}
% \vspace{-2ex}
% \end{figure}

Protein design has witnessed significant progress through generative models focusing on either sequence or structure.
Sequence generation often relies on autoregressive models~\citep{madani2023large,ferruz2022protgpt2} or discrete diffusion~\citep{alamdari2023protein,wang2024diffusion}, trained on large datasets.
For protein backbones, diffusion models have shown remarkable success, with seminal works like Chroma~\citep{ingraham2022chroma} and RFDiffusion~\citep{watson2023rfdiffusion}. 
Subsequent works employ diffusion or flow matching on frame-based representations~\citep{yim2023framediff,bose2024foldflow,yim2023frameflow,wang2024proteus,huguet2024foldflow2}, while other works apply diffusion to $C_\alpha$ coordinates~\citep{lin2023genie1,trippe2023diffusion}. 
Scaling data and model size in Genie2~\citep{lin2024genie2} and \basemodel~\citep{geffner2025proteina} has led to near-perfect backbone designability metrics.
% Recent extensions like FoldFlow2~\citep{huguet2024foldflow2} and Genie2~\citep{lin2024genie2} increased training data with the AlphaFoldDB and the latest state-of-the-art method \basemodel~\citep{geffner2025proteina} scale up the sizes of the training dataset and the model, achieving nearly perfect designability for backbone generation. 
These methods showcase diverse parameterizations and architectures within the broader diffusion/flow matching framework.

However, these single-modality generation methods typically decouple sequence and structure. 
They either generate a sequence first and then fold it with ESMFold~\citep{lin2023esm2} or AlphaFold2~\citep{jumper2019alphafold2}, or generate a structure and then infer a sequence with  ProteinMPNN~\citep{dauparas2022robust}.
In contrast, recent efforts have focused on co-design methods that aim to jointly model sequence and backbone structure distributions within a single generative framework, such as diffusion/flow-based ProteinGenerator~\citep{lisanza2024multistate},  MultiFlow~\citep{yim2024multiflow} and DPLM-2~\citep{wang2025dplm}, energy-based CarbonNovo~\citep{ren2024carbonnovo}, and language model-based ESM3~\citep{hayes2025simulating}. MultiFlow~\citep{yim2024multiflow} also distills synthetic training sequences and structures to boost co-generation performance, similarly to us leveraging ProteinMPNN, but at a smaller scale and without analyses of the AFDB.

Despite progress in protein co-design, achieving accurate atomistic detail remains challenging.
Early all-atom diffusion attempts like Protpardelle~\citep{chu2024protpardelle} yield poor results.
Pallatom's~\citep{qu2024pallatom} use of Atom14 representations could lead to atom-type ambiguities, hindering performance or downstream tasks~\citep{qu2024pallatom}. 
Other methods explore latent spaces~\citep{lu2025all}, modular design~\citep{chen2025allatom}, or specific tasks~\citep{ahern2025atom}.
\vspace{-1ex}

\subsection{\laprot}
More recently, \laprot~\citep{geffner2025laproteina} introduced a partially latent protein representation that combines explicit and implicit modeling. In this approach, the coarse $C_\alpha$-backbone structure is modeled explicitly as in \basemodel, while sequence and atomistic (non-$C_\alpha$) details are captured through per-residue latent variables of fixed dimensionality. This hybrid representation sidesteps the challenges associated with explicit side-chain representations, through the training of an initial autoencoder. By applying flow matching in this partially latent space, \laprot effectively models the joint distribution over sequences and full-atom structures. See paper for details~\citep{geffner2025laproteina}.
We use both \ourmodel and \laprot to explore the impact of synthetic data on all-atom protein generation.\looseness=-1

\section{Aligning Synthetic Protein Sequence and Structure}
\label{sec:data_def}
\vspace{-1mm}
Our investigation into constructing a new training dataset for explicit  all-atom protein generation was motivated by the limitations of the Foldseek-clustered AFDB dataset~\citep{vankempen2024foldseek,barrio2023clustering}, which was used for instance by Genie2~\citep{lin2024genie2} (\geniedata $\sim$0.6M). We assessed the in-silico co-designability of \geniedata by folding its sequences (length$\in$[32,256]) with ESMFold and computing the C$_\alpha$ and all-atom RMSD between the folded and original AFDB structures. Surprisingly, only 19.1\% of the dataset met the standard 2\si{\angstrom} co-designability threshold based on all-atom RMSD (Fig.~\ref{fig:afdb_data_dist}).
Further analysis using other public structure prediction models on a random subset of \geniedata revealed that even the best co-designability achieved, ${\sim}65\%$ with ColabFold using MSAs, fell short of the expected 100\% designability under AlphaFold2 (AF2)~\citep{jumper2019alphafold2, Mirdita2022}. This significant sequence-structure misalignment
% in \geniedata 
poses a substantial challenge to scaling fully atomistic protein generative models with existing sources of large-scale high-quality synthetic sequence-structure data. Furthermore, Boltz-1 obtains scores roughly the same as ESMFold when using MSAs. Without MSAs it exhibited the lowest consistent recovery. Although we do not expect ESMFold and Boltz-1 to be highly consistent with the AFDB, it is crucial to understand the limitations of relying on the AFDB for training protein design models due to the severe disagreement with other popular structure prediction models.\looseness=-1

To address this, we create a novel dataset (\ourdata) which we refer to as the Consistency Distilled Synthetic Protein Database (CDDB) that targets the joint alignment between synthetic sequence and synthetic structure, as follows: (i) For each cluster representative in \geniedata (length between 32 and 512 amino acids) with an average $\text{pLDDT}_{\text{AF2}}$ $\geq$ 0.8, (ii) we produce four sequences with ProteinMPNN, (iii) refold each
% sequence 
recording the $C_\alpha$-RMSD between the AFDB- and ESMFold-generated structures (using $C_\alpha$, as different sequences have different side chains), (iv) select the sequence with the lowest RMSD, and (v) filter the structures to include those with $\text{pLDDT}_{\text{ESMFold}}$ $\geq$ 0.8. This results in $\sim$0.46M high-quality samples.
Consequently, \ourdata identifies confident regions of overlap between folding and inverse folding models, to enabling the modeling of a better recoverable joint sequence-to-structure relationship.
In contrast to MultiFlow~\citep{yim2024multiflow}, which replaces PDB sequences with ProteinMPNN ones, we start from a large structurally diverse dataset and refold to recover side chains.\looseness=-1
% \looseness=-1

\section{\ourmodel}
\label{sec:method}
On the one hand, we use our new data to retrain \laprot~\cite{geffner2025laproteina}, see Sec.~\ref{sec:exps}. To make general conclusions and to also see the data's effect when training a model without a special latent framework, we additionally develop a novel, ``data-space'' fully-atomistic protein generator without latent variables, called \textit{\ourmodel}, which we now introduce.

\subsection{Explicit Multi-Modal Flow Matching Framework}
\label{sec:multi-modal_flow_matching}
Atomistic protein modeling can be decomposed into explicit modeling of the protein backbone, amino acid sequence, and side-chain atoms. 
A significant challenge within this breakdown lies in the modeling of side chains, primarily due to the fact that an amino acid residue and its side-chain structure encode the same underlying information in discrete and continuous forms, respectively. 
Specifically, during a generation process that involves discrete residue tokens, the set of side-chain atoms associated with a residue dynamically changes whenever the residue type is altered or unmasked.
Therefore, a robust atomistic modeling framework must effectively handle this variable number of atoms and also provide a good initialization strategy for these newly generated side-chain atoms (as detailed in Sec.~\ref{sec:model_arch}). This inherent complexity makes extending existing backbone or backbone-sequence design methods to joint fully atomistic modeling non-trivial.

To tackle the challenge posed by the variable number of atoms, we adopt the Atom37 representation for protein structures~\citep{jumper2019alphafold2}. In this representation, each potential heavy atom of a residue is assigned a unique position within a 37-dimensional array. 
This choice offers an advantage over the Atom14 representation used by Pallatom~\citep{qu2024pallatom}, as Atom37 avoids interpretation ambiguities where a single position can correspond to multiple atom types. 
% This clear correspondence is particularly crucial for tasks like atomistic motif scaffolding, where providing explicit and unambiguous atomistic conditions is essential. 
For any non-existent atoms of a given residue, their corresponding positions in the Atom37 array are set to zero and they are subsequently masked out in the model's sequence track (see Sec.~\ref{sec:model_arch}).\looseness=-1

\ourmodel achieves fully atomistic protein generation through multi-modal flow matching over $C_\alpha$ coordinates $\rvx\in\R^{L\times3}$, amino acid sequence $\rvr\in\{0,..,19\}^L$, and non-$C_\alpha$ atom coordinates $\rvz\in\R^{L\times36\times3}$, as illustrated in Fig.~\ref{fig:overview}.
% We deliberately avoid frame-based representations operating on Riemannian manifolds, opting for direct modeling of all atom positions in Euclidean space. Our framework combines modeling categorical sequence data with continuous coordinate data. 
In addition, while both $C_\alpha$ and non-$C_\alpha$ atoms are in Euclidean space, their roles differ: $C_\alpha$ define the global structure and non-$C_\alpha$ specify local residue details. This functional difference, coupled with the variable number of non-$C_\alpha$ atoms, presents a significant challenge in extending backbone and backbone-sequence models to full atomistic generation, a challenge that our multi-modal approach effectively addresses.
We now present the details of the \textit{\ourmodel} modeling framework:

\textbf{1. Flow Matching for $C_\alpha$ Atoms.}
Following \basemodel~\citep{geffner2025proteina}, we define a flow $\psi_t$ that pushes an easy-to-sample noise distribution $p_0$ to a data distribution $p_1$ through intermediate densities $p_t=[\psi]_t*p_0$, where “$*$” denotes push-forward and $t\in[0,1]$ is a time variable. This flow is parameterized by an ODE $d\rvx_t = \rvv^\theta(\rvx_t, t)dt$, defined through a learnable vector field $\rvv^\theta(\rvx_t, t)$ with parameters $\theta$, with $\rvx_0\sim p_0$ and $\rvx_1\sim p_1$. By the continuity equation, the true vector field $\rvu_t$ satisfies $\partial p_t/\partial t = -\nabla_{\rvx_t}\cdot\bigl(p_t\,\rvu_t\bigr)$, but $\rvu_t$ is intractable.
To address this, conditional flow matching (CFM) constructs for each data sample $\rvx_1$ a tractable conditional path $p_t(\rvx_t|\rvx_1)$. We draw $\rvx_0\sim p_0$ and interpolate linearly $\rvx_t = t\rvx_1 + (1-t)\rvx_0$,
so that the exact velocity $\rvx_1 - \rvx_0$ is known. The CFM objective then regresses the learnable field $\rvv^\theta(\rvx_t,t)$ onto this target across random $t$, $\rvx_0$, and $\rvx_1$. At convergence, $\rvv_t^\theta$ approximates the true $\rvu_t$, enabling generation of $C_\alpha$ coordinates. 

\textbf{2. Flow Matching for Amino-Acid Sequence.}
The flow matching framework for amino acid sequences operates in the discrete space of residue types $\{0,.., 19\}$. Following MultiFlow~\citep{yim2024multiflow}, we introduce a mask token $\mask$ and define the flow to push an all-mask prior $p_0=\delta\{\mask\}$ 
% \kk{Not sure I'd call this a ``noise'' distribution, maybe just an ``all-mask prior''?} 
toward the target sequence distribution $p_1 = \delta\{\rvr_1\}$, where $\delta\{i\}$ denotes the Kronecker delta (\emph{i.e.}, a one-hot distribution centered at token $i$).
To learn the "velocity", \emph{i.e.} the rate matrix in probability space, we define a conditional path $p_{t}(\rvr_t|\rvr_1) = t\,\delta\{\rvr_1\} + (1-t)\,\delta\{\mask\}$.
This path interpolates between the masked and target sequences. In practice, it corresponds to a simple stochastic masking scheme: each residue is independently masked with probability $1 - t$ and kept with probability $t$.

\textbf{3. Flow Matching for Non-$C_\alpha$ Atoms.}
We adopt the same flow matching formulation used for $C_\alpha$ atoms. Specifically, we define a linear interpolant $\rvz_t = t\rvz_1 + (1 - t)\rvz_0$, and train the parameterized velocity field $\mathbf{v}^\theta(\rvz_t, t)$ to match the exact velocity $\rvz_1 - \rvz_0$.
There are two key differences with the $C_\alpha$ case. First, as each residue contains only a subset of the 36 possible non-$C_\alpha$ atoms determined by its residue type, we mask out non-existent atoms during interpolation. Second, revealing the presence or absence of specific atoms may leak residue type information for masked positions in the sequence, making the sequence denoising task trivial.
To prevent this, we remove all non-$C_\alpha$ atoms for residues masked in $\rvr_t$ during training.
During generation, to align with training, we only denoise non-$C_\alpha$ atoms once its residues are unmasked. Therefore, it is crucial to provide a good initialization for the non-$C_\alpha$ coordinates when a residue is unmasked—an issue we discuss in the following sections.

% \kk{Can we maybe number the three previous paragraphs? These are the three modalities. We can make this very explicit, so it reads and flows nicely.}

\textbf{Local Coordinate Modeling for Non-$C_\alpha$ Atoms.}
% \dr{motif is local trans, base is local frame. So can soften this to say we can do both and how}
Non-$C_\alpha$ atoms are structurally organized around their corresponding $C_\alpha$ atoms. 
To leverage this property, we offer two local coordinate modeling strategies, simplifying the learning task by predicting offsets rather than global coordinates and facilitating better initialization of non-$C_\alpha$ atoms.
The first approach calculates the relative position of non-$C_\alpha$ atoms directly with respect to their corresponding $C_\alpha$ atom: $\rvz_i^\text{local} = \rvz_i - \rvx_i$.
The second strategy, inspired by related work~\citep{lin2024genie2}, constructs a residue-centric local coordinate frame $(\rvt_i, \mathbf{R}_i)$ with frame translations $\rvt_i$ and frame rotations $\mathbf{R}_i$ using the $C_\alpha$ coordinates of three neighboring residues ($\rvx_{i-1}$, $\rvx_i$, $\rvx_{i+1}$) via the Gram-Schmidt process. Non-$C_\alpha$ coordinates $\rvz_i$ are then transformed to local ones via $\rvz_i^\text{local\_frame} = \mathbf{R}_i^{-1}(\rvz_i - \rvt_i)$. 
Notably, while local coordinate transformations are a common technique in structure prediction models~\citep{jumper2019alphafold2,lin2023esm2}, their application in atomistic structure generation remains underexplored~\citep{costa2023ophiuchus}.
% \kk{This is also kind of novel and cool. It may be too late for that, but this could be nicely visualized in a little side figure. Also, do we have ablations over this? Emphasizing that this is critical for performance would be nice, but we'd have to back it up. These things also don't appear in MultiFlow, because it's just c-alpha, so a good opportunity to highlight unique complexities here, too.}
% \dr{we have ablations from my runs local better than frame better than global}

% \input{figures_tex/overview_figure}
% \vspace{-3ex}
\begin{figure}[t]
% \vspace{-8mm}
    \centering
    \includegraphics[width=\linewidth]{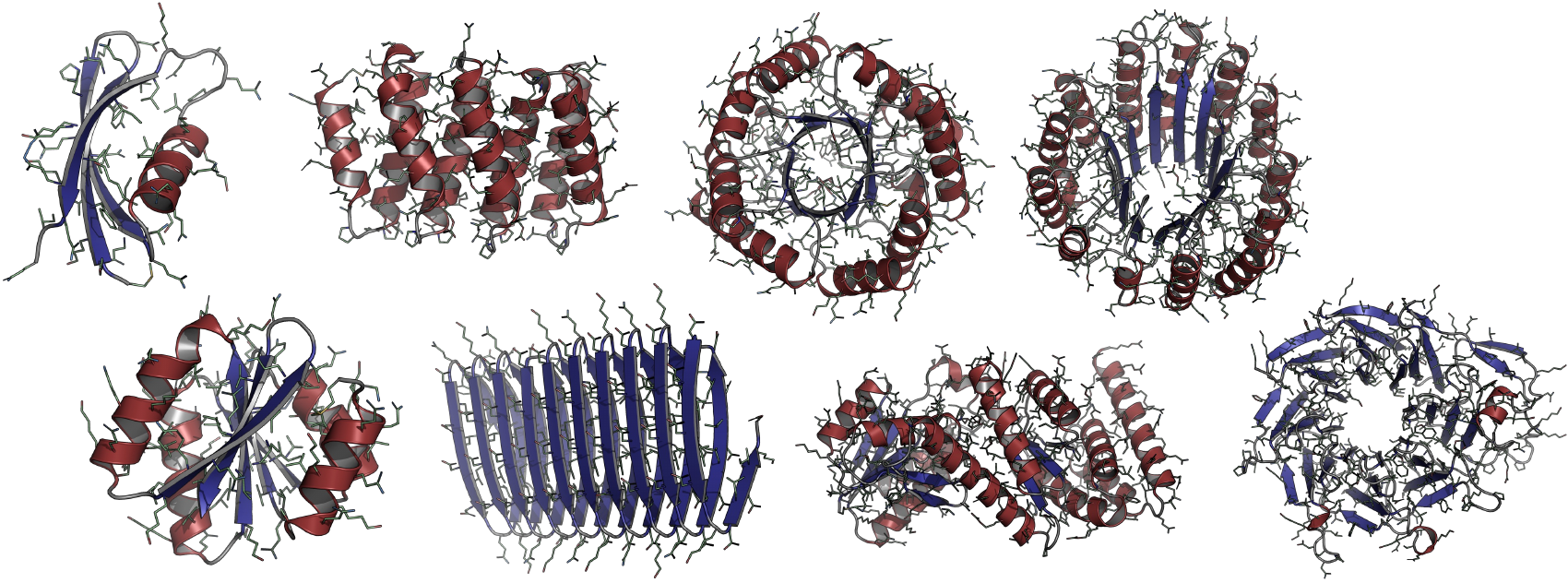}
    \vspace{-4mm}
    \caption{\small \textbf{\ourmodel samples}, ranging from 100 to 400 residues. All shown samples co-designable.
    }
    \label{fig:len400_samples}
    % \vspace{-5mm}
\end{figure}
\vspace{-1mm}
\subsection{\ourmodel Architecture}
\vspace{-1mm}
\label{sec:model_arch}

The \ourmodel architecture consists of two primary components: a core residue-level Transformer trunk and an atom-level Transformer encoder-decoder (\fref{fig:model_arch}). The residue-level trunk is responsible for the global backbone processing and is a high-capacity, non-equivariant architecture that leverages a stack of biased self-attention layers to predict the vector field for flow-based generation from noisy inputs. To address the complexities of atomistic modeling, our atom-level Transformer modules are designed to tackle three key challenges: handling the variable number of atoms, generalizing to atomistic representations, and initializing the fully masked non-$C_\alpha$ atoms of masked residues. We elaborate on our approach to these challenges in the subsequent paragraphs.

% For model parameterization, we extend the \basemodel Transformer to the atomistic level, as shown in Fig.~\ref{fig:model_arch}. Originally designed for generating protein $C_\alpha$ coordinates, the \basemodel Transformer is a high-capacity, non-equivariant architecture that processes noisy inputs through a stack of biased self-attention layers to predict the vector field for flow-based generation. We adapt this architecture to jointly model $C_\alpha$, non-$C_\alpha$, and sequence modalities.
% There are three main challenges in this extension: handling the variable number of atoms, generalizing the architecture to atomistic representations, and predicting a good initialization for the fully masked non-$C_\alpha$ atoms of masked residues. We address these challenges in the following paragraphs.
  
\textbf{Atom Sequence Expansion.}
Each residue does not possess all 36 possible non-$C_\alpha$ atoms, resulting in empty dimensions in $\rvz \in \R^{L \times 36 \times 3}$ that cannot be directly featurized. To address this, we expand the Atom37 representations into an atom sequence containing only existing atoms, following a default atom order.
For masked residues, where all non-$C_\alpha$ atoms are absent, we represent them with a pseudo-atom token $[\mask]$ in the atom sequence as a special atom type and set its coordinate to zero, and residue type to a mask token.
For instance:\looseness=-1
\begin{align*}
    \underbrace{\text{[ALA]-[\textcolor{red}{\mask}]-[CYS]}}_{\text{amino-acid sequence}} \Longrightarrow \underbrace{\text{[N-CA-C-O-CB]-[CA-\textcolor{red}{\mask}]-[N-CA-C-O-CB-SG]}}_{\text{atom sequence}}
\end{align*}
We then expand all associated residue-level features to match the atom sequence, allowing us to treat the atom sequence similarly to the residue sequence and reuse architectural modules. The Transformer's ability to handle variable-length inputs resolves the varying atom number problem.\looseness=-1

\textbf{Atom-Level Encoding and Decoding.}
Inspired by AlphaFold3~\citep{abramson2024alphafold3}, we encode atom-level information using atom encoders followed by a cross-attention layer that integrates residue and atom features before the main backbone processing trunk.
We note unlike prior methods our models do not use any triangle update layers.
% We simplify \basemodel's backbone processing trunk by removing triangle update layers.
After this trunk, another cross-attention layer updates both representations, followed by atom decoders for further atom-level refinement~\citep{abramson2024alphafold3}.
At the output stage, residue-level representations are used to predict $C_\alpha$ vector field $\rvv_{\rvx,t}^{\theta}$ and the residue type probability logits $\smash{\rvc_{1|t}^{\theta}}$, while atom-level representations are used to predict the non-$C_\alpha$ vector field $\smash{\rvv_{\rvz,t}^{\theta}}$.

\textbf{Initialization Prediction for Masked Residues.}
To predict the structure of non-$C_\alpha$ atoms of masked residues, we introduce a prediction head that leverages the pseudo-atom token's learned representation and context from neighboring atoms and residues. 
% TODO comment that this becomes more clear in less noise due to the contxt via self attention in atom level and cross attention to the residue level. input only CA features
Our initial experiments revealed that, as expected, directly predicting the clean coordinates $\rvz$ is challenging as the number of atoms to predict is unknown. 
To address this, we propose learning an initialization $\rvz_{\text{init},t}^{\theta}$ through an augmented objective. 
We refer to this as an initialization as it is used only when the residue transitions from a masked to a non-mask state (see Alg.~\ref{alg:sample}).
During training, this initialization head is regressed towards $\rvz - \boldsymbol{\epsilon}_{\rvz}$, where $\boldsymbol{\epsilon}_{\rvz}$ is a randomly sampled Gaussian noise vector. Notably, the standard conditional flow matching objective relies on learning a vector field conditioned on noisy inputs; however, for side-chain initialization, there is no noisy input available, as the residue type is unknown. As a result, the model effectively learns to predict the expected clean state $\rvz$, representing an average side-chain structure across the 20 possible residue types. This initialization is refined into a realistic atomistic structure in the remaining denoising iterations during inference. 
Note that the initialization becomes easier to learn as the denoising process progresses, as more context is available and the remaining structure is less noisy, aligning with our choice of schedules for each explicit modality (\fref{fig:sample_t}).
This approach also aligns the magnitude of the training target with that of vector fields, facilitating the training process. At generation time, this initialization serves as a reasonable approximation for the initial structure of non-$C_\alpha$ atoms in initially masked residues.
% TODO the inits become more clean further into the denoising process 
%  in contrast to the silultaneously prediction future work
% refer to the acutal ALGO 
See Appendix~\sref{app:side_chain_init} for more details.\looseness=-1

\vspace{-1mm}

\begin{table}[b]
% \vspace{-6mm}
\centering
\caption{\small \textbf{\ourmodel and \laprot de novo fully atomistic protein generation performance} when trained on $\mathcal{D}_\mathrm{SYN-ours}$ compared to baselines. All models generate 100 proteins for lengths $\in$ [50, 400] with step size 50. We report multimodal sampling configurations that generate the (i) most all-atom co-designable (codes), (ii) most diverse samples (div), and (iii) an optimal trade-off (opt). The best values are bolded.}
\scalebox{0.79}{
\begin{tabular}{lccccc}
\toprule
Method & CODES-AA (\%)$\,\uparrow$ & DES-M1 (\%)$\,\uparrow$ & DIV-AA$\,\uparrow$ & NOV-PDB-AA$\,\downarrow$ & NOV-AFDB-AA$\,\downarrow$ \\
\midrule
ProteinGenerator  & 10.0 & 57.1 & 28 & 0.75 & 0.78 \\ 
Protpardelle  & 13.6 & 62.8 & 25 & 0.74 & 0.76 \\ 
PLAID  & 22.3 & 34.9 & 63 & 0.85 & 0.88 \\
Pallatom  & 51.6 & 62.5 & 282 & \textbf{0.66} & \textbf{0.71} \\
\laprot (\geniedata)  & 70.6 & 85.5 & 314 & 0.77 & 0.84 \\
\midrule
% \ourmodel$_\mathrm{opt}$ & 83.1 & 85.8 & 321 & x & x \\
\ourmodel$_\mathrm{codes}$ & 87.8 & 88.1 & 263 & 0.77 & 0.81 \\ 
\ourmodel$_\mathrm{opt}$ & 83.1 & 85.8 & 321 & 0.76 & 0.80 \\
\ourmodel$_\mathrm{div}$ & 71.6 & 72.0 & 333 & 0.75 & 0.80 \\ 
% \midrule
\laprot\xspace$_{\mathrm{codes}, \mathcal{D}_\mathrm{SYN-ours}}$ & \textbf{92.6} & \textbf{92.5}& 418 & 0.75 & 0.83  \\
% laproteina_160M_codesdata512_ae512scratch_470kema_4nbs6-inference400-2025_08_06_13_59_34 ROW 9
\laprot\xspace$_{\mathrm{div}, \mathcal{D}_\mathrm{SYN-ours}}$ & 87.8 & 87.4 & \textbf{475} & 0.74 & 0.82  \\
% laproteina_160M_codesdata512_ae512scratch_470kema_4nbs6-inference400-2025_08_06_13_59_34 ROW 4
\bottomrule
\end{tabular}
}
% \caption{\textbf{Performance of \ourmodel on de novo all atom generation} compared to baselines. All models generate 100 proteins for lengths $\in$ [50, 400] with setup size 50. We report the three multimodal sampling configurations that generate the (i) most all atom codesignable (codes), (ii) most diverse samples (div), and (iii) an optimal tradeoff. The best values are bolded.}
\label{tab:aa_400_results}
% \vspace{-3mm}
\end{table}
\vspace{-2mm}
\section{Experiments}\label{sec:exps}
\vspace{-1mm}
\label{sec:experiment}
We trained two 200M parameter unconditional \ourmodel models,  for lengths (i) 32-400 and (ii) 32-256 using local coordinates without frames to align with prior baselines~\citep{qu2024pallatom} (alternative coordinate modeling schemes are ablated in~\tref{tab:model_hparams}).
% We also train an 80M parameter all-atom motif conditional model (60M for the $C_{\alpha}$ backbone and 20M for the atom sequence). Architecture and sampling details in Appendix. %~\sref{app:sec:model_arch}.
For \laprot we train an autoencoder from scratch and a subsequent flow matching model according to the procedure described in \laprot~\citep{geffner2025laproteina} for lengths 32-500. The only difference between the original and our \laprot is the change of training data. We emphasize that data is a critical hyperparameter in all prior de novo protein design methods. While we show \ourmodel and \laprot to be state-of-the-art performers in Sec.~\ref{sec:exp1}, we also analyze the impact of \ourdata specifically, in Sec.~\ref{sec:exp2}. As all included baselines leverage different datasets or combinations of AFDB/PDB/UniRef/etc., we intend for the public release of \ourdata to offer another alternative that can be leveraged for its synthetic consistency. We further ablate our new explicit data-space method by comparing against recent backbone-only and backbone-sequence (no side chain) models in Appendix Tables~\ref{tab:unconditional_backbone_generation}-\ref{tab:arch_codes_abl}, including a no-side-chain version of \ourmodel itself (see Appendix Sec.~\ref{app:proteina-co-des}).

We evaluate our models using standard de novo protein design metrics, extending them to backbone-sequence co-design and all-atom (AA) contexts, following prior work~\citep{geffner2025proteina, qu2024pallatom, yim2024multiflow}.
De novo success metrics include \textbf{Designability (DES)}, the ability to inverse fold the generated protein backbone with ProteinMPNN and refold the generated sequences~\citep{watson2023rfdiffusion}, with variants DES-M1 (single-shot) and DES-M8 (standard for backbone-only; best of 8 sequences); \textbf{Co-designability (CODES)}, similar to DES-M1 but using the model's output sequence; and \textbf{All-Atom Co-designability (CODES-AA)}, an extension of CODES using all-atom scRMSD. CODES and CODES-AA are reported for models that produce backbone and sequence, and atomistic side-chain structures, respectively.
We also report structural \textbf{Diversity} and \textbf{Novelty} of the (co-)designable samples, for $C_{\alpha}$ design (M8 and M1), backbone-sequence co-design, and all-atom contexts. For metric details see Geffner et al.~\cite{geffner2025proteina}.

\begin{figure*}[t!]
    % \vspace{-9mm}
    \begin{minipage}[c]{0.72\textwidth}
    \centering
    \includegraphics[width=0.73\linewidth]{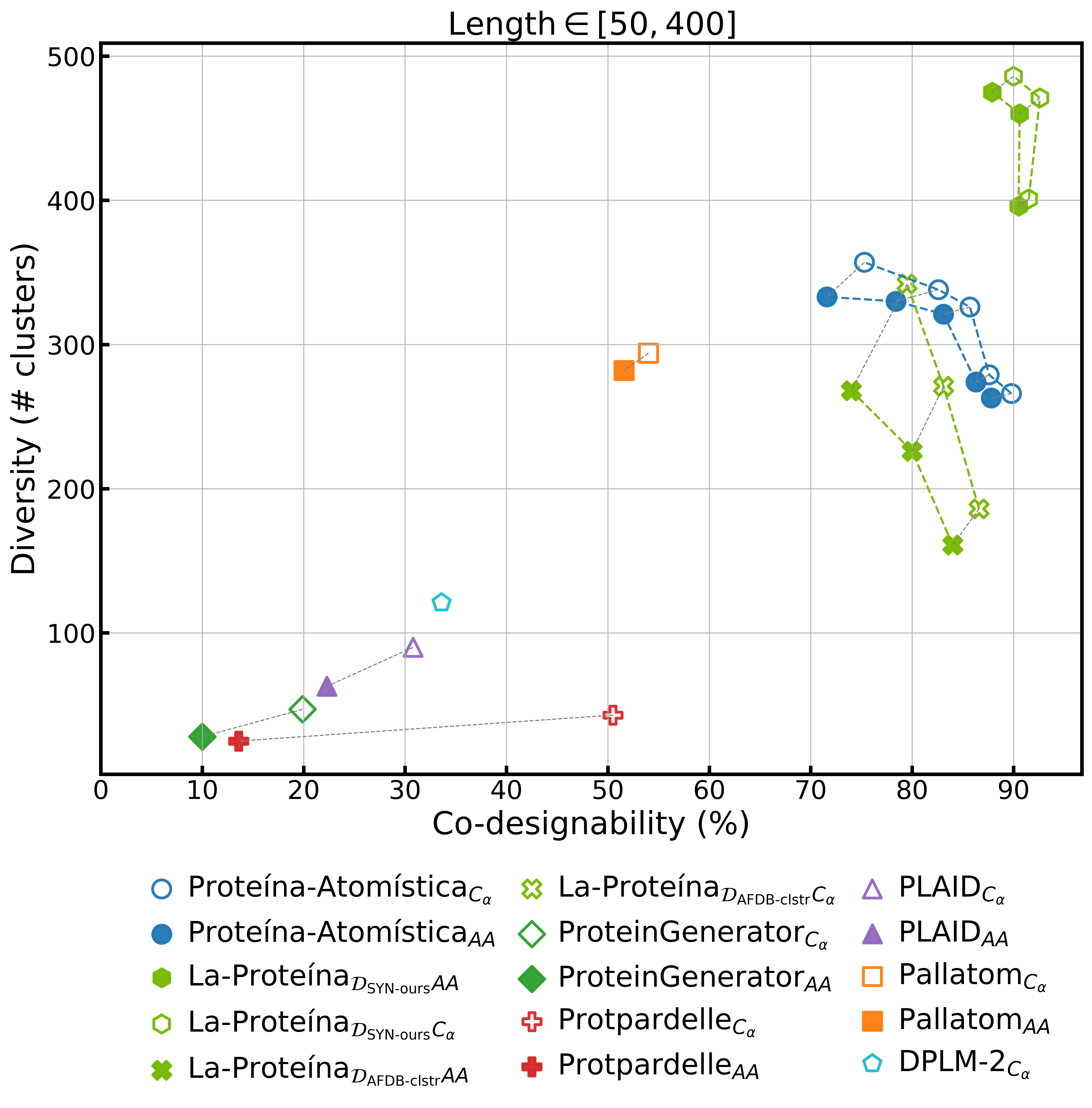}
    \end{minipage}\hfill
    \begin{minipage}[c]{0.28\textwidth}
    % \vspace{-1mm}
    \caption{\small \textbf{Pareto frontier of the co-designability-diversity trade-off of \ourmodel and \laprot} for proteins with length $\in [50, 400]$. Solid and hollow markers represent metrics calculated on all-atom and $C_{\alpha}$ basis, respectively. For atomistic models, the all-atom and $C_{\alpha}$ scores for the same generated proteins are connected by gray dashed line.   
    % \kk{Put the legend into a white box, so this doesn't overlap with the grid and doesn't look as messy. Also, the legend is really tiny -- method subscripts are important but almost invisible. Let's improve this.} \kk{also here, write that for the atomistic model, we eval both atomstic and backbone codes, and connect the values by line. The reader may not realize by just looking at the plot and the tiny legends what exactly is going on here. Hence, explain.} \dr{just realized that the num clusters here is not exactly comparable since it depends on the numbers of samples taken. Should we have the both plots have their own ranges to be clear that you cannot compare across plots}
    }
    \label{fig:pareto}
    \end{minipage}
    % \vspace{-7mm}
\end{figure*}

% \begin{figure*}[t!]
%   \vspace{-0.7cm}
%   \begin{minipage}[c]{0.68\textwidth}
%   \vspace{-1mm}
%   \centering
%     \includegraphics[width=\textwidth]{figures/pipeline.pdf}
%   \end{minipage}\hfill
%   \begin{minipage}[c]{0.3\textwidth}
%   \vspace{1mm}
%     \caption{\small \textbf{\ourmodel.} We use a multimodal flow matching framework to learn a mapping from noise distributions of $C_\alpha$ atoms ($\rvx_t$), amino acid sequences ($\rvr_t$), and non-$C_\alpha$ atoms ($\rvz_t$) to realistic atomistic structures. We prevent leakage by initiating the generation of non-$C_\alpha$ atoms $\rvz_t$ only after their corresponding residues in the sequence $\rvr_t$ are unmasked.}
%     \label{fig:overview}
%   \end{minipage}
%   \vspace{-4mm}
% \end{figure*}
\vspace{-1mm}
\subsection{De Novo All-Atom Protein Generation}
\label{sec:exp1}
\vspace{-1mm}
In~\tref{tab:aa_400_results}, we compare \ourmodel and \laprot trained on \ourdata to recent fully atomistic generative models. Using multimodal low temperature sampling, both \ourmodel and \laprot leverage the known trade-off~\citep{geffner2025proteina,lin2024genie2} between designability and diversity. We also plot the Pareto frontier for both all-atom and backbone-only co-designability in~\fref{fig:pareto}

Notably, \ourmodel generates highly designable and diverse structures (Fig.~\ref{fig:len400_samples}) while achieving competitive novelty scores, indicating that our model does not overfit to PDB or AFDB. %, based on whose cluster representatives our training data is constructed.
% \ourmodel samples are shown in~\fref{fig:len400_samples}.
These improvements are further surpassed by \laprot when trained on our \ourdata, which obtains state-of-the-art performance with all-atom co-designability of 87.8\% and 475 clusters when steered towards structural diversity via low temperature sampling. Furthermore, both \basemodel-based models on average generate 66-70\% $\alpha$-helices and 6-10\% $\beta$-sheets.  We further demonstrate that both \ourmodel and \laprot obtain comparable geometric side chain accuracy metrics in Appendix~\fref{fig:sc_eval}.
% The impact of synthetic consistency up to length 400 can be seen in the comparison of \laprot with \geniedata and \ourdata. We see a best improved all-atom co-designability of 31\% and diversity of 51\%, respectively. In ~\sref{sec:exp2} we show that these gains are model specific and that while still overall better, different models are impacted by the synthetic data consistency in different ways.
The impact of synthetic consistency up to length 400 is evident in the comparison of \laprot with \geniedata and \ourdata, where we observe a best-case improvement in all-atom co-designability of 31\% and diversity of 51\%, respectively, establish new state-of-the-art results. We further discuss the generalization of performance gains due to synthetic consistent data in Sec.~\ref{sec:exp2}.\looseness=-1

\vspace{-1mm}
\subsection{Understanding the Impact of Synthetic Data}
\label{sec:exp2}
\vspace{-1mm}
\begin{table}[t]
% \vspace{-7mm}
% \vspace{-4ex}
\centering
\caption{\textbf{Impact of Synthetic Data.} All models generate 100 proteins for lengths $\in$ [50, 100, 150, 200, 250]. 
Training on ESMFold structures or filtering for ESMFold designability hurts performance unless those synthetic ESMFold structures are coupled with recoverable sequences.}
\scalebox{0.85}{
\begin{tabular}{lccc}
    \toprule
    \textbf{Model} & CODES-AA (\%)$\,\uparrow$ & DES-M1 (\%)$\,\uparrow$ & DIV-AA$\,\uparrow$ \\
    \midrule
    \ourmodel$_{\mathcal{D}_\mathrm{AFDB-clstr}}$ & 76.8 & 87.6 & 154 \\
    \ourmodel$_{\mathcal{D}_\mathrm{ESMFold}}$ & 71.0 & 86.0 & 132 \\ 
    \ourmodel$_{\mathcal{D}_\mathrm{Des}}$ & 72.2 & 87.2 & 120 \\ 
    \laprot$_{\mathcal{D}_\mathrm{AFDB-clstr}}$ & 81.0 & 89.8 & 213 \\ 
    \midrule
    \ourmodel$_{\mathcal{D}_\mathrm{SYN-ours}}$ & 81.2 & 82.4 & 267\\
    \laprot$_{\mathcal{D}_\mathrm{SYN-ours}}$ & \textbf{92.2} & \textbf{93.2} & \textbf{283} \\ 
    \bottomrule
\end{tabular}
% \vspace{-7mm}

}
% \vspace{-1em}

\label{tab:training_data_choice}

\end{table}

% \begin{wraptable}{r}{7.9cm}
% \vspace{-5mm}
%         \caption{\small \textbf{Synthetic Data Ablation.} Performance of \ourmodel trained on different datasets. Numbers are calculated over 500 proteins (100 for length $\in$ [50, 100, 150, 200, 250]).
%         }
%     \label{tab:training_data_choice}
%     % \vspace{-5mm}
% % \end{minipage}\hfill
% % \begin{minipage}[c]{0.74\textwidth}
%     \centering
%     \scalebox{0.65}{
%     % \rowcolors{2}{gray!15}{white}
%     \begin{tabular}{l c c c}
    % \toprule
    % \textbf{Model} & CODES-AA (\%)$\,\uparrow$ & DES-M1 (\%)$\,\uparrow$ & DIV-AA$\,\uparrow$ \\
    % \midrule
    % \ourmodel$_{\mathcal{D}_\mathrm{Genie2}}$ & 76.8 & \textbf{87.6} & 154 \\
    % \ourmodel$_{\mathcal{D}_\mathrm{ESMFold}}$ & 71.0 & 86.0 & 132 \\ 
    % \ourmodel$_{\mathcal{D}_\mathrm{Des}}$ & 72.2 & 87.2 & 120 \\ 
    % \laprot$_{\mathcal{D}_\mathrm{Genie2}}$ & - & - & - \\ 
    % \midrule
    % \ourmodel$_{\mathcal{D}_\mathrm{codes}}$ & 81.2 & 82.4 & 267\\
    % \laprot$_{\mathcal{D}_\mathrm{codes}}$ & \textbf{92.2} & \textbf{93.2} & \textbf{283} \\ 
    % \bottomrule
%     \end{tabular}
%     }
%     \vspace{-4mm}
% \end{wraptable}%
To demonstrate that \geniedata is challenging for facilitating joint learning of sequence and structure, we further investigated the impact of synthetic data. % on the contributions of structure and sequence. 
To this end, we constructed two further synthetic datasets based on \geniedata: (1) $\mathcal{D}_\mathrm{ESMFold}$ and (2) $\mathcal{D}_\mathrm{des}$. In $\mathcal{D}_\mathrm{ESMFold}$, samples have the same sequences as in \geniedata but the structures are computed by ESMFold with a filter of pLDDT $\geq$ 0.8. $\mathcal{D}_\mathrm{Des}$ is a subset of \geniedata (uses direct AFDB structures) with all structures passing the DES-M8 filter. Both $\mathcal{D}_\mathrm{ESMFold}$ and $\mathcal{D}_\mathrm{Des}$ contain ${\sim}0.16$M samples.

\tref{tab:training_data_choice} demonstrates that, counterintuitively, neither using 100\% designable structures $\mathcal{D}_\mathrm{ESMFold}$ for training nor leveraging the designable subset $\mathcal{D}_\mathrm{Des}$ improves the performance of the model, even when the goal is to generate designable and diverse structures. As a side, \tref{tab:training_data_choice} also confirms that the new \ourmodel architecture trained on \ourdata, which combines AFDB's structural diversity with ProteinMPNN sequences (subsequently refolded with ESMFold to recover consistent full atomistic detail), achieves highly accurate and diverse fully atomistic generation (see~\sref{sec:data_def} for \ourdata procedure). This highlights the importance of utilizing better-aligned synthetic sequences \emph{and} structures to facilitate scalable co-design over both modalities. Furthermore by training on \ourdata \laprot sees co-designability and diversity improvements of 13.8\% and 32.9\%.

% \vspace{-1mm}
\subsection{Latent vs. Explicit Modeling of Protein Sequences}
\tref{tab:training_data_choice} shows that \laprot's latent approach better learns aligned sequence-structure co-generation compared to \ourmodel in particular when trained on \geniedata. We found that this is due to lower co-designability at longer lengths, also implying lower diversity scores (diversity is calculated among designable samples only).
\laprot's autoencoder bypasses the challenge of aligning explicit, discrete, and continuous modalities, generating more diverse and co-designable samples.
The latent variable framework avoids minimizing a complex joint continuous and discrete objective during the generative model training. Moreover, \laprot's autoencoder component effectively learns to tie together consistent sequences and structures rather than trying to learn how to explicitly match them through separate modality-based objectives. Although learning the structure-to-sequence mapping in the explicit data space is more challenging, \ourmodel establishes a strong alternative for future work that relies on direct access to explicit observables.
% While the explicit access to all data modalties during training and a generation

Switching to \ourdata significantly improves \ourmodel's co-designability, dramatically boosts diversity, and yields competitive results with \laprot. The model now learns a more empirically recoverable sequence distribution from its structures (\fref{fig:seq_dist}).
%, likely due to its explicit discrete variable conditioned on the continuous structure.
Notably, both \laprot and \ourmodel show significant improvements on \ourdata, generating more co-designable and diverse samples when compared to training on \geniedata. Furthermore, sequences generated from our models trained on \ourdata fold better into their co-generated structures than those from ProteinMPNN (Appendix~\tref{tab:codesign_results_250}; see $\geq$DES-M1). This alleviates the need for ProteinMPNN re-design of generated backbones, a common component in design pipelines.\looseness=-1

\textit{Please see our Appendix for ablation studies, experiment, dataset, and model architecture details.}
% \vspace{-1ex}
% \tref{tab:training_data_choice} highlights a limitation of \ourmodel's explicit data space modeling approach, contrasting with \laprot's latent approach. On \geniedata, \ourmodel struggles to learn the sequence-structure mapping, resulting in reduced diversity scores due to having lower co-designability at longer lengths.
% In contrast, \laprot's robust autoencoder, trained on data pairs, bypasses the challenge of aligning explicit discrete and continuous modalities, generating more diverse and co-designable samples at longer lengths. This is because the autoencoder avoids learning a joint continuous and discrete objective over the entire protein during the generative model training.

% Switching to \ourdata, our consistent dataset, resolves \ourmodel's co-designability drop-off issue and yields competitive results with \laprot. \ourmodel learns a better sequence distribution, empirically recoverable from its structures, likely due to its explicit discrete variable conditioned on the continuous structure.
% Notably, both \laprot and \ourmodel show significant improvements on \ourdata, generating more co-designable and diverse samples. Furthermore, sequences generated with \ourdata fold better into their structures than those from ProteinMPNN (see Appendix~\tref{tab:codesign_results_250}).
% \looseness=-1
% \vspace{-1mm}
% \input{tables_tex/syn_data_choice}
% \vspace{-1mm}
\vspace{-1mm}
\section{Conclusions}
\vspace{-2mm}
Our study finds that AFDB structures are not recoverable with publicly available protein structure predictions models, which motivated us to create a carefully curated, yet diverse dataset of aligned sequences and structures. We also introduce and successfully validate \ourmodel, a new unified multi-modal flow-based framework for de novo atomistic protein design that represents sequence, backbone, and side chains explicitly, without latent variables. Training both \ourmodel and \laprot on \ourdata dramatically improves their performance, achieving new state-of-the-art results. This demonstrates the critical importance of consistent and recoverable sequence-structure training data for atomistic protein design. Future work could address the consistent generation of longer atomistic proteins and analyze the importance of aligned sequence-structure data in the context of conditional tasks such as motif scaffolding and binder design.

\newpage
% %%%%%%%%%%%%%%%%%%%%%%%%%%%%%%%%%%%%%%%%%%%%%%%%%%%%%%%%%%%%
\bibliographystyle{plainnat}
\bibliography{reference}

@article{geffner2025laproteina,
    title={La-Proteina: Atomistic Protein Generation via Partially Latent Flow Matching},
    author={Geffner, Tomas and Didi, Kieran and Cao, Zhonglin and Reidenbach, Danny and Zhang, Zuobai and Dallago, Christian and Kucukbenli, Emine and Kreis, Karsten and Vahdat, Arash},
    journal={arXiv preprint arXiv:2507.09466},
    year={2025}
}

@article{wohlwend2024boltz1,
  author = {Wohlwend, Jeremy and Corso, Gabriele and Passaro, Saro and Getz, Noah and Reveiz, Mateo and Leidal, Ken and Swiderski, Wojtek and Atkinson, Liam and Portnoi, Tally and Chinn, Itamar and Silterra, Jacob and Jaakkola, Tommi and Barzilay, Regina},
  title = {Boltz-1: Democratizing Biomolecular Interaction Modeling},
  year = {2024},
  doi = {10.1101/2024.11.19.624167},
  journal = {bioRxiv}
}

@article{richardson1989protein,
title = {The de novo design of protein structures},
journal = {Trends in Biochemical Sciences},
volume = {14},
number = {7},
pages = {304-309},
year = {1989},
author = {Janes S. Richardson and David C. Richardson}
}

@article{kuhlman2019protein,
	author = {Brian Kuhlman and Philip Bradley },
	title = {Advances in protein structure prediction and design},
	year = {2019},
	journal = {Nat. Rev. Mol. Cell Biol.},
        volume = {20},
        pages = {681–697}
}

@article{huang2016protein,
	author = {Po-Ssu Huang and Scott E. Boyken and David Baker},
	title = {The coming of age of de novo protein design},
	year = {2016},
	journal = {Nature},
        volume = {537},
        pages = {320–327}
}

@article{Mirdita2022,
  author = {Mirdita, Milot and Schütze, Konstantin and Moriwaki, Yoshitaka and Heo, Lim and Ovchinnikov, Sergey and Steinegger, Martin},
  title = {ColabFold: making protein folding accessible to all},
  journal = {Nature Methods},
  volume = {19},
  number = {6},
  pages = {679-682},
  year = {2022},
  doi = {10.1038/s41592-022-01488-1},
  url = {https://doi.org/10.1038/s41592-022-01488-1},
  issn = {1548-7105}
}

@article{ingraham2022chroma,
	author = {John Ingraham and Max Baranov and Zak Costello and Vincent Frappier and Ahmed Ismail and Shan Tie and Wujie Wang and Vincent Xue and Fritz Obermeyer and Andrew Beam and Gevorg Grigoryan},
	title = {Illuminating protein space with a programmable generative model},
	year = {2023},
	journal = {Nature},
        volume = {623},
        pages = {1070-1078}
}

@article{watson2023rfdiffusion,
	author = {Joseph L. Watson and David Juergens and Nathaniel R. Bennett and Brian L. Trippe and Jason Yim and Helen E. Eisenach and Woody Ahern and Andrew J. Borst and Robert J. Ragotte and Lukas F. Milles and Basile I. M. Wicky and Nikita Hanikel and Samuel J. Pellock and Alexis Courbet and William Sheffler and Jue Wang and Preetham Venkatesh and Isaac Sappington and Susana V{\'a}zquez Torres and Anna Lauko and Valentin De Bortoli and Emile Mathieu and Regina Barzilay and Tommi S. Jaakkola and Frank DiMaio and Minkyung Baek and David Baker},
	title = {De novo design of protein structure and function with RFdiffusion},
	year = {2023},
	journal = {Nature},
        volume = {620},
        pages = {1089–1100}
}

@inproceedings{yim2023framediff,
  title = 	 {{SE}(3) diffusion model with application to protein backbone generation},
  author={Jason Yim and Brian L. Trippe and Valentin De Bortoli and Emile Mathieu and Arnaud Doucet and Regina Barzilay and Tommi Jaakkola},
  booktitle = 	 {Proceedings of the 40th International Conference on Machine Learning (ICML)},
  year = 	 {2023},
}

@article{madani2023large,
  title={Large language models generate functional protein sequences across diverse families},
  author={Madani, Ali and Krause, Ben and Greene, Eric R and Subramanian, Subu and Mohr, Benjamin P and Holton, James M and Olmos Jr, Jose Luis and Xiong, Caiming and Sun, Zachary Z and Socher, Richard and others},
  journal={Nature biotechnology},
  volume={41},
  number={8},
  pages={1099--1106},
  year={2023},
  publisher={Nature Publishing Group US New York}
}

@article{lisanza2024multistate,
  title={Multistate and functional protein design using RoseTTAFold sequence space diffusion},
  author={Lisanza, Sidney Lyayuga and Gershon, Jacob Merle and Tipps, Samuel WK and Sims, Jeremiah Nelson and Arnoldt, Lucas and Hendel, Samuel J and Simma, Miriam K and Liu, Ge and Yase, Muna and Wu, Hongwei and others},
  journal={Nature biotechnology},
  pages={1--11},
  year={2024},
  publisher={Nature Publishing Group US New York}
}

@inproceedings{ren2024carbonnovo,
  title={Carbonnovo: Joint design of protein structure and sequence using a unified energy-based model},
  author={Ren, Milong and Zhu, Tian and Zhang, Haicang},
  booktitle={Forty-first International Conference on Machine Learning},
  year={2024}
}

@article{ferruz2022protgpt2,
  title={ProtGPT2 is a deep unsupervised language model for protein design},
  author={Ferruz, Noelia and Schmidt, Steffen and H{\"o}cker, Birte},
  journal={Nature communications},
  volume={13},
  number={1},
  pages={4348},
  year={2022},
  publisher={Nature Publishing Group UK London}
}

@article{alamdari2023protein,
  title={Protein generation with evolutionary diffusion: sequence is all you need},
  author={Alamdari, Sarah and Thakkar, Nitya and van den Berg, Rianne and Tenenholtz, Neil and Strome, Bob and Moses, Alan and Lu, Alex Xijie and Fusi, Nicolo and Amini, Ava Pardis and Yang, Kevin K},
  journal={BioRxiv},
  pages={2023--09},
  year={2023},
  publisher={Cold Spring Harbor Laboratory}
}

@inproceedings{wang2024diffusion,
  title={Diffusion Language Models Are Versatile Protein Learners},
  author={Wang, Xinyou and Zheng, Zaixiang and Ye, Fei and Xue, Dongyu and Huang, Shujian and Gu, Quanquan},
  booktitle={International Conference on Machine Learning},
  year={2024}
}

@inproceedings{bose2024foldflow,
title={{SE}(3)-Stochastic Flow Matching for Protein Backbone Generation},
author={Joey Bose and Tara Akhound-Sadegh and Guillaume Huguet and Kilian Fatras and Jarrid Rector-Brooks and Cheng-Hao Liu and Andrei Cristian Nica and Maksym Korablyov and Michael M. Bronstein and Alexander Tong},
booktitle={The Twelfth International Conference on Learning Representations (ICLR)},
year={2024},
}

@article{jumper2019alphafold2,
	author = {John Jumper and Richard Evans and Alexander Pritzel and Tim Green and Michael Figurnov and Olaf Ronneberger and Kathryn Tunyasuvunakool and Russ Bates and Augustin Zidek and Anna Potapenko and Alex Bridgland and Clemens Meyer and Simon A. A. Kohl and Andrew J. Ballard and Andrew Cowie and Bernardino Romera-Paredes and Stanislav Nikolov and Rishub Jain and Jonas Adler and Trevor Back and Stig Petersen and David Reiman and Ellen Clancy and Michal Zielinski and Martin Steinegger and Michalina Pacholska and Tamas Berghammer and Sebastian Bodenstein and David Silver and Oriol Vinyals and Andrew W. Senior and Koray Kavukcuoglu and Pushmeet Kohli and Demis Hassabis},
	title = {Highly accurate protein structure prediction with AlphaFold},
	year = {2021},
	journal = {Nature},
        volume = {596},
        pages = {583–589}
}

@article{pacesa2024bindcraft,
  title={BindCraft: one-shot design of functional protein binders},
  author={Pacesa, Martin and Nickel, Lennart and Schellhaas, Christian and Schmidt, Joseph and Pyatova, Ekaterina and Kissling, Lucas and Barendse, Patrick and Choudhury, Jagrity and Kapoor, Srajan and Alcaraz-Serna, Ana and others},
  journal={bioRxiv},
  pages={2024--09},
  year={2024},
  publisher={Cold Spring Harbor Laboratory}
}

@article{lin2023esm2,
author = {Zeming Lin  and Halil Akin  and Roshan Rao  and Brian Hie  and Zhongkai Zhu  and Wenting Lu  and Nikita Smetanin  and Robert Verkuil  and Ori Kabeli  and Yaniv Shmueli  and Allan dos Santos Costa  and Maryam Fazel-Zarandi  and Tom Sercu  and Salvatore Candido  and Alexander Rives },
title = {Evolutionary-scale prediction of atomic-level protein structure with a language model},
journal = {Science},
volume = {379},
number = {6637},
pages = {1123-1130},
year = {2023},
}

@inproceedings{lin2023genie1,
  title = 	 {Generating Novel, Designable, and Diverse Protein Structures by Equivariantly Diffusing Oriented Residue Clouds},
  author =       {Lin, Yeqing and Alquraishi, Mohammed},
  booktitle = 	 {Proceedings of the 40th International Conference on Machine Learning (ICML)},
  year = 	 {2023},
}

@article{lin2024genie2,
      title={Out of Many, One: Designing and Scaffolding Proteins at the Scale of the Structural Universe with Genie 2}, 
      author={Yeqing Lin and Minji Lee and Zhao Zhang and Mohammed AlQuraishi},
      year={2024},
      journal={arXiv preprint arXiv:2405.15489},
}

@article{huguet2024foldflow2,
      title={Sequence-Augmented SE(3)-Flow Matching For Conditional Protein Backbone Generation}, 
      author={Guillaume Huguet and James Vuckovic and Kilian Fatras and Eric Thibodeau-Laufer and Pablo Lemos and Riashat Islam and Cheng-Hao Liu and Jarrid Rector-Brooks and Tara Akhound-Sadegh and Michael Bronstein and Alexander Tong and Avishek Joey Bose},
      year={2024},
      journal={arXiv preprint arXiv:2405.20313},
}

@article{abramson2024alphafold3,
	author = {Josh Abramson and Jonas Adler and Jack Dunger and Richard Evans and Tim Green and Alexander Pritzel and Olaf Ronneberger and Lindsay Willmore and Andrew J. Ballard and Joshua Bambrick and Sebastian W. Bodenstein and David A. Evans and Chia-Chun Hung and Michael O’Neill and David Reiman and Kathryn Tunyasuvunakool and Zachary Wu and Akvilė Zemgulyte and Eirini Arvaniti and Charles Beattie and Ottavia Bertolli and Alex Bridgland and Alexey Cherepanov and Miles Congreve and Alexander I. Cowen-Rivers and Andrew Cowie and Michael Figurnov and Fabian B. Fuchs and Hannah Gladman and Rishub Jain and Yousuf A. Khan and Caroline M. R. Low and Kuba Perlin and Anna Potapenko and Pascal Savy and Sukhdeep Singh and Adrian Stecula and Ashok Thillaisundaram and Catherine Tong and Sergei Yakneen and Ellen D. Zhong and Michal Zielinski and Augustin Zidek and Victor Bapst and Pushmeet Kohli and Max Jaderberg and Demis Hassabis and John M. Jumper},
	title = {Accurate structure prediction of biomolecular interactions with AlphaFold 3},
	year = {2024},
	journal = {Nature},
        volume = {630},
        pages = {493–500}
}

@article{varadi2021afdb,
author = {Varadi, Mihaly and Anyango, Stephen and Deshpande, Mandar and Nair, Sreenath and Natassia, Cindy and Yordanova, Galabina and Yuan, David and Stroe, Oana and Wood, Gemma and Laydon, Agata and Žídek, Augustin and Green, Tim and Tunyasuvunakool, Kathryn and Petersen, Stig and Jumper, John and Clancy, Ellen and Green, Richard and Vora, Ankur and Lutfi, Mira and Velankar, Sameer},
year = {2021},
pages = {D439–D444},
title = {AlphaFold Protein Structure Database: Massively expanding the structural coverage of protein-sequence space with high-accuracy models},
volume = {50},
journal={Nucleic Acids Research},
}

@article{barrio2023clustering,
	author = {Inigo Barrio-Hernandez and Jingi Yeo and J\"{u}rgen J\"{a}nes and Milot Mirdita and Cameron L. M. Gilchrist and Tanita Wein and Mihaly Varadi and Sameer Velankar and Pedro Beltrao and Martin Steinegger},
	title = {Clustering predicted structures at the scale of the known protein universe},
	year = {2023},
	journal = {Nature},
        volume = {622},
        pages = {637–645}
}

@article{vankempen2024foldseek,
	author = {Michel van Kempen and Stephanie S. Kim and Charlotte Tumescheit and Milot Mirdita and Jeongjae Lee and Cameron L. M. Gilchrist and Johannes S\"{o}ding and Martin Steinegger},
	title = {Fast and accurate protein structure search with Foldseek},
	year = {2024},
	journal = {Nat Biotechnol.},
        volume = {42},
        pages = {243–246}
}

@article{yim2023frameflow,
      title={Fast protein backbone generation with SE(3) flow matching}, 
      author={Jason Yim and Andrew Campbell and Andrew Y. K. Foong and Michael Gastegger and José Jiménez-Luna and Sarah Lewis and Victor Garcia Satorras and Bastiaan S. Veeling and Regina Barzilay and Tommi Jaakkola and Frank Noé},
      year={2023},
      journal={arXiv preprint arXiv:2310.05297},
}

@inproceedings{
peng2025path,
title={Path Planning for Masked Diffusion Models with Applications to Biological Sequence Generation},
author={Fred Zhangzhi Peng and Zachary Bezemek and Sawan Patel and Jarrid Rector-Brooks and Sherwood Yao and Alexander Tong and Pranam Chatterjee},
booktitle={ICLR 2025 Workshop on Deep Generative Model in Machine Learning: Theory, Principle and Efficacy},
year={2025},
url={https://openreview.net/forum?id=fFuVPKpSt0}
}

@misc{tang2022improved,
    title={Improved Vector Quantized Diffusion Models},
    author={Zhicong Tang and Shuyang Gu and Jianmin Bao and Dong Chen and Fang Wen},
    year={2022},
    eprint={2205.16007},
    archivePrefix={arXiv},
    primaryClass={cs.CV}
}

@article{wang2024proteus,
	author = {Chentong Wang and Yannan Qu and Zhangzhi Peng and Yukai Wang and Hongli Zhu and Dachuan Chen and Longxing Cao},
	title = {Proteus: Exploring Protein Structure Generation for Enhanced Designability and Efficiency},
      year={2024},
      journal={bioRxiv},
	eprint = {https://www.biorxiv.org/content/10.1101/2024.02.10.579791v3.full.pdf},
}

@inproceedings{yim2024multiflow,
  title = 	 {Generative Flows on Discrete State-Spaces: Enabling Multimodal Flows with Applications to Protein Co-Design},
  author =       {Campbell, Andrew and Yim, Jason and Barzilay, Regina and Rainforth, Tom and Jaakkola, Tommi},
  booktitle = 	 {Proceedings of the 41st International Conference on Machine Learning (ICML)},
  year = 	 {2024},
}

@article{chu2024protpardelle,
author = {Alexander E. Chu  and Jinho Kim  and Lucy Cheng  and Gina El Nesr  and Minkai Xu  and Richard W. Shuai  and Po-Ssu Huang },
title = {An all-atom protein generative model},
journal = {Proceedings of the National Academy of Sciences},
volume = {121},
number = {27},
pages = {e2311500121},
year = {2024},
}

@article{hayes2025simulating,
  title={Simulating 500 million years of evolution with a language model},
  author={Hayes, Thomas and Rao, Roshan and Akin, Halil and Sofroniew, Nicholas J and Oktay, Deniz and Lin, Zeming and Verkuil, Robert and Tran, Vincent Q and Deaton, Jonathan and Wiggert, Marius and others},
  journal={Science},
  pages={eads0018},
  year={2025},
  publisher={American Association for the Advancement of Science}
}

@inproceedings{trippe2023diffusion,
title={Diffusion Probabilistic Modeling of Protein Backbones in 3D for the motif-scaffolding problem},
author={Brian L. Trippe and Jason Yim and Doug Tischer and David Baker and Tamara Broderick and Regina Barzilay and Tommi S. Jaakkola},
booktitle={The Eleventh International Conference on Learning Representations (ICLR)},
year={2023},
}

@article{dauparas2022robust,
  title={Robust deep learning--based protein sequence design using ProteinMPNN},
  author={Dauparas, Justas and Anishchenko, Ivan and Bennett, Nathaniel and Bai, Hua and Ragotte, Robert J and Milles, Lukas F and Wicky, Basile IM and Courbet, Alexis and de Haas, Rob J and Bethel, Neville and others},
  journal={Science},
  volume={378},
  number={6615},
  pages={49--56},
  year={2022},
  publisher={American Association for the Advancement of Science}
}

@article{kim2023foldcomp,
    author = {Kim, Hyunbin and Mirdita, Milot and Steinegger, Martin},
    title = "{Foldcomp: a library and format for compressing and indexing large protein structure sets}",
    journal = {Bioinformatics},
    volume = {39},
    number = {4},
    pages = {btad153},
    year = {2023},
    month = {03},
}

@article{paszke2019pytorch,
  title={Pytorch: An imperative style, high-performance deep learning library},
  author={Paszke, Adam and Gross, Sam and Massa, Francisco and Lerer, Adam and Bradbury, James and Chanan, Gregory and Killeen, Trevor and Lin, Zeming and Gimelshein, Natalia and Antiga, Luca and others},
  journal={Advances in neural information processing systems},
  volume={32},
  year={2019}
}

@misc{chen2025allatom,
    title={An All-Atom Generative Model for Designing Protein Complexes},
    author={Ruizhe Chen and Dongyu Xue and Xiangxin Zhou and Zaixiang Zheng and Xiangxiang Zeng and Quanquan Gu},
    year={2025},
    eprint={2504.13075},
    archivePrefix={arXiv},
    primaryClass={cs.LG}
}

@inproceedings{lu2025all,
  title={All-Atom Protein Generation with Latent Diffusion},
  author={Lu, Amy X and Yan, Wilson and Robinson, Sarah A and Kelow, Simon and Yang, Kevin K and Gligorijevic, Vladimir and Cho, Kyunghyun and Bonneau, Richard and Abbeel, Pieter and Frey, Nathan C},
  year={2025},
  booktitle={ICLR 2025 Workshop on Generative and Experimental Perspectives for Biomolecular Design}
}

@article{ahern2025atom,
  title={Atom level enzyme active site scaffolding using RFdiffusion2},
  author={Ahern, Woody and Yim, Jason and Tischer, Doug and Salike, Saman and Woodbury, Seth and Kim, Donghyo and Kalvet, Indrek and Kipnis, Yakov and Coventry, Brian and Altae-Tran, Han and others},
  journal={bioRxiv},
  pages={2025--04},
  year={2025},
  publisher={Cold Spring Harbor Laboratory}
}

@article{qu2024pallatom,
  title={P (all-atom) Is Unlocking New Path For Protein Design},
  author={Qu, Wei and Guan, Jiawei and Ma, Rui and Zhai, Ke and Wu, Weikun and Wang, Haobo},
  journal={bioRxiv},
  pages={2024--08},
  year={2024},
  publisher={Cold Spring Harbor Laboratory}
}

@inproceedings{geffner2025proteina,
    title={Proteina: Scaling Flow-based Protein Structure Generative Models},
    author={Tomas Geffner and Kieran Didi and Zuobai Zhang and Danny Reidenbach and Zhonglin Cao and Jason Yim and Mario Geiger and Christian Dallago and Emine Kucukbenli and Arash Vahdat and Karsten Kreis},
    booktitle={International Conference on Learning Representations (ICLR)},
    year={2025}
}

@inproceedings{wang2025dplm,
  title={Dplm-2: A multimodal diffusion protein language model},
  author={Wang, Xinyou and Zheng, Zaixiang and Ye, Fei and Xue, Dongyu and Huang, Shujian and Gu, Quanquan},
  booktitle={International Conference on Learning Representations (ICLR)},
  year={2025}
}

@article{huang2022backbone,
  title={A backbone-centred energy function of neural networks for protein design},
  author={Huang, Bin and Xu, Yang and Hu, Xiuhong and Liu, Yongrui and Liao, Shanhui and Zhang, Jiahai and Huang, Chengdong and Hong, Jingjun and Chen, Quan and Liu, Haiyan},
  journal={Nature},
  volume={602},
  number={7897},
  pages={523--528},
  year={2022},
  publisher={Nature Publishing Group UK London}
}

@article{anishchenko2021novo,
  title={De novo protein design by deep network hallucination},
  author={Anishchenko, Ivan and Pellock, Samuel J and Chidyausiku, Tamuka M and Ramelot, Theresa A and Ovchinnikov, Sergey and Hao, Jingzhou and Bafna, Khushboo and Norn, Christoffer and Kang, Alex and Bera, Asim K and others},
  journal={Nature},
  volume={600},
  number={7889},
  pages={547--552},
  year={2021},
  publisher={Nature Publishing Group UK London}
}

@article{huang2011rosettaremodel,
  title={RosettaRemodel: a generalized framework for flexible backbone protein design},
  author={Huang, Po-Ssu and Ban, Yih-En Andrew and Richter, Florian and Andre, Ingemar and Vernon, Robert and Schief, William R and Baker, David},
  journal={PloS one},
  volume={6},
  number={8},
  pages={e24109},
  year={2011},
  publisher={Public Library of Science San Francisco, USA}
}

@article{korendovych2020novo,
  title={De novo protein design, a retrospective},
  author={Korendovych, Ivan V and DeGrado, William F},
  journal={Quarterly reviews of biophysics},
  volume={53},
  pages={e3},
  year={2020},
  publisher={Cambridge University Press}
}

@article{davis2007molprobity,
  title={MolProbity: all-atom contacts and structure validation for proteins and nucleic acids},
  author={Davis, Ian W and Leaver-Fay, Andrew and Chen, Vincent B and Block, Jeremy N and Kapral, Gary J and Wang, Xueyi and Murray, Laura W and Arendall III, W Bryan and Snoeyink, Jack and Richardson, Jane S and others},
  journal={Nucleic acids research},
  volume={35},
  number={suppl\_2},
  pages={W375--W383},
  year={2007},
  publisher={Oxford University Press}
}

@misc{costa2023ophiuchus,
    title={Ophiuchus: Scalable Modeling of Protein Structures through Hierarchical Coarse-graining SO(3)-Equivariant Autoencoders},
    author={Allan dos Santos Costa and Ilan Mitnikov and Mario Geiger and Manvitha Ponnapati and Tess Smidt and Joseph Jacobson},
    year={2023},
    eprint={2310.02508},
    archivePrefix={arXiv},
    primaryClass={cs.LG}
}

@article{delAlamo2025.05.09.653228,
	author = {del Alamo, Diego and Frick, Rahel and Truan, Daphne and Karpiak, Joel},
	title = {Adapting ProteinMPNN for antibody design without retraining},
	elocation-id = {2025.05.09.653228},
	year = {2025},
	doi = {10.1101/2025.05.09.653228},
	publisher = {Cold Spring Harbor Laboratory},
	URL = {https://www.biorxiv.org/content/early/2025/05/15/2025.05.09.653228},
	eprint = {https://www.biorxiv.org/content/early/2025/05/15/2025.05.09.653228.full.pdf},
	journal = {bioRxiv}
}

\newpage
\newpage

\newpage
\appendix

\renewcommand\ptctitle{}
\addcontentsline{toc}{section}{Appendix} % Add the appendix text to the document TOC
\part{Appendix} % Start the appendix part
% \vspace{6mm}
\parttoc % Insert the appendix TOC

\newpage
\section{Additional \ourmodel Sample Visualizations}
\begin{figure}[ht!]
% \vspace{-8mm}
    \centering
    \includegraphics[width=1.0\linewidth]{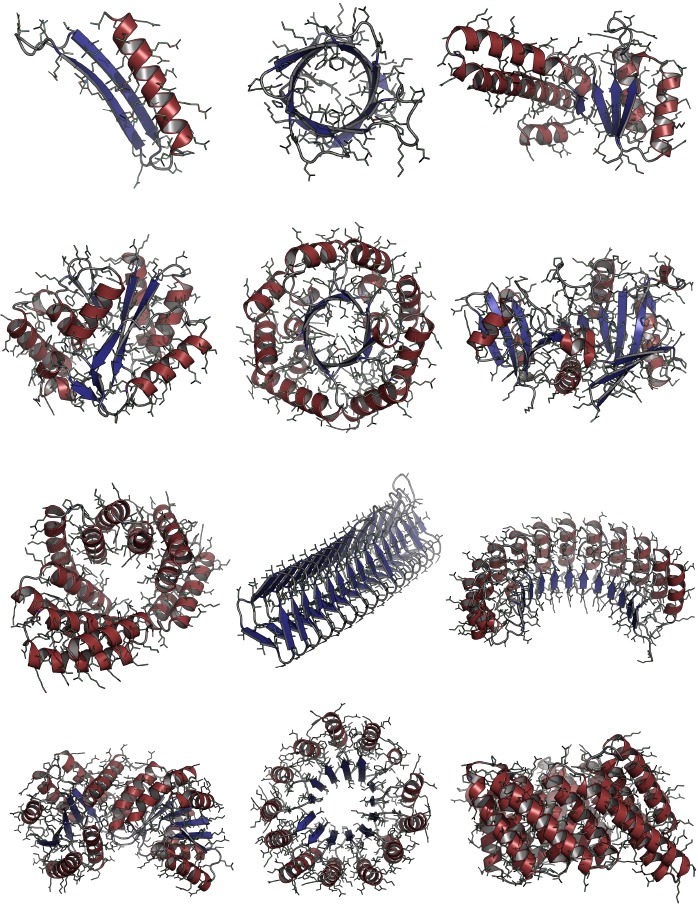}
    % \vspace{-4mm}
    \caption{\small \textbf{\ourmodel Samples}. Additional fully atomistic proteins generated by our model, ranging from 100 to 400 residues, including side chains. All shown samples are co-designable.
    }
    \label{fig:len400_samples_appendix}
    % \vspace{-4mm}
\end{figure}

In \fref{fig:len400_samples_appendix}, we show additional fully atomistic proteins generated by \ourmodel. Our model outputs diverse (co-)designable samples, including realistic side chain structures.
\begin{figure}[!t]
    % \vspace{-8mm}
    \centering
    \includegraphics[width=0.9\linewidth]{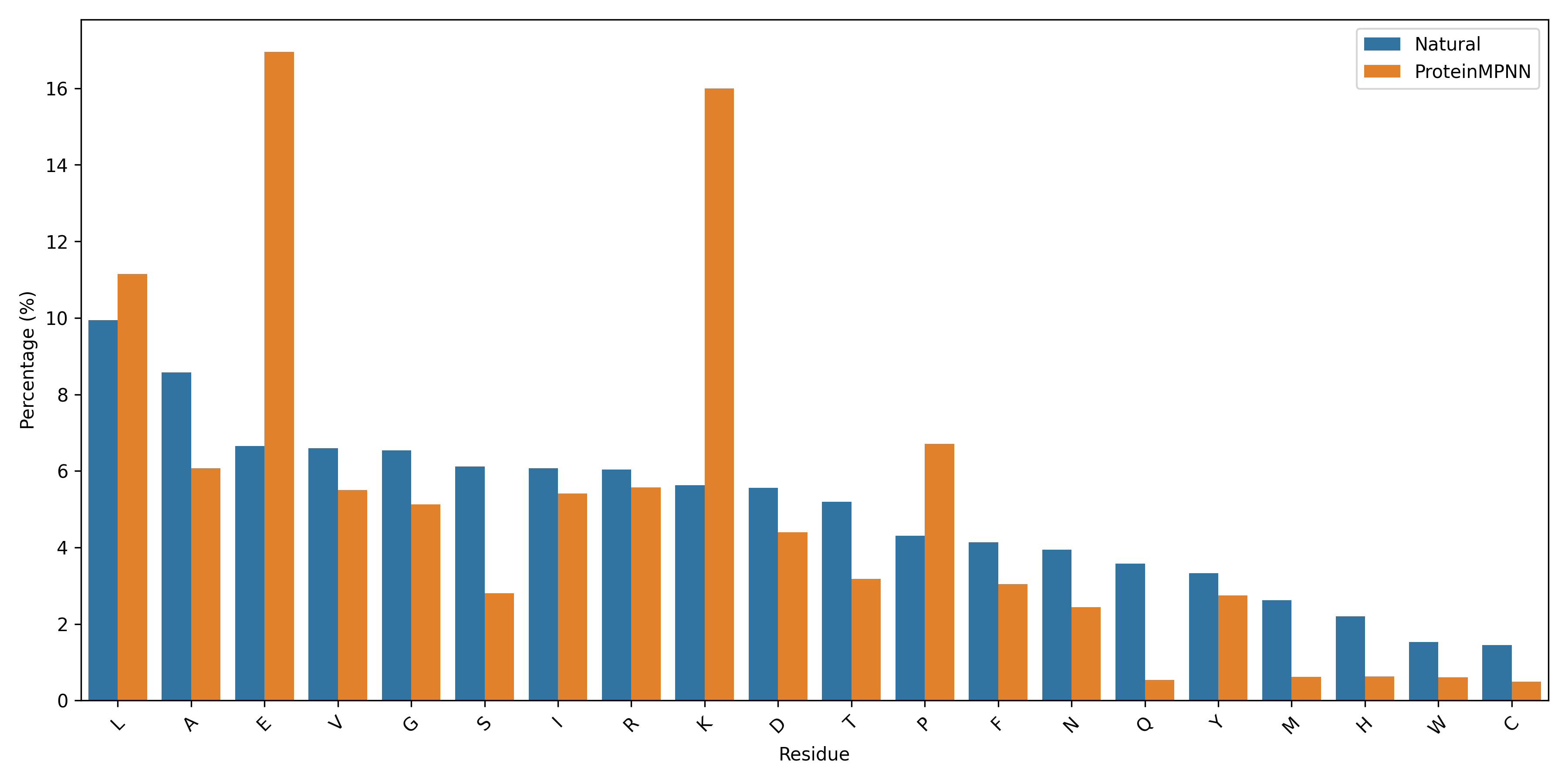}
    \vspace{-1mm}
    \caption{Amino acid sequence distribution for \geniedata (natural) and \ourdata (synthetic) for sequence length $\in [32, 256]$.
    }
    \label{fig:seq_dist}
    % \vspace{-5mm}
\end{figure}
\section{Dataset Details} \label{app:sec_dataset}
% \subsection{\geniedata and its shortcoming on training multimodal protein generation model}
\subsection{Dealing with the inconsistency between structure and sequence data of AFDB}
\geniedata is the dataset used in Genie2~\citep{lin2024genie2} for training a protein backbone generative model. It is a subset of the AlphaFold Database (AFDB)~\citep{varadi2021afdb} containing proteins clustered with both MMseqs2~\citep{barrio2023clustering} based on sequence similarity and Foldseek~\citep{vankempen2024foldseek} based on structure similarity. \geniedata only contains one structure per cluster (the cluster representative). After filtering with $N_\mathrm{residue}\in[32, 256]$ and pLDDT$\geq$80, \geniedata contains 588,318 structures. 

We analyzed the co-designability of structure-sequence pairs in \geniedata by folding the sequences with ESMFold and checking if the lowest RMSD between folded structure and the AFDB structure is less than 2\si{\angstrom}. We discovered that only $26.6\%$ of \geniedata are co-designable by $C_{\alpha}$ RMSD and even less by all-atom RMSD (\fref{fig:afdb_data_dist}). The low co-designability of \geniedata poses a significant challenge to multimodal protein generation model training: even if the model fit both sequence and structure distributions of the dataset very well, the generated protein structure will not be consistent with the generated sequence. To address this challenge, we explored three other synthetic datasets based on downstream augmentations of \geniedata.

\subsection{Datasets with designable structures}
\label{app:des_data_fail}
Our initial explorations began with targeting the structure-sequence inconsistency at the structural level. We hypothesized that refolding \geniedata with ESMFold would, by definition, yield 100\% co-designable samples, and that training on these designable samples would improve model performance. To test this hypothesis, we created two datasets:

1. $D_\mathrm{ESMFold}$: We took sequences in the original \geniedata and folded them with ESMFold. We applied a filter of $\mathrm{pLDDT}_\mathrm{ESMFold}\geq80$ on the folded structures. As a result, all remaining structures should be co-designable with a high confidence score. $D_\mathrm{ESMFold}$ contained 163,552 samples. 

2. $D_\mathrm{Des}$: We took sequences in the original \geniedata and folded them with ESMFold. We computed the all-atom RMSD between the ESMFold-folded structure and the original AFDB structure. We then filtered \geniedata based on the all-atom RMSD with 2\si{\angstrom} cutoff to create the $D_\mathrm{des}$ dataset containing 155,957 samples. Hence, in contrast to $D_\mathrm{ESMFold}$, here we are relying on the original \geniedata structures and filtering them to a designable subset.

We trained \textit{\ourmodel} on both $D_\mathrm{ESMFold}$ and $D_\mathrm{des}$. Both datasets reduced the model's performance on all metrics (\tref{tab:training_data_choice}) and caused sequence overfitting (structure losses remained unaffected) despite the sequences for both datasets remaining unchanged. The results suggest that jointly modeling natural sequences and synthetic structures (predicted by either ESMFold or AF2) remains challenging when those structures cannot be easily recovered with both ESMFold and AF2\footnote{ColabFold with MSAs yields ${\sim}65\%$ co-designability on a random subset of 100 samples from \geniedata.}. While \geniedata provides a diverse sequence and structure space, realigning the structures to the sequences using ESMFold and filtering by its confidence score may inadvertently reduce both data diversity and volume. This process may also contribute to the observed overfitting.

\subsection{Details of Our Synthetic Data $\mathcal{D}_\mathrm{SYN-ours}$}
Given that a sequence must reasonably fold into its given structure to be co-designable, and enforcing co-designability at the structural level worsened all models, we shifted our focus to the sequences. 
We observed that ProteinMPNN-based sequence resampling can significantly improve designability (see ~\sref{app:agree_data} for detailed discussion), prompting us to create a dataset with synthetic sequences to target the central issue of inconsistent sequence-structure pairs.
This choice is further motivated by the fact that ProteinMPNN is widely used for the \textit{inverse folding} step in the standard multi-step \textit{backbone generation}\textemdash\textit{inverse folding}\textemdash\textit{forward folding} pipeline employed by most de novo protein generative models, owing to its validated performance in wet-lab experiments~\citep{watson2023rfdiffusion, dauparas2022robust, delAlamo2025.05.09.653228}.
% , making it a meaningful tool to ensure co-designability in our situation.

It is crucial to note that, since we are modeling fully atomistic protein structures, we cannot utilize the given AFDB structures if there are any residue changes in the predicted synthetic sequences. This is because a change in sequence implies a different side-chain structure, potentially with a different number of atoms. Consequently, to use synthetic sequences, we refold the new sequences to ensure all-atom compatibility. We visualize both the natural and synthetic sequence distributions in~\fref{fig:seq_dist}.

% Our solution to the aforementioned problems is to create a dataset focusing on jointly aligning \textit{synthetic} sequences and structures. Most of the current de novo protein generative models use the standard multi-step "backbone generation"-"inverse fold"-"forward fold" pipeline, in which the "inverse fold" step is done by ProteinMPNN due to its established wet-lab validation~\citep{dauparas2022robust}. Leveraging the power of ProteinMPNN,

% Therefore to address the aformentioned problems while also preversing scale and diveristy we created $\mathcal{D}_\mathrm{codes}$ with the following steps: (i) generating four ProteinMPNN sequences for each \geniedata structure with lengths between 32 and 400, (ii) folding the ProteinMPNN sequences with ESMFold for its computational efficiency, being $\sim$60x faster than AF2, (iii) selecting the sequence-structure pair with the lowest $C_\alpha$ RMSD to the original AFDB structure. This is to preserve the structure diversity as the original AFDB structures are cluster representatives. After filtering out samples with average ESMFold pLDDT below 0.8, our curated dataset distilled knowledge from ProteinMPNN, ESMFold, and AF2, resulting in $\sim$0.43M high-quality samples.
To address the problems discussed in~\sref{app:des_data_fail} while preserving scale and diversity, we created $\mathcal{D}_\mathrm{codes}$ through the following steps: (i) generating four ProteinMPNN sequences for each \geniedata structure with lengths between 32 and 512, (ii) folding the ProteinMPNN sequences with ESMFold due to its computational efficiency, being $\sim60\times$ faster than AF2, and (iii) selecting the sequence-structure pair with the lowest $C_\alpha$ RMSD to the original AFDB structure to preserve the structural diversity, as the original AFDB structures are cluster representatives. After filtering out samples with an average ESMFold pLDDT below 80, our curated dataset, which combines knowledge from ProteinMPNN and the confident predictions of both ESMFold and AF2, results in 455,473 high-quality samples. Furthermore, rather than relying on redesigning the sequence after structure-based generation and regenerating the side chains each time, $\mathcal{D}_\mathrm{codes}$ enables learning a consistent sequence-structure distribution, facilitating accurate single-step, fully atomistic design. It is worth mentioning that FoldComp \citep{kim2023foldcomp} was used to store and access all datasets we prepared efficiently.

We present the amino acid residue distribution of all training samples, ranging in length from 32 to 256, in \fref{fig:seq_dist} for both the natural \geniedata sequences and those generated using ProteinMPNN in \ourdata. We chose ProteinMPNN for its robust wetlab validation~\citep{dauparas2022robust, delAlamo2025.05.09.653228}. However, it does overrepresent certain residue types, particularly charged species (E, K). While this overrepresentation is not inherently problematic for de novo design, as it allows the model to generate fully atomistic structures with high fidelity without redesign, it is still an important consideration for downstream usage.

\section{Architecture Details}
\begin{table*}[h!]
\caption{Hyperparameters for \ourmodel model training. Rows highlighed in grey are specfic to the all-atom architecture. We denote two versions of \ourmodel the one trained on shorter lengths up to 256 and the standard model trained to max length 400.}
\label{tab:train_config}
\centering
\scalebox{0.77}{
\begin{tabular}{l | l | l | l | l | l }
\toprule
\textbf{Model}  &  \multicolumn{5}{c}{\textbf{Prote\'{i}na-}} \\
& \small \textbf{Co-design}  & \small \textbf{Atom\'{i}stica (256)} & \small \textbf{Atom\'{i}stica (400)} & \small \textbf{Atom\'{i}stica Motif} & \small \textbf{Atom\'{i}stica-tri}\\
\midrule
\textbf{Architecture Component} & & & & & \\
initialization & random & random & random & random & random\\
% (a) sequence representation
sequence repr dim & 512 & 768 & 768 & 512 & 768\\
\# registers & 10 & 10 & 10 & 10 & 10\\
% (b) sequence conditioning
sequence cond dim & 128 & 512 & 512 & 128 & 512\\
$t$ sinusoidal enc dim & 196 & 256 & 512 & 196 & 512\\
idx. sinusoidal enc dim & 196 & 128 & 256 & 196 & 256 \\
% (c) pair representation
pair repr dim & 196 & 512 & 256 & 196 & 256\\
seq separation dim & 128 & 128 & 128 & 128 & 128\\
% self cond. seq. dists. dim. & 64 & 64 & 64 & 64 \\
pair distances dim ($\rvx_t$) & 64 & 64 & 64 & 64 & 64 \\
pair distances dim ($\hat \rvx(\rvx_t)$) & 128 & 128 & 128 & 128 & 128\\
pair distances min (Å) & 1 & 1 & 1 & 1 & 1 \\
pair distances max (Å) & 30 & 30 & 30 & 30 & 30\\
residue type embedding dim & 196 & 512 & 512 & 196 & 512 \\
% (d) neural network
\# attention heads & 12 & 12 & 12 & 12 & 12\\
\# transformer layers & 12 & 15 & 15 & 12 & 15\\
\# triangle layers & 0 & 0 & 0 & 0 & 3 \\
\rowcolor{gray!15}
\# number of atom layers & 0 & 5 & 5 & 5 & 5 \\
\rowcolor{gray!15}
atom cond dim & 0 & 128 & 128 & 128 & 128 \\
\rowcolor{gray!15}
atom dim & 0 & 128 & 128 & 128 & 128 \\
\rowcolor{gray!15}
atom type embedding dim & 0 & 128 & 128 & 128 & 128 \\
\rowcolor{gray!15}
\# atom attention heads dim & 0 & 8 & 8 & 8 & 8 \\
\rowcolor{gray!15}
\# atom cross attention heads& 0 & 8 & 8 & 8 & 8 \\
\rowcolor{gray!15}
side chain coords & N/A & local trans & local frame & local trans & local frame \\
\# trainable parameters & 59.3M & 221M & 222M & 73.6M & 226M \\
\midrule
\textbf{Training Details} & & & &  \\
\# train steps (length$\in$[32, 256]) & 100K & 190k & 210k & 100K & 145k \\
\# finetune steps (length$\in$[32, 400]) & N/A & N/A & 100k & N/A & N/A \\
% learning rate & $10^{-4}$ & $10^{-4}$ & $10^{-4}$ & $10^{-4}$ \\
train batch size per GPU & 28 & 8 & 12 & 8 & 4\\
finetune batch size per GPU & N/A & N/A & 1 & N/A & N/A\\
\# GPUs & 96 & 96 & 96 & 96 & 96 \\
\# grad. acc. steps & 1 & 1 & 1 & 1 & 1\\
\% forward folding & 10 & 5 & 10 & 5 & 10 \\
\% inverse folding & 10 & 5 & 10 & 5 & 10 \\
\% side chain packing & 0 & 0 & 5 & 0 & 5 \\
\bottomrule
\end{tabular}
}
\end{table*}
\begin{figure}[t]
\vspace{-8mm}
    \centering
    \includegraphics[width=\linewidth]{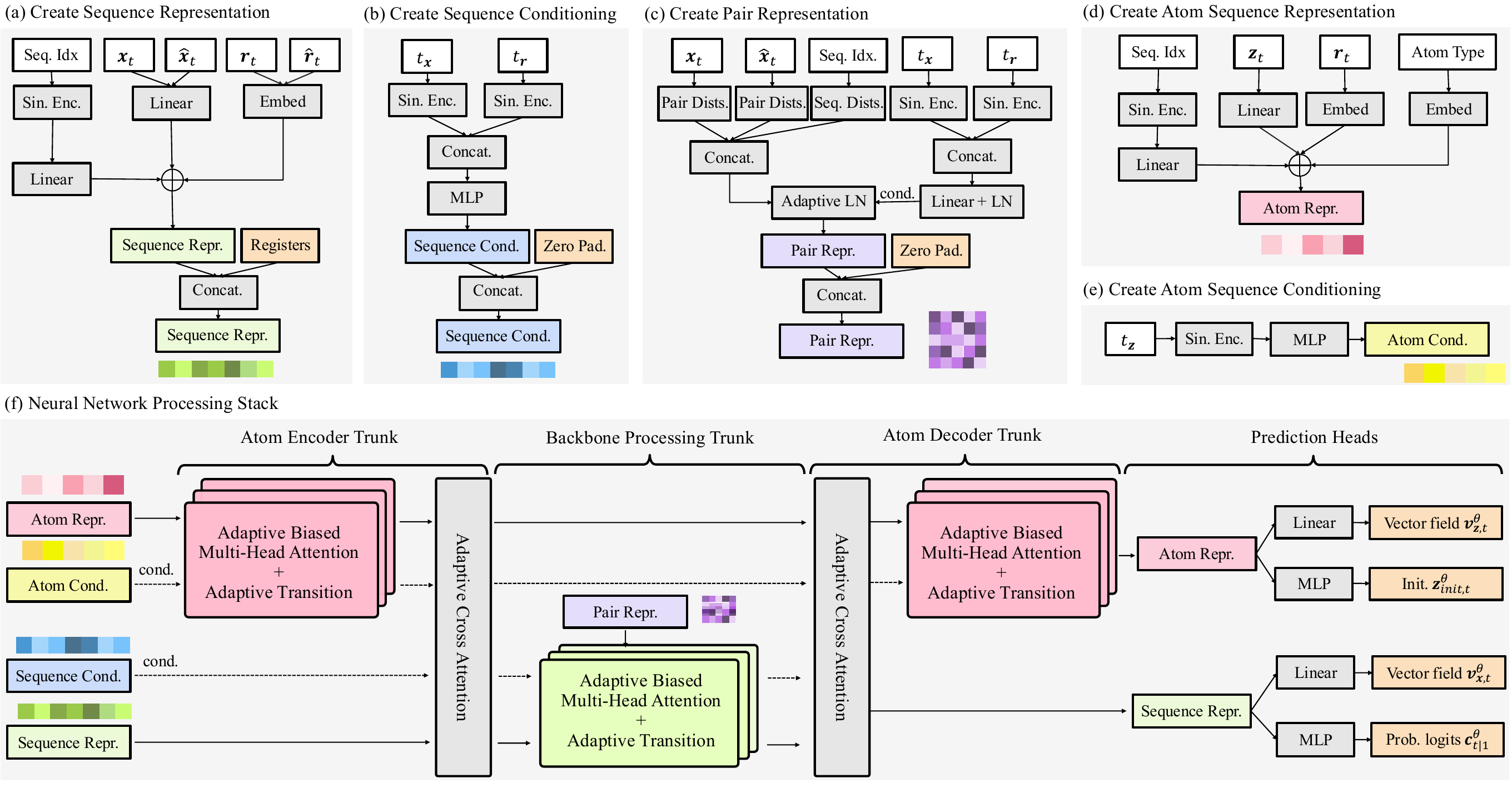}
    \caption{\small \textbf{\ourmodel's transformer architecture.} \textit{(a)-(c)} First generate an initial sequence representation, sequence conditioning features, and a pair representation. \textit{(d)-(e)} Create atom representations and atom conditioning features for the expanded atom sequence. \textit{(f)} Process these representations iteratively through trunks, moving from atom-level to sequence-level and back to atom-level. Each trunk incorporates conditioned multi-head attention layers, biased by the pair representation. Adaptive cross-attention is employed between trunks to update atom and sequence representations (see Appendix). 
    % The final sequence representations predict the vector field for $C_\alpha$ atoms and the probability logits for amino acid types, while the atom representations are used to predict the vector field and initial structure for non-$C_\alpha$ atoms.
    }
    \label{fig:model_arch}
    \vspace{-2ex}
\end{figure}

Here we introduce the model versions in order of complexity. Starting with \basemodel we add discrete sequence co-generation to create \basemodel-Co-Design. We then extend this with side-chain co-generation to yield the full \ourmodel framework.

\subsection{\basemodel-Co-Design}\label{app:proteina-co-des}
For the co-design setting, we start from the $\sim60M$ \basemodel architecture configuration that shows an optimal balance of accuracy and speed in the backbone-only setting (See Appendix C.2 of Geffner et al.~\citep{geffner2025proteina} for \basemodel speed analysis). To enable joint backbone-sequence modeling from a pure backbone model, we add three features: 
\begin{enumerate}
    \item[(i)] residue type index embeddings
    \item[(ii)] argmax residue type index predictions for self-conditioning,
    \item[(iii)] the independent residue type time variable, which dictates how much noise or, in this case, the percentage of tokens to be replaced with MASK tokens
\end{enumerate}
We note that both the $C_\alpha$ coordinates and residue types leverage self-conditioning, where in 50\% of the training iterations, we run a first model forward pass to obtain predictions of the current structure and sequence and use those as additional inputs to the model during a second forward pass. This is a common technique for improving diffusion models and can be viewed as a form of recycling employed by AlphaFold2~\citep{geffner2025proteina, jumper2019alphafold2}.

For the Co-design task only, we sample the sequence time from $\mathcal{B}(1.0, 2.5)$, where $\mathcal{B}(\cdot,\cdot)$ is the Beta distribution. This is a severely left-skewed distribution, which gives more weight to noisy times (sequences with a higher masking rate). For reference, we found that this did not make an impact in the all-atom task. Instead, we used the standard uniform distribution, given that we were directly modeling the structure-sequence duality with residue types and their structures. For co-design training, $10\%$ of the batch iterations are used for forward and inverse folding, respectively. This was done to pin the two independent schedules so that when both structure and sequence time reach one, the structure and sequences are trained to align. Please see~\tref{tab:train_config} for complete model configurations and compute resources used.

\subsection{\ourmodelnoit}
\begin{figure}[t!]
% \vspace{-8mm}
    \centering
    \includegraphics[width=\linewidth]{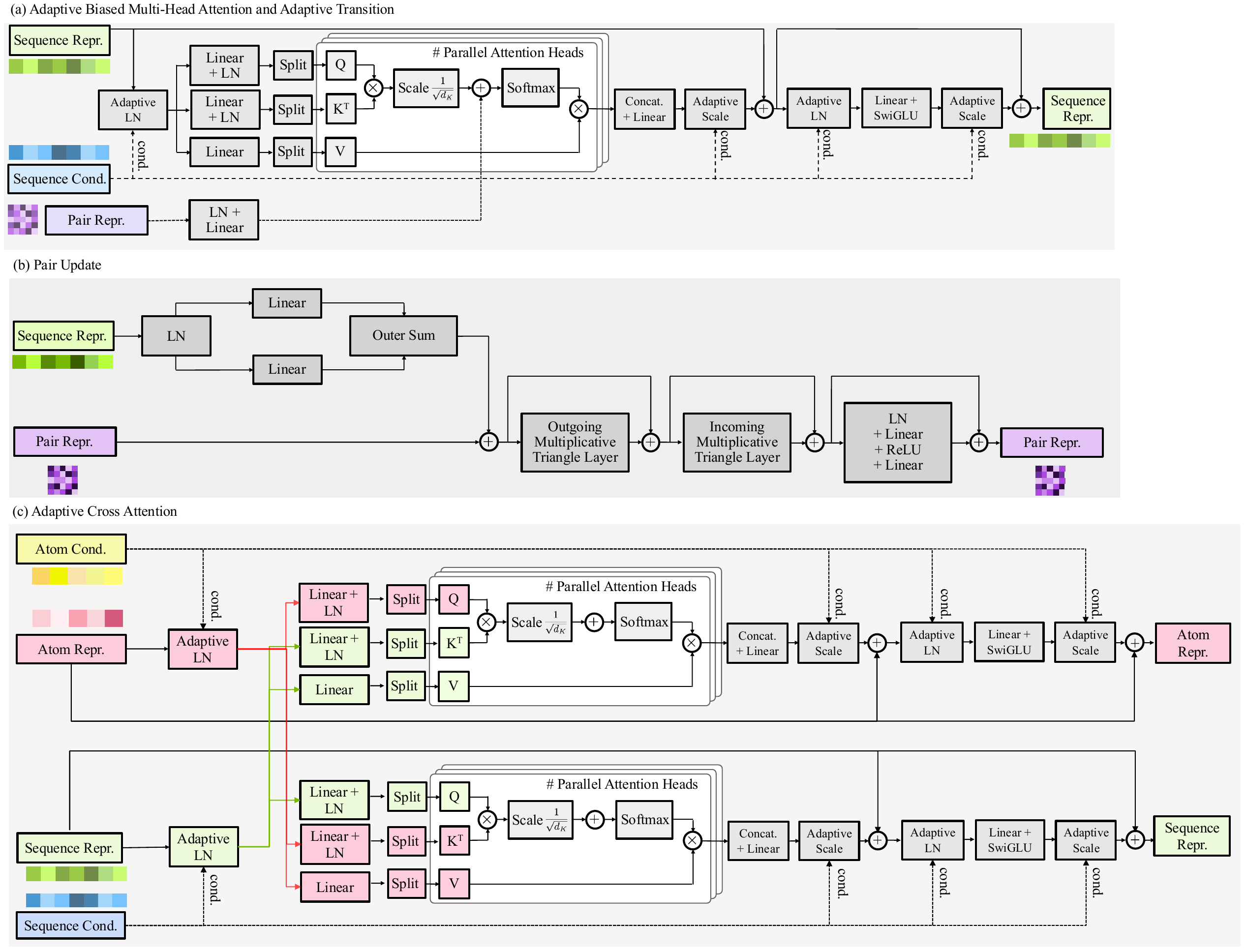}
    \caption{\small \textbf{Additional modules of \ourmodel transformer architecture.} (a) Adaptive attention and transition. (b) Optional pair representation update with triangle multiplicative layers. (c) Adaptive cross attention.
    % The final sequence representations predict the vector field for $C_\alpha$ atoms and the probability logits for amino acid types, while the atom representations are used to predict the vector field and initial structure for non-$C_\alpha$ atoms.
    }
    \label{fig:cross_attn}
    % \vspace{-1ex}
\end{figure}
More architectural components are illustrated in Fig.~\ref{fig:cross_attn}.

\subsection{Optional Triangle Multiplicative Updates}
\label{app:tri_details}
% \dr{@Zhonglin}
% \dr{@Zuobai to add figure?}
In addition to the highly scalable \textit{\ourmodel} demonstrated by \fref{fig:model_arch}, we trained another variant, \textit{\ourmodel-tri}, with triangle multiplicative layers, which were used to update the pair representation. Fig.~\ref{fig:cross_attn}(b) shows how triangle multiplicative layers are used in the \textit{\ourmodel} architecture. During training, the pair representation was updated every 5 backbone processing layers, where the backbone processing layers are the core transformer layers shown in Fig.~\ref{fig:cross_attn}(a), resulting in 3 updates in total and $\sim4$M parameters in triangle multiplicative layers. \tref{tab:aa_400_tri_results} demonstrates that \textit{\ourmodel-tri} exhibits improved performance on all metrics, especially the all-atom diversity. Considering that the triangle multiplicative layers are highly memory-intensive, we keep them as an optional and sparse add-on to our model architecture. 
\section{\ourmodel Training and Inference Details}
\begin{figure}[t]
\centering
\vspace{-8mm}
\scalebox{0.8}{
\begin{minipage}[t]{0.5\linewidth}
\begin{algorithm}[H]
\begin{small}
   \caption{\method Training}
   \label{alg:train}
\begin{algorithmic}[1]
    \WHILE{not converged}
    \STATE Sample protein $(\rvx, \rvr, \rvz)$ from dataset
    \STATE Sample time steps $t_{\rvx}$, $t_{\rvr}$, $t_{\rvz}$ for each modality
    \STATE Convert global $\rvz$ to local coordinates using $\rvx$
    \STATE Sample noisy input $\rvx_t,\rvr_t,\rvz_t$ for each modality
    \STATE Zero out $\rvz_t$ for masked residues in $\rvr_t$
    
    \STATE
    \STATE Predict $\rvv_{\rvx,t}^{\theta}$, $\rvv_{\rvz,t}^{\theta}$, $\rvz_{\text{init},t}^{\theta}$ and $\rvc_{1|t}^{\theta}$
    \STATE Compute loss across modalities
    \STATE $\gL_{\rvx} \gets \|\rvv^{\theta}_{\rvx,t}-(\rvx-\boldsymbol{\epsilon}_{\rvx})\|_2^2$
    \STATE $\gL_{\rvr} \gets \text{CrossEntropy}(\rvc_{1|t}^{\theta},\rvr)$
    \FOR{each residue $i$}
    \STATE $\gL_{\rvz,i}\gets
  \|\rvv^{\theta}_{\rvz,t,i}-(\rvz_i-\boldsymbol{\epsilon}_{\rvz,i})\|_2^2, \text{if}\,\rvr_{t,i} \neq \mask$
    \STATE $\gL_{\rvz,i}\gets
  \|\rvz_{\text{init},t,i}^{\theta}-(\rvz_i-\boldsymbol{\epsilon}_{\rvz,i})\|_2^2, \text{if}\,\rvr_{t,i} = \mask$
    \ENDFOR
    \STATE $\gL \gets \frac{1}{L} \left(\gL_{\rvx}+\gL_{\rvr}+\gL_{\rvz}\right)$
    \STATE Calculate gradient and update model parameters
    \ENDWHILE
\end{algorithmic}
\end{small}
\end{algorithm}
\end{minipage}
}
\hspace{0.5em}
\scalebox{0.95}{
\begin{minipage}[t]{0.47\linewidth}
\begin{algorithm}[H]
\begin{small}
   \caption{\method Sampling}
   \label{alg:sample}
\begin{algorithmic}[1]
    \STATE Initialize $\rvx$, $\rvr$, $\rvz$ from noise distribution
\FOR{$i = 0$ {\bfseries to} $N-1$}
    \STATE Predict $\rvv_{\rvx,t}^{\theta}$, $\rvv_{\rvz,t}^{\theta}$, $\rvz_{\text{init},t}^{\theta}$, $\rvc^{\theta}_{1|t}$
    \IF{${\textrm{d}t}_{\rvx} > 0$}
        \STATE Update $\rvx$ with Eq. (\ref{eq:main_sde_ode_ca})
    \ENDIF
    \IF{${\textrm{d}t}_{\rvr} > 0$}
        \STATE Unmask $\rvr$ with prob.\ ${\textrm{d}t}_{\rvr}\cdot\frac{1 + \eta t_{\rvr}}{1 - t_{\rvr}}$
        \STATE Remask $\rvr$ with prob.\ ${\textrm{d}t}_{\rvr}\cdot\eta$
    \ENDIF
    \IF{${\textrm{d}t}_{\rvz} > 0$}
        \FOR{each residue $j$}
            \STATE If unmasked: update $\rvz_j$ with Eq. (\ref{eq:main_sde_ode_nonca})
            \STATE If newly unmasked: set $\rvz_j \leftarrow \rvz_{\text{init},t}^{\theta}$
            \STATE If masked: set $\rvz_j \leftarrow 0$
        \ENDFOR
    \ENDIF
\ENDFOR
\end{algorithmic}
\end{small}
\end{algorithm}
\end{minipage}
}
\vspace{-4mm}
\end{figure}

\subsection{\ourmodel Training}
% \input{figures_tex/algorithm_box}

% \textbf{Training Objective.}
% The training process is outlined in Alg.~\ref{alg:train}.
% We begin by sampling noisy inputs for each modality as described in Sec.~\ref{sec:multi-modal_flow_matching}, and feed them into the model.
% The training objective for $C_\alpha$ atoms follows that of \basemodel, while the amino acid sequence is trained using a standard cross-entropy loss.
% For non-$C_\alpha$ atoms, we use two objectives: (i) for unmasked residues, we apply the flow matching loss to their existing atoms, analogous to the $C_\alpha$ case; (ii) for masked residues, we regress the predicted pseudo-velocity $\rvw_{\text{init},t}^{\theta}$ toward the true velocity. Note that this is called pseudo-velocity, as it is based on representations from pseudo-atom token without knowing the input noisy coordinates. This pseudo-velocity enables a one step side chain noisy initialization in the event of unmasking a residue token. Intuitively, in our framework if the residue type is unknown we cannot model side chain structure. In the unmasking event we cannot use $\rvv_{\rvz,t}^{\theta}$ as that is conditioned on $\rvz_t$. Thus $\rvw_{\text{init},t}^{\theta}$ is required to bridge the gap betweenn having no side chain structure and the conditional flow.
The training process is outlined in Alg.~\ref{alg:train}. We start by sampling time steps to create noisy inputs for each modality (Sec.~\ref{sec:multi-modal_flow_matching}) and feeding them into the model. 
% \kk{What are the distributions from which the three t's are sampled from by the way?}\textcolor{red}{It's described right after this comment.}. 
Both $C_\alpha$ and non-$C_\alpha$ sample time from the mixed uniform-beta distribution from \basemodel~\cite{geffner2025proteina} and the sequence time is sampled from $\mathcal{U}(0,1)$.
The training objectives are as follows: for $C_\alpha$ atoms, we use the standard conditional flow matching objective, while for amino acid sequences, we use a standard cross-entropy loss. For non-$C_\alpha$ atoms (i) for unmasked residues, we apply the flow matching loss to existing atoms, similar to $C_\alpha$ atoms; (ii) for masked residues, we regress the predicted pseudo-velocity $\rvz_{\text{init},t}^{\theta}$ towards an augmented objective as discussed in Sec.~\ref{sec:model_arch}.
See Appendix for further training details. 

% This pseudo-velocity is necessary for one-step side chain noisy initialization when a residue token is unmasked. Since when the residue type is unknown, we cannot model side chain structure. When unmasked, we cannot use $\rvv_{\rvz,t}^{\theta}$ as it is conditioned on $\rvz_t$. Thus, $\rvw_{\text{init},t}^{\theta}$ bridges the gap between having no side chain structure and the conditional flow. Empirically, learning the pseudo-velocity is more stable than clean data prediction when both do not have access to the noisy input data.

%~\sref{app:training_details}. \kk{If we change the appendix structure next week, this reference won't be accurate. Let's just generally refer to Appendix.}
% \textcolor{red}{@Danny, how can we explain this design?}
% Additionally, we apply self-conditioning with probability 0.5, conditioning the model on its own clean predictions for each modality.

% 10\% of training data batch to forward folding ($t_{\rvr} = 1$), 10\% to inverse folding ($t_{\rvx} = 1$), and 5\% to side-chain packing ($t_{\rvx} = 1$ and $t_{\rvr} = 1$). \dr{will check its either 10/10/5 or 5/5/0. Its both depending on motif or not so I have rephrased}

% \begin{align}
%     \gL_{\rvz,i} = 
% \begin{cases}
%   \left\|\rvv^{\theta}_{\rvz,t,i}-(\rvz_i-\boldsymbol{\epsilon}_{\rvz,i})\right\|_2^2, & \quad\rvr_i \neq \mask \\
%   \left\|\rvz^{\theta}_{\text{init},i}-\rvz_i\right\|_2^2, & \quad\rvr_i = \mask
% \end{cases}
% \end{align}

\vspace{-1mm}
\subsection{\ourmodel Sampling}
\vspace{-1mm}
We sample $C_\alpha$ atoms by simulating the learned flow via an SDE. 
% \kk{This is confusing. We say in the next sentence that we don't use ODE, but SDE. Let's reformulate a bit.} 
Since our flow is Gaussian, it relates to the score function as: $\rvs_{\rvx,t}^{\theta} = (t\rvv^\theta_{\rvx,t} - \rvx_t)/(1 - t)$.
This allows us to define an SDE for sampling
% \kk{Langevin is only the last two terms. Let's just call it ``generative SDE for sampling'' or such}:
\begin{equation} \label{eq:main_sde_ode_ca}
\textrm{d}\rvx_t = \rvv^\theta_{\rvx,t}\,\textrm{d}t + g_{\rvx}(t)\rvs_{\rvx,t}^\theta\,\textrm{d}t + \sqrt{2g_{\rvx}(t)\gamma_{\rvx}}\,\textrm{d}\mathcal{W}_t,
\end{equation}
with noise scale $\gamma_{\rvx}$ and Wiener process $\mathcal{W}_t$. Setting $\gamma_{\rvx}{=}1$ produces the model's marginal distribution, while reducing $\gamma_{\rvx}$ can boost designability by lowering noise during generation, at the cost of diversity.

Following MultiFlow~\cite{yim2024multiflow}, for sequence sampling, we effectively perform iterative unmasking and remasking. Starting with a fully masked sequence $[\mask]^L$, at each timestep $t$, we predict residue type logits $\smash{\rvc^\theta_{1|t}}$ and sharpen the distribution using a temperature $\tau$ to obtain probabilities $p_{1|r}(\rvr) = \text{softmax}(\rvc^\theta_{1|t} / \tau)$. Each masked residue is then unmasked with probability $\smash{{\textrm{d}t}\cdot (1 + \eta t) / (1 - t)}$, where $\eta$ controls sampling stochasticity, and its type is sampled from $p_{1|r}(\rvr)$. To maintain balance, each unmasked residue is subsequently remasked with probability ${\textrm{d}t}\cdot\eta$. We also explore recent advances in discrete diffusion sampling algorithms~\citep{tang2022improved, peng2025path}.
% For further details please see Appendix. %~\sref{app:discrete_diffusion}. \kk{Better general Appendix ref.}

The generation of non-$C_\alpha$ atoms depends on the sequence generation process. Following the flow matching framework in Sec.~\ref{sec:multi-modal_flow_matching}, we begin generating non-$C_\alpha$ atoms for a residue only after it is unmasked. Accordingly, the generation process falls into three cases: (1) If the residue is already unmasked, we update its non-$C_\alpha$ coordinates using the same SDE as for $C_\alpha$ atoms: 
\begin{equation} \label{eq:main_sde_ode_nonca}
\textrm{d}\rvz_t = \rvv^\theta_{\rvz,t}\,\textrm{d}t + g_{\rvz}(t)\rvs_{\rvz,t}^\theta\,\textrm{d}t + \sqrt{2g_{\rvz}(t)\gamma_{\rvz}}\,\textrm{d}\mathcal{W}_t;
\end{equation}
(2) if the residue is newly unmasked at the current step, we initialize its atom coordinates using a single step Euler integration using the the predicted initialization $\rvz^\theta_{\text{init},t}$; (3) if the residue is still masked or has been remasked, we set $\rvz$ to zero. 
This framework enables the concurrent generation of side chains alongside backbones and sequences, contrasting with methods that generate side chains after backbone and sequence generation. This simultaneous approach allows for an increased influence of side chains on the local structure while retaining the flexibility to alter sequence identities. 
% Furthermore, it updates side-chain coordinates more frequently than Protpardelle~\cite{chu2024protpardelle}, which only updates coordinates based on the current residue type.

% In contrast to Protpardelle~\cite{chu2024protpardelle}, which models all possible side chains simultaneously, our three cases were designed to facilitate the modeling of a single side chain at a time, while accommodating the variable length of side chain atoms. 
% We believe that by explicitly modeling a single side chain that can better influence the structure around it while still preserving the ability to change sequence identities.
% \kk{Anything we can say why we believe our approach is not only different, but also better? I.e. answer the ``why?''.}

The sampling process is detailed in Alg.~\ref{alg:sample}. As we use distinct time schedules for the three modalities, we denote their respective timesteps with corresponding subscripts and use $N$ to denote the number of timesteps. 
Our flexible and general framework, in principle, allows for sampling modalities in any order by adjusting these time schedules. 
% Further details can be found in the Appendix.
% \kk{Could consider saying that our flexible and general framework in principle allows to sample in any order, any modalities.} %~\sref{app:sec:multi-modal_flow_matching} for more details. \kk{fix App ref}

% \textbf{multi-modal Inference Time Schedule.}
% \textcolor{red}{@Zuobai, to be decided whether to include this or not.}
% multi-modal controllability by adjusting time step.
% Vanilla case: use the same number of timesteps and use 
% Fine-grained refinement: log schedule for structures
% When there are multiple modalities, three separate schedules are introduced for three modalities.
% Here we introduce two schemes.

% Time padding: refine one modality after the generation of another. Refine structure after sequence or align sequence with the generated structure

% Time dilation: refine one modality during the generation of the other one.

\subsection{Defining Noise Schedules via the Time Distribution}
\label{app:sec:training_time_distribution}
\textbf{Training Time Sampling Distribution.}
A key design choice in diffusion and flow matching models is the time sampling distribution $p(t)$, which effectively controls how the training objective is weighted across different stages of the generative process. 
Here, since we consider three distinct modalities, we sample time steps independently for each.
\ourmodel proposes to bias sampling toward later timesteps ($t\approx1$) to encourage the model to allocate more capacity to generating fine-grained local structure. %, which is refined near the end of generation.
Specifically, for flow matching in Euclidean space—i.e., for $\rvx$ and $\rvz$—we use a mixed Beta distribution~\citep{geffner2025proteina} for $t_{\rvx}$ and $t_{\rvz}$.
\begin{equation} \nonumber
p(t) = 0.02\,\mathcal{U}(0,1) + 0.98\,\mathcal{B}(1.9,1.0),
\end{equation}
where $\mathcal{B}(\cdot,\cdot)$ is the Beta distribution.
For the discrete modality $\rvr$, we sample $t_{\rvr}$ from $\mathcal{U}(0,1)$.
Additionally, following \cite{yim2024multiflow}, we give the options to allocate a small percentage of each training batch to forward folding ($t_{\rvr} = 1$), inverse folding ($t_{\rvx} = 1$) and also extend to side chain packing ($t_{\rvx} = 1$ and $t_{\rvr} = 1$). Please see~\tref{tab:model_hparams} for specific ratios for each model configuration.

\subsection{Side Chain Initialization}
\label{app:side_chain_init}
% In Fig.~\ref{fig:model_arch}, the initialization $\rvz^\theta_{\text{init},t}$ is predicted from atom representations that are also used for the vector field $\rvv^\theta_{\rvz,t}$. 
% We observe that directly predicting clean coordinates is challenging due to both the high variance of coordinate values and our model's non-equivariant nature. To address this, we introduce an augmented objective that predicts $\rvz-\boldsymbol{\epsilon}_{\rvz}$, where $\boldsymbol{\epsilon}_{\rvz}$ is standard Gaussian noise not visible to the model. This approach is effective for two key reasons: (1) it maintains consistency with the vector field objective for existing atoms, (2) since $\boldsymbol{\epsilon}_{\rvz}$ is hidden from the model, the optimal prediction becomes the average value $\E[\rvz-\boldsymbol{\epsilon}_{\rvz}]=\rvz-\E[\boldsymbol{\epsilon}_{\rvz}]=\rvz$. This formulation ensures the prediction converges to the clean coordinates $\rvz$, which then properly initialize newly unmasked side chain atoms. 

In Fig.~\ref{fig:model_arch}, the initialization $\rvz^\theta_{\text{init},t}$ is predicted from atom representations that are also used for the vector field $\rvv^\theta_{\rvz,t}$.  
Notably, the model does not have access to $\rvz_t$ for masked residues due to the structures being undefined for unknown residue types. This differs from the standard flow matching objective, which predicts a vector field conditioned on the noisy input. We have found that separating the initialization from the standard structure-to-structure vector field works best in practice (\tref{tab:init_aa_abl}).

In \tref{tab:init_aa_abl} we empirically observed that directly predicting clean coordinates is challenging due to their high variance and our model's non-equivariant nature. To address this, we introduce an auxiliary objective that predicts $\rvz - \boldsymbol{\epsilon}_{\rvz}$, where $\boldsymbol{\epsilon}_{\rvz}$ is standard Gaussian noise not visible to the model. This formulation is effective for two reasons: (1) it aligns with the vector field objective for existing atoms, and (2) since $\boldsymbol{\epsilon}_{\rvz}$ is not known to the model, the optimal prediction converges to $\E[\rvz - \boldsymbol{\epsilon}_{\rvz}] = \E[\rvz] - \E[\boldsymbol{\epsilon}_{\rvz}] = \E[\rvz]$, ensuring that the prediction converges to the average clean coordinates. This, in turn, properly initializes newly unmasked side chain atoms.

An alternative interpretation of this augmented objective is that we aim to learn an augmented vector field that transforms a random starting point with average $\E[\boldsymbol{\epsilon}_{\rvz}] = 0$ to the average clean data $\E[\rvz]$. During generation, we can then obtain an initialization by performing a single-step Euler integration from noise ($t_\text{pre-init} = 0$) towards ``clean data'' ($t_\text{init} = 1$) using the learned vector field. 
We assume the side-chain structures for masked residues originate from zero. This conceptually means the side-chain coordinates are initially hidden behind the $C_\alpha$ atoms (in local coordinates) before being unmasked.

% This can be interpreted as learning an initialization vector field, where a single-step Euler integration is performed from $t_\text{pre-init} = 0$ to $t_\text{init} = 1$, yielding $\rvz_t = \rvz_\text{init} + \rvv^\theta_{\rvz,t} (t_\text{init} - t_\text{pre-init})$. We set $\rvz_\text{init} = 0$, as side chain structures are only defined when the residue type is known. In other words, the side chain structure of a MASK residue is zero. \dr{everyone to read. we need some sort of analogy to explain this comment left in the main body}

\subsection{Two Stage Training}
We used a training + finetuning strategy to train \textit{\ourmodel} on \ourdata. The model was first trained on a subset of \ourdata containing proteins with lengths ranging from 32 to 256. The model is then finetuned on the full \ourdata with protein lengths ranging from 32 to 400. The model with triangle multiplicative layers (\textit{\ourmodel-tri}) was only trained on protein lengths ranging from 32 to 256. We recorded the number of steps and learning rate in both training and finetuning stages for each variant of the model in~\tref{tab:model_hparams}.

\subsection{Details in Multimodal Flow Matching}
\label{app:sec:multimodal_flow_matching}
We present the detailed version of our training algorithm in Alg.~\ref{alg:training_detail}.
Here, SampleTimestep() is the function to sample timesteps for each modality based on the training time distributions in Sec.~\ref{app:sec:training_time_distribution} and Global2Local() is the function to transform global coordinates to local coordinates, where the transformation scheme (local translations or local frames) is chosen as a hyperparameter.

\textbf{Global2Local:} Non-$C_\alpha$ atoms are structurally organized around their corresponding $C_\alpha$ atoms. 
We offer two local coordinate modeling strategies to leverage this property, simplifying the learning task by predicting offsets rather than global coordinates and facilitating better initialization of non-$C_\alpha$ atoms.
The first approach calculates the relative position of non-$C_\alpha$ atoms directly with respect to their corresponding $C_\alpha$ atom: $\rvz_i^\text{local} = \rvz_i - \rvx_i$.
The second strategy, inspired by related work~\citep{lin2024genie2}, constructs a residue-centric local coordinate frame $(\rvt_i, \mathbf{R}_i)$ using the $C_\alpha$ coordinates of three neighboring residues ($\rvx_{i-1}$, $\rvx_i$, $\rvx_{i+1}$) via the Gram-Schmidt process. Non-$C_\alpha$ coordinates $\rvz_i$ are then transformed to local ones via $\rvz_i^\text{local\_frame} = \mathbf{R}_i^{-1}(\rvz_i - \rvt_i)$. 
In the following sections, models denoted by \textit{local trans} employ the local translation parameterization, while those denoted by \textit{local frame} utilize the frame-based parameterization.

\begin{algorithm}[h]
   \caption{\method Training}
   \label{alg:training_detail}
\begin{algorithmic}[1]
    \STATE {\bfseries Input:} $C_\alpha$ atom $\rvx\in \R^{L\times3}$, amino-acid sequence $\rvr\in\{0,...,19\}^L$, non-$C_\alpha$ atom $\rvz\in \R^{L\times36\times3}$ 

    \STATE
    \WHILE{not converged}
    \STATE \# \textbf{Step 1: Noising Process}
    \STATE
    $t_{\ca}, t_{\seq}, t_{\nonca} \gets \text{SampleTimestep}()$
    \STATE $\rvr_t \sim t_{\seq}\delta\{\rvr\} + {1-t_{\seq}}\delta(\mask)$
    \STATE $\boldsymbol{\epsilon}_{\ca}\sim \boldsymbol{\gN}(\boldsymbol{0}, \mI)\in \R^{L\times3}, \boldsymbol{\epsilon}_{\nonca}\sim \boldsymbol{\gN}(\boldsymbol{0}, \mI)\in \R^{L\times36\times3}$
    \STATE $\rvx_t \gets t_{\ca}\rvx+(1-t_{\ca})\boldsymbol{\epsilon}_{\ca}$
    \STATE $\rvz \gets \text{Global2Local}(\rvz,\rvx)$ \quad\quad \# if using local coordinates
    \STATE $\rvz_t \gets t_{\nonca}\rvz+(1-t_{\nonca})\boldsymbol{\epsilon}_{\nonca}$
    \STATE Zero out non-existing atoms in $\rvz_t$ based on $\rvr_t$
    \STATE
    
    \STATE \# \textbf{Step 2: Neural Network}
    \STATE $\rvv_{\rvx,t}^{\theta},\rvv_{\rvz,t}^{\theta},\rvz_{\text{init},t}^{\theta},\rvc^{\theta}_{1|t} \gets \text{Transformer}(\rvx_t,\rvr_t,\rvz_t,\emptyset,\emptyset,\emptyset,t_{\ca},t_{\seq},t_{\nonca})$
    \IF{$\text{rand(0, 1)}>0.5$}
    \STATE $\bar{\rvr} \gets \arg\max \rvc^{\theta}_{1|t}$
    \STATE $\bar{\rvx} \gets \rvx_t + (1-t_{\ca})\rvv_{\ca,t}^{\theta}$
    \STATE $\bar{\rvz} \gets \rvz_t + (1-t_{\nonca})\rvv_{\nonca,t}^{\theta}$
    \STATE $\rvv_{\rvx,t}^{\theta},\rvv_{\rvz,t}^{\theta},\rvz_{\text{init},t}^{\theta},\rvc^{\theta}_{1|t} \gets \text{Transformer}(\rvx_t,\rvr_t,\rvz_t,\text{sg}(\bar{\rvx}),\text{sg}(\bar{\rvr}),\text{sg}(\bar{\rvz}),t_{\ca},t_{\seq},t_{\nonca})$
    \ENDIF
    
    \STATE
    \STATE \# \textbf{Step 3: Loss Calculation}
    \STATE $\gL_{\seq} \gets \text{CrossEntropy}(\rvc^{\theta}_{1|t},\rvr)$
    \STATE $\gL_{\ca} \gets \frac{1}{L}||\rvv^{\theta}_{\ca,t}-(\rvx-\boldsymbol{\epsilon}_{\ca})||_2^2$
    \FOR{each residue $i$}
    \STATE $\gL_{\rvz,i}\gets
  \|\rvv^{\theta}_{\rvz,t,i}-(\rvz_i-\boldsymbol{\epsilon}_{\rvz,i})\|_2^2, \text{if}\,\rvr_{t,i} \neq \mask$
    \STATE $\gL_{\rvz,i}\gets
  \|\rvz_{\text{init},t,i}^{\theta}-(\rvz_i-\boldsymbol{\epsilon}_{\rvz,i})\|_2^2, \text{if}\,\rvr_{t,i} = \mask$
    \ENDFOR
    \STATE $\gL \gets \frac{1}{L} \left(\gL_{\rvx}+\gL_{\rvr}+\gL_{\rvz}\right)$
    \STATE
    % \STATE $\theta \gets \theta - \alpha \nabla (\gL_{\seq}+\gL_{\ca}+\gL_{\nonca})$
    \STATE Calculate gradient and update model parameters
    \ENDWHILE
\end{algorithmic}
\end{algorithm}

\section{Inference Details and Hyperparameters}
We present a detailed version of the sampling algorithm in Alg.~\ref{alg:sampling_detail}.
\begin{figure}[htp]
\centering
\vspace{-8mm}
\scalebox{0.95}{
\begin{minipage}[t]{\linewidth}
\begin{algorithm}[H]
   \caption{\method Multimodal Sampling}
   \label{alg:sampling_detail}
\begin{algorithmic}[1]
    \STATE {\bfseries Input:} discretized timesteps for three modalities $\{t_{\rvx,i}\}_{0..N}$, $\{t_{\rvr,i}\}_{0..N}$, and $\{t_{\rvz,i}\}_{0..N}$, stochasticity schedules $g_{\rvx}(t)$ and $g_{\rvz}(t)$, noise scales $\gamma_{\rvx},\gamma_{\rvz},\eta$, sequence temperature $\tau$
    \STATE {\bfseries Output:} generated proteins $(\rvx,\rvr,\rvz)$

    \STATE $\rvx \sim \boldsymbol{\gN}(\boldsymbol{0}, \mI)\in \R^{L\times3}$
    \STATE $\rvz \sim \boldsymbol{\gN}(\boldsymbol{0}, \mI)\in \R^{L\times36\times3}$
    \STATE $\rvr \gets [\mask]^L$
    
    \FOR{$i=0$ {\bfseries to} $N-1$}
    \STATE $\rvv_{\rvx,t}^{\theta},\rvv_{\rvz,t}^{\theta},\rvz_{\text{init},t}^{\theta},\rvc^{\theta}_{1|t} \gets \text{Transformer}(\rvx,\rvr,\rvz,\emptyset,\emptyset,\emptyset,t_{\rvx,i},t_{\rvr,i},t_{\rvz,i})$

    \IF{self-condition}
    \STATE $\bar{\rvr} \gets \arg\max \rvc^{\theta}_{1|t}$
    \STATE $\bar{\rvx} \gets \rvx_t + (1-t_{\ca,i})\rvv_{\ca,t}^{\theta}$
    \STATE $\bar{\rvz} \gets \rvz_t + (1-t_{\nonca,i})\rvv_{\nonca,t}^{\theta}$
    \STATE $\rvv_{\rvx,t}^{\theta},\rvv_{\rvz,t}^{\theta},\rvz_{\text{init},t}^{\theta},\rvc^{\theta}_{1|t} \gets \text{Transformer}(\rvx_t,\rvr_t,\rvz_t,\bar{\rvx},\bar{\rvr},\bar{\rvz},t_{\ca,i},t_{\seq,i},t_{\nonca,i})$
    \ENDIF

    \STATE
    \STATE \# Update CA Atoms
    \STATE ${\textrm{d}t}_{\rvx} = t_{\rvx,i+1}-t_{\rvx,i}$
    \IF{ ${\textrm{d}t}_{\rvx}>0$ }
    \STATE $\hat{\rvx} \gets \rvx + \rvv^{\theta}_{\rvx,t} \textrm{d}t_{\rvx} + g_{\rvx}(t_{\rvx,i})\rvs_{\rvx,t}^{\theta} \textrm{d}t_{\rvx} + \sqrt{2g_{\rvx}(t_{\rvx,i})\gamma_{\rvx}\,\textrm{d}\mathcal{W}_t}$
    \ENDIF

    \STATE
    \STATE \# Update Amino-Acid Sequence
    % \STATE ${\textrm{d}t}_{\rvr} = t_{\rvr,i+1}-t_{\rvr,i}$
    % \IF{ ${\textrm{d}t}_{\rvr}>0$ }
    % \STATE $\hat{\rvr}_1 \sim \text{Softmax}(\rvc^{\theta}_{1|t} / \tau)$
    % \STATE $p_{\text{unmask}} \gets {\textrm{d}t}_{\rvr} \cdot (1 + \eta t_{\rvr,i}) / (1 - t_{\rvr,i})$
    % \STATE $p_{\text{remask}} \gets {\textrm{d}t}_{\rvr} \cdot \eta$
    % % \STATE $\hat{\rvr} \gets \rvr$
    % \FOR{$j=1$ {\bfseries to} $L$}
    % \IF{$\rvr_j=\mask$}
    % \STATE $\hat{\rvr}_j \sim (1-p_{\text{unmask}})\delta\{\mask\}+p_{\text{unmask}}\delta\{\hat{\rvr}_{1,j}\}$
    % \ELSE
    % \STATE $\hat{\rvr}_j \sim (1-p_{\text{remask}})\delta\{\rvr_j\}+p_{\text{remask}}\delta\{\mask\}$
    % \ENDIF
    % \ENDFOR
    % \ENDIF

    \STATE ${\textrm{d}t}_{\rvr} = t_{\rvr,i+1}-t_{\rvr,i}$
    \IF{ ${\textrm{d}t}_{\rvr}>0$ }
    \IF{sampling\_alg = PURITY}
    \STATE $\hat{\rvr} \gets \text{purity\_sample}(\rvc^{\theta}_{1|t}, {\textrm{d}t}_{\rvr}, \eta, \tau, \hat{\rvr} )$ (\aref{alg:purity-def})
    \ELSIF{sampling\_alg = P2}
    \STATE $\hat{\rvr} \gets \text{p2\_sample}(\rvc^{\theta}_{1|t}, {\textrm{d}t}_{\rvr}, \eta, \tau, \hat{\rvr} )$ (\aref{alg:p2-def})
    \ELSE
    \STATE $\hat{\rvr}_1 \sim \text{Softmax}(\rvc^{\theta}_{1|t} / \tau)$
    \STATE $p_{\text{unmask}} \gets {\textrm{d}t}_{\rvr} \cdot (1 + \eta t_{\rvr,i}) / (1 - t_{\rvr,i})$
    \STATE $p_{\text{remask}} \gets {\textrm{d}t}_{\rvr} \cdot \eta$
    \FOR{$j=1$ {\bfseries to} $L$}
    \IF{$\rvr_j=\mask$}
    \STATE $\hat{\rvr}_j \sim (1-p_{\text{unmask}})\delta\{\mask\}+p_{\text{unmask}}\delta\{\hat{\rvr}_{1,j}\}$
    \ELSE
    \STATE $\hat{\rvr}_j \sim (1-p_{\text{remask}})\delta\{\rvr_j\}+p_{\text{remask}}\delta\{\mask\}$
    \ENDIF
    \ENDFOR
    \ENDIF
    \ENDIF

    \STATE
    \STATE \# Update Non-CA Atoms
    \STATE ${\textrm{d}t}_{\rvz} = t_{\rvz,i+1}-t_{\rvz,i}$
    \IF{ ${\textrm{d}t}_{\rvz}>0$ }
    % \STATE $\hat{\rvz} \gets \rvz$
    \FOR{$j=1$ {\bfseries to} $L$}
    \IF{$\rvr_j\neq\mask\,\text{and}\,\hat{\rvr}_j\neq \mask$}
    \STATE $\hat{\rvz}_j \gets \rvz_j + \rvv^{\theta}_{\rvz,t,j} \textrm{d}t_{\rvz} + g_{\rvz}(t_{\rvz,i})\rvs_{\rvz,t,j}^{\theta} \textrm{d}t_{\rvz} + \sqrt{2g_{\rvz}(t_{\rvz,i})\gamma_{\rvz}\,\textrm{d}\mathcal{W}_t}$
    \ELSIF{$\rvr_j=\mask\,\text{and}\,\hat{\rvr}_j\neq \mask$}
    \STATE $\rvz_j\gets\rvz_{\text{init},t,j}^{\theta}$
    \ELSE
    \STATE $\rvz_j=0$
    \ENDIF
    \ENDFOR
    \ENDIF
    \STATE $\rvx, \rvr, \rvz \gets \hat{\rvx}, \hat{\rvr}, \hat{\rvz}$
    \ENDFOR
    
    \STATE {\bfseries Return} $(\rvx,\rvr,\rvz)$
\end{algorithmic}
\end{algorithm}
\end{minipage}
}
\vspace{-4mm}
\end{figure}
\subsection{Inference Time Schedules}
\begin{figure}[htb]
    % \vspace{-8mm}
    \centering
    \includegraphics[width=0.55\linewidth]{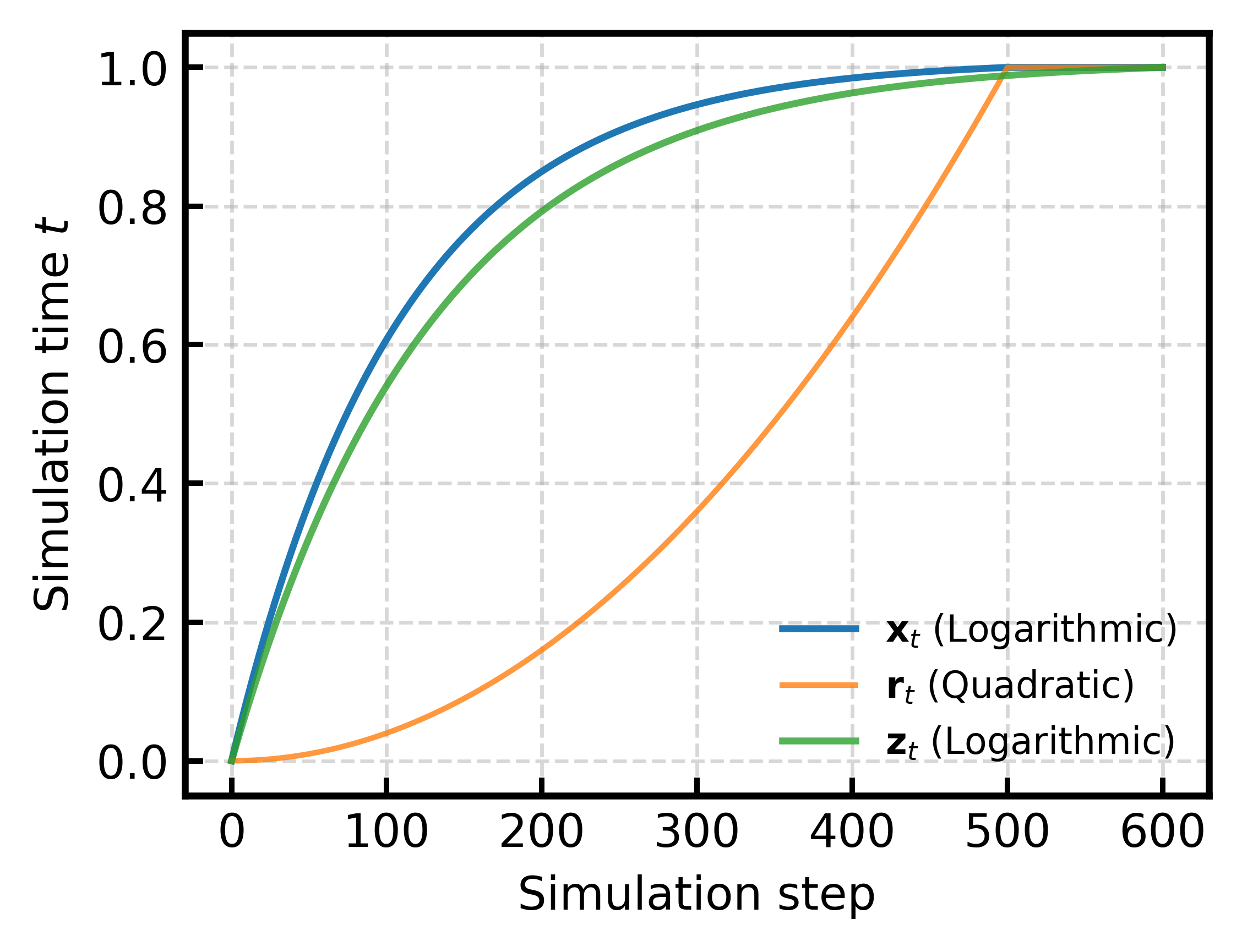}
    \vspace{-1mm}
    \caption{\small \textbf{Discretized $t$ schedule during sampling} for backbone $C_\alpha$ ($\rvx_t$), amino acid residue types ($\rvr_t$), and all non-$C_\alpha$ atoms ($\rvz_t$). The last 100 steps of $\rvx_t$ and $\rvr_t$ are padded with 1.}
    \label{fig:sample_t}
    % \vspace{-5mm}
\end{figure}

We sample from \method following Alg.~\ref{alg:sample} for the $C_\alpha$ coordinates, residue types, and non-$C_\alpha$ backbone and side chain atoms, integrating from $t=0$ to $t=1$. For the coordinates of $C_\alpha$ ($\mathbf{x}$) and non-$C_\alpha$ atoms ($\mathbf{z}$), we simulate the SDE (\eref{eq:main_sde_ode_ca} and \eref{eq:main_sde_ode_nonca}) with the following definition for $g(t)$:
\begin{equation*}
\begin{cases}
    g(t)= 1/(t+0.01),\quad t\in[0, 0.99] \\
    g(t) = 0,\quad t\in(0.99, 1)
\end{cases}    
\end{equation*}
We use $N_{\mathbf{x}}=500$ and $N_{\mathbf{z}}=600$ steps to discretize the unit interval into logarithmically spaced points. The PyTorch~\citep{paszke2019pytorch} code snippet to generate the logarithmic discretization is as follows:
\begin{equation*}
\begin{aligned}
t &= 1.0 - \text{torch.logspace}(-2, 0, \text{nsteps} + 1).\text{flip}(0) \\
t &= t - \text{torch.min}(t) \\
t &= t / \text{torch.max}(t)
\end{aligned}
\end{equation*}
% \verb|                    t = 1.0 - torch.logspace(-2, 0, nsteps + 1).flip(0)|\\
% \verb|                    t = t - torch.min(t)|\\
% \verb|                    t = t / torch.max(t)|\\
which ensures that $t\in[0, 1]$. For the sampling of residue types ($\mathbf{r}$), we use the $N_{\mathbf{r}}=500$ steps to discretize the unit interval into quadratically spaced points. $\rvx_t$ and $\rvr_t$ are padded with ones for extra 100 steps to match the total number of 600 steps of the simulation. \fref{fig:sample_t} visualizes the discretized $t$-schedule of different modalities during sampling. 

% ZC: Moved the multimodal sampling algo to an individual .tex file
% ZC: Moved discrete sampling algos to individual .tex files
% ZC: those algo files are now in algo_tex

\subsection{Backbone-Sequence Co-Design}
For both the backbone-sequence co-design and all-atom generation tasks, our models perform best with a combination of the offset schedules defined above and various low-temperature settings across the explicit data modalities that the model has access to (backbone $C_\alpha$, residue types, other backbone and side chain atoms). Here we detail the specific parameters and how they impact the generated results.

\subsubsection{Low Temperature Sampling for Backbone $C_{\alpha}$}
The stochasticity of backbone sampling is handled in the same way as done in \basemodel~\cite{geffner2025proteina}. This means we have a single noise scale to decrease the impact of the noise in relation to the score and vector field contributions. This can be seen in \eref{eq:main_sde_ode_ca} where $\gamma_{\rvx}$ refers to the backbone noise scale shown in~\tref{tab:model_hparams}.

\subsubsection{Discrete Diffusion Sampling for Residue Types}
\label{app:discrete_diffusion}
As detailed in~\aref{alg:sampling_detail}, we investigate two discrete diffusion sampling algorithms to refine the incorporation of stochasticity in the masking and unmasking processes. We adopt purity sampling~\citep{tang2022improved} (\aref{alg:purity-def}), which prioritizes unmasking tokens with high confidence, and self-path planning (P2) sampling~\citep{peng2025path} (\aref{alg:p2-def}), which instead encourages remasking tokens with low confidence. 
% Although the implementations of these algorithms are similar, the sorting direction and the formulation of remasking via added stochasticity has a significant impact on the distribution of generated amino acid residue types.
% Notably, we identify hyperparameter settings that enable both methods to achieve state-of-the-art results. Interestingly, P2 sampling exhibits an inherent temperature annealing-like behavior (\emph{i.e.} slowly decreasing temperature as a function of time), leading to a substantial reduction ($\sim\frac{2}{3}$) in the overrepresentation of Alanine when training on AFDB, a common issue in sequence-based protein generative models. This finding underscores the importance of carefully designing sampling strategies to mitigate biases in generated sequences.
\begin{algorithm}[!h]
   \caption{Purity Sampling}
   \label{alg:purity-def}
\begin{algorithmic}[1]
\STATE \textbf{Input:} predicted logits $\rvc^{\theta}_{1|t}$, time step $t$, time delta $\text{d}t$, sampling temperature $\tau$, stochasticity $\eta$, current sequence $\rvr_t$
\STATE \textbf{Output:} updated sequence $\hat{\rvr}$

\STATE $\text{probs} \gets \text{Softmax}(\rvc^{\theta}_{1|t} / \tau)$ 
\STATE $\text{MaxLogProb} \gets \max(\log(\text{probs}), \text{dim}=-1)$
\STATE $\text{MaxLogProb} \gets \text{MaxLogProb} - (\rvr_t \neq \mask) \cdot \infty$
\STATE $\text{SortedIndices} \gets \text{ArgSort}(\text{MaxLogProb}, \text{descending}=\text{True})$

\STATE $p_{\text{unmask}} \gets \min(1, \text{d}t \cdot \frac{1 + \eta t}{1 - t})$
\STATE $\text{ToUnMask} \gets (\text{Uniform}(0,1) \leq p_{\text{unmask}}) \land (\rvr_t = \mask)$
\STATE $\text{NumToUnmask} \gets \sum(\text{ToUnMask})$

\STATE $\hat{\rvr}_{\text{samples}} \sim \text{Categorical}(\text{probs})$
\STATE $\hat{\rvr} \gets \rvr_t$
\FOR{$i=1$ \TO $\text{NumToUnmask}$}
    \STATE $\text{idx} \gets \text{SortedIndices}[i]$
    \STATE $\hat{\rvr}_{\text{idx}} \gets \hat{\rvr}_{\text{samples},\text{idx}}$
\ENDFOR

\STATE $p_{\text{remask}} \gets \text{d}t \cdot \eta$
\STATE $\text{ToReMask} \gets (\text{Uniform}(0,1) < p_{\text{remask}}) \land (t + \text{d}t < 1)$
\STATE $\hat{\rvr} \gets \hat{\rvr} \cdot (1 - \text{ToReMask}) + \mask \cdot \text{ToReMask}$

\STATE \textbf{return} $\hat{\rvr}$
\end{algorithmic}
\end{algorithm}%
\begin{algorithm}[htp]
   \caption{P2 Sampling}
   \label{alg:p2-def}
\begin{algorithmic}[1]
\STATE \textbf{Input:} predicted logits $\rvc^{\theta}_{1|t}$, sampling temperature $\tau$, stochasticity $\eta$, current sequence $\rvr_t$
\STATE \textbf{Ouput:} updated sequence $\hat{\rvr}$
\STATE $\kappa(t) = 1 - t$
\STATE $\eps\sim\text{Gumbel}(0,1)$
\STATE $\text{logprob}, \hat{\rvr_1} = \text{LogSoftmax}(\rvc^{\theta}_{1|t} / \tau + \eps)\text{.max(dim=-1)}$
\STATE $\text{score} \gets \text{logprob}$
\STATE $\text{score}[\rvr_t \neq M] \gets \text{score}[\rvr_t \neq M]*\eta$
\STATE ToMask $\gets \text{Top-K-Lowest}_{\kappa(t)}(\text{score})$
\STATE $\hat{\rvr} \gets \hat{\rvr}_1$
% \STATE \textbf{Denoise:}
\FOR{$j \in \text{ToMask}$ }
\IF{$[\rvr]_j \neq \mask$}
\STATE $[\rvr]_j \gets \mask$
\ENDIF
\ENDFOR
\FOR{$j \notin \text{ToMask}$}
\IF{$[\rvr]_j =\mask$}
\STATE $[\rvr]_j \gets [\hat{\rvr}_1]_j$
\ENDIF
\ENDFOR
\end{algorithmic}
\end{algorithm}

In total, three hyperparameters define the discrete flow matching sampling: (1) the sampling algorithm type, either purity or P2; (2) the temperature $\tau$, which directly controls the predicted logits, with $\tau < 1$ emphasizing likely residue types and $\tau > 1$ acting as a distribution smoothing parameter; and (3) the stochasticity ($\eta$), which governs the remasking ratio.

We test both purity and P2 sampling for all models and select the optimal settings based on the empirical results.

\subsection{Stochastic Side Chain and Non-$C_\alpha$ Sampling}

% The stochasticity of side chain sampling is handled in the same manner as the backbones once we have an unmasked residue. 
Once a residue is unmasked, side chains and non-$C_\alpha$ atoms are updated with the same stochastic SDE as used for backbone $C_\alpha$ generation.
This can be seen in \eref{eq:main_sde_ode_nonca} where $\gamma_{\rvz}$ refers to the side chain and non-$C_\alpha$
noise scale shown in~\tref{tab:model_hparams}.
When the residue is masked, we do not update the side chain structure as there is not yet a residue to predict the structure of. In the event of a residue being unmasked we follow the steps described in~\sref{app:side_chain_init}

% Our flow matching design is constructed specifically to handle dynamic changes of residue types during generation and, as a result, allow a change of sequence length of the number of atoms as a function of time.
Our flow matching design is tailored to accommodate dynamic changes in residue types during generation, enabling flexible atom sequence lengths, \emph{i.e.} a variable number of side chain atoms as a function of time.
For example, consider a single residue during generation where 0 < $t_\alpha < t_\beta < t_\omega < 1$. At $t_\alpha$, the current residue is alanine with associated structures $\rvx_{t_{\alpha}}, \rvz_{t_{\alpha}}$. At $t_\beta$, the alanine is replaced with a mask token, and the side chain and non-$C_\alpha$ backbone structure $\rvz_{t_{\alpha}}$ is zeroed out since the structure of an unknown residue is undefined. Then, at $t_\omega$, the residue token is unmasked to lysine, requiring the initialization of the side chain structure with a different number of atoms than previously used at $t_\alpha$.

% \input{appendix/K_release}
% \newpage
\section{Ablation Studies}

% \begin{table*}[!h]
% \caption{Inference Parameters for \ourmodel and \basemodel-Co-design with (max trained length, local side chain coord parameterization).}
% \label{tab:model_hparams}
% \centering
% \scalebox{0.6}{
% \begin{tabular}{l | ccccc}
% \toprule
% \textbf{Model}  &  \textbf{Backbone Noise Scale}  & \textbf{Sequence Algo} & \textbf{Sequence Temperature} & \textbf{Sequence Noise} & \textbf{Sidechain Noise Scale} \\
% \midrule
% \ourmodel$_{\mathrm{div}}$ (256, local trans) & 0.62 & P2 & 0.20 & 5.0 & 0.45 \\ % trans 132
% \ourmodel$_{\mathrm{codes}}$ (256, local trans) & 0.20 & P2 & 0.45 & 5.0 & 0.60 \\ %trans 19
% \ourmodel$_{\mathrm{div}}$ (256, local frame) & 0.60 & Purity & 0.20 & 0 & 0.60 \\ %zc 125
% \ourmodel$_{\mathrm{codes}}$ (256, local frame) & 0.30 & Purity & 0.20 & 0 & 0.45 \\ %zc 28
% \midrule
% \ourmodel$_\mathrm{div}$ (400, local frame) & x & x & x & x & x \\
% \ourmodel$_\mathrm{opt}$ (400, local frame)  & x & x & x & x & x \\
% \ourmodel$_\mathrm{codes}$ (400, local frame)  & x & x & x & x & x \\
% \midrule
% \ourmodel-tri (400, local frame)  & x & x & x & x & x \\
% \midrule
% \basemodel-Co-design$_{\mathrm{div}}$ (256) & 0.60 & Purity & 0.20 & 5.0 & - \\ % row 49
% \basemodel-Co-design$_{\mathrm{codes}}$ (256) & 0.30 & Purity & 0.20 & 5.0 & - \\ % row 22
% \bottomrule
% \end{tabular}
% }
% \end{table*}
\begin{table*}[!h]
\caption{Inference Parameters for \ourmodel, \codesmodel, and \laprot for backbone $C_\alpha$, sequence, and side chain/non-$C_\alpha$ modalities. For \laprot we report the local latent temperature as side chain noise scale.}
\label{tab:model_hparams}
\centering
\scalebox{0.67}{
\begin{tabular}{l | c | c | c | c | c c c | c}
\toprule
& & & & \textbf{Backbone} & \multicolumn{3}{c|}{\textbf{Sequence}} & \textbf{Side Chain} \\
\textbf{Model}  & \textbf{Where} & \textbf{Train Length} & \textbf{Local Coord} & \textbf{Noise Scale}  & \textbf{Algo.} & \textbf{Temperature} & \textbf{Noise Scale} & \textbf{Noise Scale} \\
\midrule %aa_400_tri_results
\ourmodel$_{\mathrm{div}}$ & \tref{tab:codesign_results_250}, ~\ref{tab:codesign_results_all_atom} & 256 & trans & 0.62 & P2 & 0.20 & 5.0 & 0.45 \\ 
\ourmodel$_{\mathrm{codes}}$ &  \tref{tab:codesign_results_250}, ~\ref{tab:codesign_results_all_atom} & 256 & trans & 0.20 & P2 & 0.45 & 5.0 & 0.60 \\ 
\ourmodel$_{\mathrm{div}}$ &  \tref{tab:codesign_results_250}, ~\ref{tab:codesign_results_all_atom} & 256 & frame & 0.60 & Purity & 0.20 & 0 & 0.60 \\ 
\ourmodel$_{\mathrm{codes}}$ &  \tref{tab:codesign_results_250}, ~\ref{tab:codesign_results_all_atom} & 256 & frame & 0.30 & Purity & 0.20 & 0 & 0.45 \\ 
\midrule
\ourmodel$_{\mathrm{div}}$ &\tref{tab:aa_400_tri_results} & 400 & frame & 0.60 & Purity & 0.45 & 0 & 0.1 \\
\ourmodel$_{\mathrm{opt}}$ & \tref{tab:aa_400_tri_results} & 400 & frame & 0.45 & Purity & 0.30 & 0 & 0.1 \\
\ourmodel$_{\mathrm{codes}}$ & \tref{tab:aa_400_tri_results} & 400 & frame & 0.35 & Purity & 0.30 & 0 & 0.1 \\
\midrule
\ourmodel-tri$_{\mathrm{opt}}$ & \tref{tab:aa_400_tri_results} & 256 & frame & 0.5 & Purity & 0.45 & 0 & 0.3 \\
\midrule
\codesmodel$_{\mathrm{div}}$ & \tref{tab:codesign_results_250} & 256 & N/A & 0.60 & Purity & 0.20 & 5.0 & N/A \\ 
\codesmodel$_{\mathrm{codes}}$ & \tref{tab:codesign_results_250} & 256 & N/A & 0.30 & Purity & 0.20 & 5.0 & N/A \\ 
\midrule
\laprot$_{\mathrm{div}}$ &  \tref{tab:codesign_results_250}, ~\ref{tab:codesign_results_all_atom} & 512 & N/A & 0.30 & N/A & N/A & N/A & 0.1 \\ 
\laprot$_{\mathrm{codes}}$ &  \tref{tab:codesign_results_250}, ~\ref{tab:codesign_results_all_atom} & 512 & N/A & 0.10 & N/A & N/A & N/A & 0.1 \\ 
\bottomrule
\end{tabular}
}
\end{table*}

\textbf{Model Specifications.}
\tref{tab:model_hparams} details the specific hyperparameters for the \ourmodel and \textit{\codesmodel} variants used in~\tref{tab:aa_400_results}, ~\fref{fig:pareto_app},~\tref{tab:codesign_results_250}, and \tref{tab:codesign_results_all_atom}. All models follow a similar paradigm: the lower the noise scale, the lower the diversity and the higher the designability. The choices between the various discrete flow matching parameters were based on settings that yielded optimal results, focusing on a strong trade-off between diversity and designability. Please see the end of~\sref{sec:multi-modal_flow_matching} for precise definitions of (i) local translational, and (ii) local frame-based coordinates.

\textbf{Ablation Strategy.} Since \ourmodel generates backbone $C_\alpha$ atoms, amino acid residues, and atomistic side chains with non-$C_\alpha$ atoms simultaneously, our ablation study follows a hierarchical structure, incrementally integrating data modalities, starting from the simplest representation: (1) $C_\alpha$ only (\sref{app:bb_only}), (2) backbone-sequence co-design (\sref{app:cod_only}), and (3) fully atomistic proteins (\sref{app:at_only}).

\subsection{Backbone $C_\alpha$ Only Design}
\label{app:bb_only}
We start with comparing architectures designed for backbone $C_\alpha$ generation, and analyzing the role synthetic sequences play on structure-based benchmarks.
\tref{tab:unconditional_backbone_generation} presents the primary subset of de novo backbone design benchmarks for recent models, following Geffner et al.~\cite{geffner2025proteina}. We include both M1 (single-shot) and M8 (standard best of 8 ProteinMPNN samples) variants to quantify the impact of ProteinMPNN and its learned sequence distribution on structure-based protein design.

The overall ranking of models in ~\tref{tab:unconditional_backbone_generation} changes depending on whether M1 or M8 is used to evaluate designability and diversity. Notably, a large portion of undesignable samples can be made designable by resampling the possible sequence, as shown by the percentage of designable samples, which increases by 10-28\% when moving from M1 to M8. 

We also introduce Genie2-Flow in ~\tref{tab:unconditional_backbone_generation}, a variant of the Genie2 model that replaces the backbone diffusion process with the flow matching training and inference procedure of \basemodel. Genie2-Flow achieves the best balance between designability and diversity at both the M1 and M8 levels.

\begin{table}[h]
\centering
\caption{\textbf{Performance of de novo backbone generation for C$_\alpha$ only models.} All models generate 100 proteins for lengths $\in$ [50, 100, 150, 200, 250]. Genie2-Flow uses the Genie2 architecture with the conditional flow matching training of \basemodel. Here M8 refers results gain through best of 8 ProteinMPNN sequences. M1 denotes using the first ProteinMPNN sequence.}
\renewcommand{\arraystretch}{1.2} % Increase row spacing
\scalebox{0.9}{
\begin{tabular}{lcccccc}
\toprule
Method & Dataset & DES-M8 (\%)$\,\uparrow$ & DES-M1 (\%)$\,\uparrow$ & DIV-M8$\,\uparrow$ & DIV-M1$\,\uparrow$ & NOV-PDB$\,\downarrow$ \\
\midrule
FrameFlow & PDB & 88.6 & 61.2 & 236 (0.53) & 160 (0.52) & 0.69 \\
RFDiffusion & PDB & 94.4 & 77.8 & 217 (0.46) & 158 (0.34) & 0.71 \\
Genie2 & AFDB & 95.2 & 74.3 & 281 (0.59) & 233 (0.49) & 0.63 \\
FoldFlow-2 & PDB & 97.4 & 83.2 & 239 (0.49) & 200 (0.48) & 0.68 \\
FoldFlow-2 (reft) & PDB & 81.6 & 53.2 & 218 (0.53) & 131 (0.49) & 0.65 \\
\basemodel$_{\gamma = 0.25}^{\tiny\text{60M no tri}}$ & Genie2 & \textbf{98.4} & \textbf{87.8} & 139 (0.28) & 127 (0.29) & 0.75 \\
\basemodel$_{\gamma = 0.45}^{\tiny\text{60M no tri}}$ & Genie2 & 95.8 & 79.2 & 250 (0.52) & 203 (0.51) & 0.70 \\
Genie2-Flow$_{\gamma = 0.25}$ & Genie2 & 96.6 & 78.2 & \textbf{359 (0.74)} & \textbf{284 (0.73)} & \textbf{0.62} \\
\bottomrule
\end{tabular}
}

\label{tab:unconditional_backbone_generation}
\end{table}

\subsection{Backbone-Sequence Co-Design}
\label{app:cod_only}
\begin{table}[b!]
\centering
\caption{\textbf{Ablation of popular architectures for codesign on AFDB}. Results for Multiflow base without distillation are taken from their original paper. We trained Multiflow and Genie2-flow-codesign, and evaluated all models by generating 100 proteins for lengths $\in$ [50, 100, 150, 200, 250]. }
\scalebox{1.0}{
\begin{tabular}{lcc}
\toprule
Method & CODES-CA (\%)$\,\uparrow$ & DIV-CA$\,\uparrow$ \\
\midrule
MultiFlow (PDB) & 42.0 & 72 \\
MultiFlow (PDB \& distilled) & 86.7 & 160 \\
MultiFlow ($\mathcal{D}_\mathrm{AFDB-clstr}$) & 40.0 & 52 \\
\midrule
Genie2-Flow-Co-design ($\mathcal{D}_\mathrm{AFDB-clstr}$)& 83.0 & 79 \\
\midrule
\codesmodel ($\mathcal{D}_\mathrm{AFDB-clstr}$) & 86.4 & 153 \\
\codesmodel ($\mathcal{D}_\mathrm{SYN-ours}$) & 87.0 & 226 \\
\bottomrule
\end{tabular}
}

\label{tab:arch_codes_abl}
\end{table}

Now, we examine the backbone-sequence co-design task to gain a deeper understanding of how sequences influence the generated structures during explicit joint learning.
% \textcolor{red}{We now turn to... one or two sentences.}

\subsubsection{Extended Discussion of Explicit Co-Design}

First, we see that \codesmodel and \ourmodel generate more consistent designable proteins compared to having a separate ProteinMPNN step. This is shown by the $\geq$ DES-M1 column of~\tref{tab:codesign_results_250}, where the sequences generated by our models yield higher designability than a separate ProteinMPNN call. This is important because it demonstrates that we have an accurate model that can operate without the need for always trying to redesign a more fitting sequence (and side chain structure by definition) to the already generated structure, as done in standard multi-stage design pipelines. DES-M8 is always higher than both CODES and DES-M1, signifying that many sequences can fold into similar structures, which we know to be true fundamentally. 
% Our model does not completely rule out ever needing inverse folding-based post-optimization if the goal is maximization of M8 scores, but with \ourmodel, we achieve high single-shot accuracy with best-in-class side chain structures (\fref{fig:sc_eval}). 
While our model does not eliminate the potential need for inverse folding-based post-optimization to maximize M8 scores, it achieves high single-shot accuracy with superior side chain structures (\fref{fig:sc_eval}), setting a strong foundation for further optimization.
Although DES-M8 is higher than M1, finding the best of eight different sequences would require redesigning the side chain structures afterwards. In contrast, \ourmodel generates accurate fully atomistic structures with an aligned sequence in one go.

Second, the success of \ourmodel and \codesmodel is not just due to solving the consistency issues present in using AFDB for fully atomistic training. In~\tref{tab:arch_codes_abl}, we see that when we take three prominent model architectures (\basemodel, MultiFlow/FrameFlow, Genie2) and train them on the same data \geniedata, our \codesmodel outperforms them significantly. Furthermore, we observe that when MultiFlow is trained with its distilled data (comprising PDB and model-generated structures, all with ProteinMPNN sequences), \codesmodel trained on \geniedata achieves competitive performance. Additionally, we find that removing the adversarial or inconsistent structure-sequence pairs and replacing them with in-silico consistent ones (i.e. training on \ourdata) increases the accuracy for both co-designability and diversity. As a result, we demonstrate that both architecture and framework as well as data make non-trivial contributions.
This result mirrors the behavior observed in~\tref{tab:unconditional_backbone_generation}, where instead of relying on noisy best-of-8 ProteinMPNN sampling, here we can learn a diverse and consistent structure-sequence distribution.

\subsubsection{What led us to build a more consistent dataset?}
\label{app:agree_data}
We observed that ProteinMPNN-based sequence resampling can significantly improve designability, as evident from the disparity in designability between M1 and M8 in~\tref{tab:unconditional_backbone_generation}. Notably, up to 28\% of the generated backbones can transition from undesignable to designable simply by resampling the sequence and selecting the best of 8. This suggests that suitable sequences exist for these novel de novo structures, but generating them in a single shot is non-trivial. Moreover, even with ProteinMPNN, the most likely sequence is not guaranteed to be the best, highlighting the need for low-temperature sampling in many of its applications~\citep{yim2023frameflow, delAlamo2025.05.09.653228}.

The observed disparity, combined with the fact that the clustered AFDB is only 19.1\% co-designable-all-atom (\fref{fig:afdb_data_dist}), led us to investigate the role of ProteinMPNN in enabling consistency in modeling the joint distribution of protein structure and sequence.
Given that finding the proper sequence significantly affects sequence-free model performance (\tref{tab:unconditional_backbone_generation}), training on largely non-co-designable data seemed problematic.

We emphasize that simply aligning the structures to known sequences (\emph{i.e.}, training on ESMFold structures) is insufficient and even hurts performance (\fref{tab:training_data_choice}). To clarify, although we aim to push our models to generate the best designability possible, training on a large amount of diverse and 100\% designable structures hurts performance compared to a largely non-designable dataset. To gain a deeper understanding, we investigated the effects of architecture and data on explicitly learning the joint backbone-sequence distribution in the de novo co-design setting (\tref{tab:arch_codes_abl}).

Also see related discussions in \sref{app:sec_dataset}.

\subsubsection{Backbone success does not always translate to multi-modal tasks}
\tref{tab:arch_codes_abl} shows that while Genie2-Flow sets new state-of-the-art results for backbone design, it performs poorly when extended to backbone-sequence co-design. Specifically, Genie2-Flow exhibits a 3.6x diversity drop when comparing ProteinMPNN single-shot (M1) diversity to that of the model-generated sequences (CA). We note that \basemodel, Genie2, Genie2-Flow, and \codesmodel were trained on identical datasets, with \codesmodel being identical to the 60M \basemodel but with sequence features and discrete flow matching training.

Furthermore, we found that \codesmodel, trained on the unaltered clustered AFDB, matches MultiFlow's performance when trained on PDB and model-generated structures with distilled ProteinMPNN sequences. In contrast, training MultiFlow on the same Genie2 data resulted in co-designability and diversity collapse compared to its distilled form. This highlights the core \basemodel transformer's accurate and robust usage for both backbone and backbone-sequence co-design, across natural and synthetic sequence datasets.

% Notably, \codesmodel outperforms all prior explicit-sequence models on natural sequences, even matching the performance of prior distilled models. When trained on our newly introduced more aligned dataset, we observed a boost in diversity without sacrificing co-designability. This result mirrors the behavior observed in~\tref{tab:unconditional_backbone_generation}, where instead of relying on noisy best-of-8 ProteinMPNN sampling, here we can learn a diverse and aligned structure-sequence distribution.

\subsubsection{Extended Co-Design Results}
\tref{tab:codesign_results_250} presents the full benchmark performance of the models captured in~\fref{fig:pareto_app}. 
% We also compare our model to DPLM-2~\citep{wang2025dplm}, which released their code after the main paper submission deadline. 
Overall, \codesmodel outperforms all prior baselines. Furthermore, how we model the side chains and non-$C_\alpha$ atoms with respect to their central $C_\alpha$ (local vs. frame) greatly impacts the diversity metric.
Lastly, by comparing our backbone $C_\alpha$-sequence co-design model, \codesmodel, to \ourmodel, we observe that significant backbone diversity can be achieved through the incorporation of all-atom modeling (non-$C_\alpha$ backbone atoms and side chains). Here both models are trained on \ourdata for fair comparisons.

\begin{table}[!t]
\centering
\caption{\small \textbf{Backbone-Sequence Co-design performance} compared to baselines. All models generate 100 proteins for lengths $\in$ [50, 100, 150, 200, 250]. We report the two multi-modal sampling configurations that generate the (i) most co-designable (codes) and (ii) most diverse samples (div). The best model for co-designability and diversity is emphasized. For parameterization definitions see~\tref{tab:model_hparams}. All \ourmodel  and \basemodel-Co-design are trained with $\mathcal{D}_\mathrm{SYN-ours}$. The $\geq$ DES-M1 column refers to models in which the co-generated sequences offer higher co-designability than ProteinMPNN redesign (1 sample).}
\scalebox{0.80}{
\begin{tabular}{lccc|cccc}
\toprule
\multicolumn{1}{l}{\textbf{Method}} & \multicolumn{3}{c}{\textbf{Backbone-Sequence Co-design}} & \multicolumn{4}{c}{\textbf{Backbone-Only Design}} \\
& $\geq$ DES-M1 & CODES (\%)$\,\uparrow$ & DIV-CA$\,\uparrow$ & \multicolumn{2}{c}{DES (\%)} & \multicolumn{2}{c}{DIV} \\
% \cmidrule(lr){5-6}\cmidrule(lr){7-8} %\cmidrule(lr){2-4}
&  &  &  & M8$\,\uparrow$ & M1$\,\uparrow$ & M8$\,\uparrow$ & M1$\,\uparrow$ \\
\midrule
ProteinGenerator  & \xmark & 32.0 & 48 & 86.2 & 73.0 & 85  & 82  \\ 
Protpardelle  & \xmark & 65.8 & 41  & 95.8 & 75.0 & 59  & 51  \\ 
PLAID  & \xmark & 34.0 & 79  & 49.2 & 36.4 & 117  & 81  \\ 
DPLM-2 (650M, co-generation) & \xmark & 40.6 & 90  & 59.0 & 42.8 & 133 & 100 \\
Multiflow  & \xmark & 86.7 & 160  & 99.6 & 92.0 & 191  & 173  \\
CarbonNovo & \checkmark & 76.0 & 161  & 89.6 & 70.4 & 201  & 148 \\
% DPLM-2 & \xmark & 40.6 & 90 (0.44) & 59.0 & 42.8 & 133 (0.45) & 100 (0.46) \\
P(all-atom) & \xmark & 80.0 & 263  & 98.2 & 95.4 & 349  & 299  \\
\midrule
\codesmodel$_{\mathrm{codes}}$ & \checkmark & 97.0 &  156& 99.2 & 97.0  & 158  & 155 \\ 
\codesmodel$_{\mathrm{div}}$ & \checkmark & 87.0 & 226 & 96.2 & 85.2 & 256  & 223  \\
\midrule
\ourmodel$_{\mathrm{codes}, \text{local frame}}$ & \checkmark & 96.2 & 162 & 99.4 & 93.8 & 166 & 162 \\ 
\ourmodel$_{\mathrm{div}, \text{local frame}}$  & \checkmark & 82.4 & 230 & 94.8 & 81.8 & 260  & 227  \\ 
\ourmodel$_{\mathrm{codes}, \text{local trans}}$  & \checkmark & 97.2 & 136 & 99.2 & 96.0 &  134 & 130\\ 
% \rowcolor{ngreen!10}
\ourmodel$_{\mathrm{div}, \text{local trans}}$  & \checkmark & 84.2 & 274 & 96.4 & 82.4 &  320 &  268 \\ 
% \midrule
\ourmodel-tri\xspace$_{\text{local frame}}$ & \checkmark & 88.6 & 236 & 97.0 & 87.8 & 257 & 227 \\
\midrule
\laprot\xspace$_{\mathrm{codes},\mathcal{D}_\mathrm{AFDB-clstr}}$ & \xmark & 88.6 & 221 & 99.0 & 95.0 & 249 & 235 \\
\laprot\xspace$_{\mathrm{div},\mathcal{D}_\mathrm{AFDB-clstr}}$ & \xmark & 84.6 & 221 & 98.6 & 89.8 & 259 & 233 \\
% \laprot-tri\xspace$_{\mathrm{opt},\mathcal{D}_\mathrm{AFDB-clstr}}$ & - & - & - & - & - & - & - & - & - \\
\laprot\xspace$_{\mathrm{codes},\mathcal{D}_\mathrm{SYN-ours}}$ & \checkmark & 96.8 & 244 & 99.6 & 96.6 & 249 & 242 \\
%laproteina_160M_codesdata512_ae512scratch_470kema_4nbs6-inference250-2025_08_06_13_15_32 ROW 2
%  chk_epoch=00000105_step=000000250000-EMA.ckpt ca 0.1 is codes 0.3 is div
\laprot\xspace$_{\mathrm{div},\mathcal{D}_\mathrm{SYN-ours}}$ & \checkmark & 93.6 & 285 & 99.2 & 93.2 & 298 & 278 \\
%laproteina_160M_codesdata512_ae512scratch_470kema_4nbs6-inference250-2025_08_06_13_15_32 ROW 4
\bottomrule
\end{tabular}
}
\label{tab:codesign_results_250}
\end{table}

\subsection{Fully Atomistic De Novo Protein Generation}
\label{app:at_only}
Building on the findings from our backbone $C_\alpha$-sequence co-design model, \codesmodel, we investigate key aspects of \ourmodel model, including its architecture, stochastic multi-modal sampling procedure, and side chain initialization method, and assess their individual impacts on model performance.

\subsubsection{Extended Atomistic Benchmarks and Side Chain Representations}
\begin{table}[!t]
\centering
\caption{\small \textbf{All Atom max length 250 performance} compared to baselines. All models generate 100 proteins for lengths $\in$ [50, 100, 150, 200, 250]. We report the two multimodal sampling configurations that generate the (i) most all atom codesignable (codes) and (ii) most diverse samples (div). For parameterization definitions see~\tref{tab:model_hparams}.}
\scalebox{0.75}{
\begin{tabular}{lccccc}
\toprule
Method & CODES-AA (\%)$\,\uparrow$ & DES-M1 (\%)$\,\uparrow$ & DIV-AA$\,\uparrow$ & NOV-PDB-AA$\,\downarrow$ & NOV-AFDB-AA$\,\downarrow$ \\
\midrule
ProteinGenerator  & 16.0 & 73.0 & 24 & 0.75 & 0.78 \\ 
Protpardelle  & 19.2 & 75.0 & 22 & 0.74 & 0.77 \\ 
PLAID    & 25.4 & 36.4 & 56 & 0.83 & 0.87 \\ 
P(all-atom)  & 76.8 & 87.2 &  251 & \textbf{0.67} & \textbf{0.73} \\
\midrule
\ourmodel$_{\mathrm{codes},\text{local frame}}$  & 95.4 & 94.0 & 163 & 0.76 & 0.81 \\ %zc best row 28 or row 17 of local trans
% \ourmodel_\delta & 81.4 & 84.0 & 0.58 (235) & x & x \\ %%row 39 GOOD_all_atom_256_local_af3_bs12_12nodes_distill_esmfold_try1-inference_try5-2025_05_09_07_56_45
% \ourmodel_\rho & 89 &  & 0.51 (228) & x & x \\ % ZC best row 113
\ourmodel$_{\mathrm{div}, \text{local frame}}$  & 77.0 & 81.8 & 215 & 0.74 & 0.80\\  % ZC row 125
\ourmodel$_{\mathrm{codes}, \text{local trans}}$  & 96.2 & 96.0 & 135 & 0.78 & 0.81\\ %trans div row 19
% \rowcolor{ngreen!10}
\ourmodel$_{\mathrm{div},\text{local trans}}$  & 81.4 & 82.4 & 267 & 0.73 & 0.79 \\ %all_atom_atomseq_200M_iclr_local_trans_af3_12n_bs8_distill-inference-2025_05_13_13_18_19
% \ourmodel  & x & x & x & x & x \\
% \midrule
\ourmodel-tri\xspace$_{ \text{local frame}}$  & 86.4 & 87.8 & 235 & 0.75 & 0.80\\  % 
\midrule
\laprot\xspace$_{\mathrm{codes}, \mathcal{D}_\mathrm{AFDB-clstr}}$ & 84.4 & 95.0 & 208 & 0.80 & 0.87 \\
\laprot\xspace$_{\mathrm{div}, \mathcal{D}_\mathrm{AFDB-clstr}}$ & 81.0 & 89.8 & 213 & 0.79 & 0.86 \\
\laprot-tri\xspace$_{\mathrm{codes},\mathcal{D}_\mathrm{AFDB-clstr}}$ & 89.2 & 95.0 & 124 & 0.81 & 0.87 \\
\laprot-tri\xspace$_{\mathrm{div},\mathcal{D}_\mathrm{AFDB-clstr}}$ & 83.6 & 90.2 & 176 & 0.78 & 0.85 \\
% \laprot\xspace$_{\mathrm{codes},\mathcal{D}_\mathrm{SYN-ours}}$ & \textbf{97.6} & \textbf{96.2} & 223 & - & - \\
% %laproteina_160M_codesdata512_ae512scratch_470kema_4nbs6-inference250-2025_08_06_13_15_32 ROW 14
\laprot\xspace$_{\mathrm{codes},\mathcal{D}_\mathrm{SYN-ours}}$ & \textbf{96.2} & \textbf{96.6} & 242 & 0.78 & 0.85 \\
%laproteina_160M_codesdata512_ae512scratch_470kema_4nbs6-inference250-2025_08_06_13_15_32 ROW 1
\laprot\xspace$_{\mathrm{div},\mathcal{D}_\mathrm{SYN-ours}}$ & 92.2 & 93.2 & \textbf{283} & 0.78 & 0.85 \\
%laproteina_160M_codesdata512_ae512scratch_470kema_4nbs6-inference250-2025_08_06_13_15_32 ROW 4
\bottomrule
\end{tabular}
}
\label{tab:codesign_results_all_atom}
% \vspace{-5mm}
\end{table}

\begin{table}[!b]
\vspace{-6mm}
\centering
\caption{\small \textbf{Max length 400 performance of \ourmodel on de novo all atom generation} compared to baselines. All models generate 100 proteins for lengths $\in$ [50, 400] with step size 50. We report the three multimodal sampling configurations that generate the (i) most all-atom co-designable (codes), (ii) most diverse samples (div), and (iii) an optimal trade-off (opt). The best values are bolded. All instances of \ourmodel here use local frames for the side chains. For parameterization definitions see~\tref{tab:model_hparams}.}
\scalebox{0.78}{
\begin{tabular}{lccccc}
\toprule
Method & CODES-AA (\%)$\,\uparrow$ & DES-M1 (\%)$\,\uparrow$ & DIV-AA$\,\uparrow$ & NOV-PDB-AA$\,\downarrow$ & NOV-AFDB-AA$\,\downarrow$ \\
\midrule
ProteinGenerator  & 10.0 & 57.1 & 28 & 0.75 & 0.78 \\ 
Protpardelle  & 13.6 & 62.8 & 25 & 0.74 & 0.76 \\ 
PLAID  & 22.3 & 34.9 & 63 & 0.85 & 0.88 \\
Pallatom  & 51.6 & 62.5 & 282 & \textbf{0.66} & \textbf{0.71} \\
\midrule
\ourmodel$_\mathrm{codes}$ & 87.8 & 88.1 & 263 & 0.77 & 0.81 \\ 
\ourmodel$_\mathrm{opt}$ & 83.1 & 85.8 & 321 & 0.76 & 0.80 \\
\ourmodel$_\mathrm{div}$ & 71.6 & 72.0 & 333 & 0.75 & 0.80 \\ 
% \midrule
\ourmodel-tri\xspace$_\mathrm{opt}$ & 87.6 & 88.3 & 396 & 0.73 & 0.77 \\
\midrule
\laprot\xspace$_{\mathrm{codes}, \mathcal{D}_\mathrm{AFDB-clstr}}$ & 76.0 & 90.1 & 308 & 0.77 & 0.85 \\
\laprot\xspace$_{\mathrm{div}, \mathcal{D}_\mathrm{AFDB-clstr}}$ & 70.6 & 85.5 & 314 & 0.77 & 0.84 \\
\laprot-tri\xspace$_{\mathrm{codes},\mathcal{D}_\mathrm{AFDB-clstr}}$ & 84.8 & 90.1 & 161 & 0.81 & 0.87 \\
\laprot-tri\xspace$_{\mathrm{div},\mathcal{D}_\mathrm{AFDB-clstr}}$ & 75.0 & 84.3 & 268 & 0.78 & 0.84 \\
% \laprot\xspace$_{\mathrm{codes}, \mathcal{D}_\mathrm{SYN-ours}}$ & \textbf{92.6} & \textbf{92.5}& 418 & - & -  \\
% laproteina_160M_codesdata512_ae512scratch_470kema_4nbs6-inference400-2025_08_06_13_59_34 ROW 9
\laprot\xspace$_{\mathrm{codes}, \mathcal{D}_\mathrm{SYN-ours}}$ & \textbf{90.6} & \textbf{91.2}& 460 & 0.75 & 0.83  \\
% laproteina_160M_codesdata512_ae512scratch_470kema_4nbs6-inference400-2025_08_06_13_59_34 ROW 3
\laprot\xspace$_{\mathrm{div}, \mathcal{D}_\mathrm{SYN-ours}}$ & 87.9 & 87.4 & \textbf{475} & 0.74 & 0.82  \\
% laproteina_160M_codesdata512_ae512scratch_470kema_4nbs6-inference400-2025_08_06_13_59_34 ROW 4
\bottomrule
\end{tabular}
}
\label{tab:aa_400_tri_results}
\vspace{-3mm}
\end{table}

% \begin{wrapfigure}{r}{7.6cm}
%   \vspace{-7mm}
%     \includegraphics[width=0.56\columnwidth]{figures/pareto.png}
%   \vspace{-7mm}
%     \caption{\small \textbf{Pareto frontier of the codesignability-diversity trade-off of \ourmodel}. We demonstrate metrics of protein with length$\in [50, 250]$ in top panel and protein with length$\in [50, 400]$ in bottom panel. \zc{Bottom panel values subject to change. waiting for some numbers now.}}
%     \label{fig:pareto}
%   \vspace{-4mm}
%   % \vspace{-3mm}
% \end{wrapfigure}

\begin{figure}[!h]
    % \vspace{-8mm}
    \centering
    \includegraphics[width=0.75\linewidth]{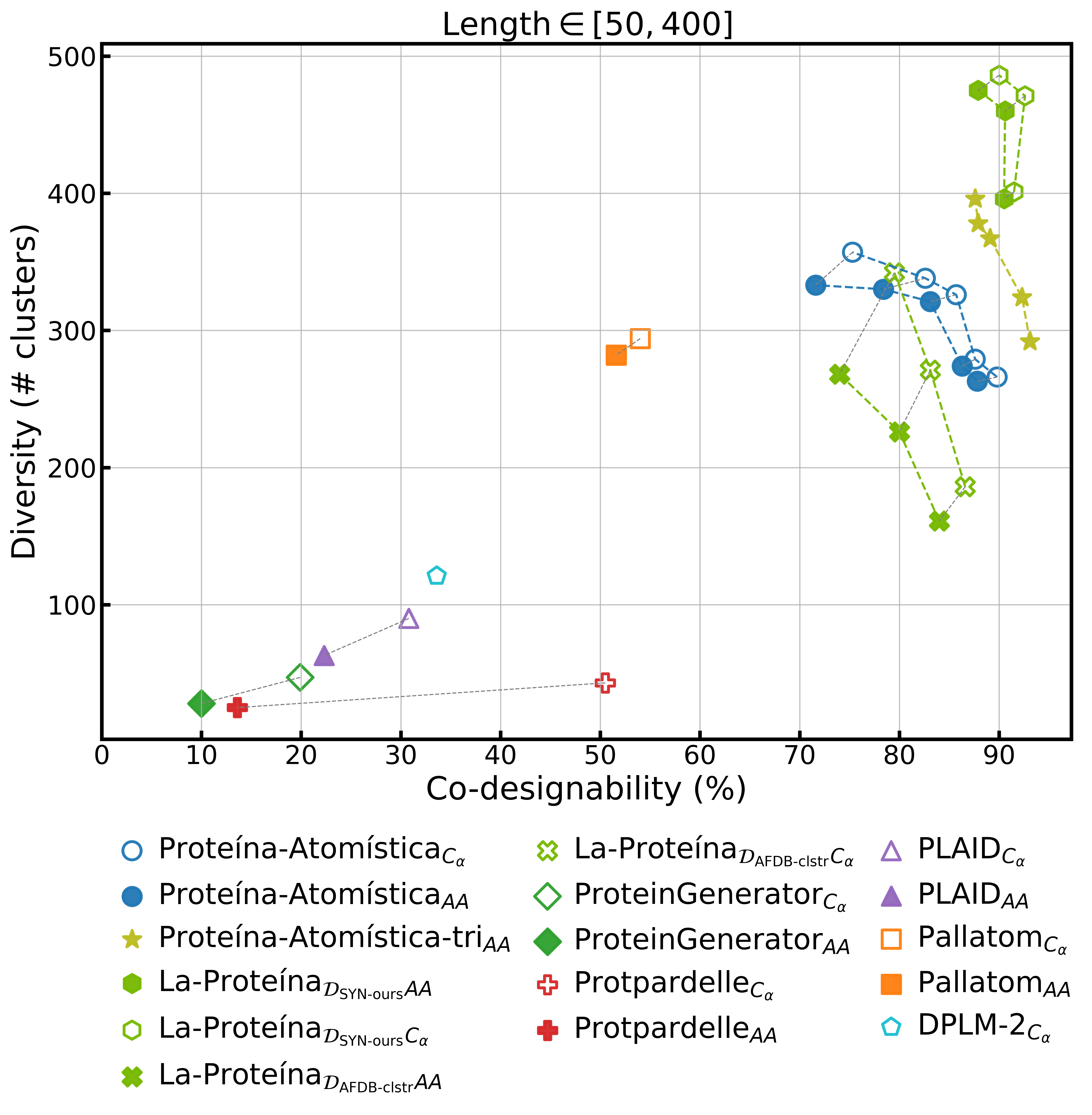}
    \vspace{-1mm}
    \caption{\small \textbf{Pareto frontier of the co-designability-diversity trade-off}. We show metrics of proteins with length $\in [50, 400]$. Solid and hollow markers represent metrics calculated on all-atom and $C_{\alpha}$ basis, respectively. For atomistic models, the all-atom and $C_{\alpha}$ scores for the same generated proteins are connected by gray dashed line, and obtained from the same model.}
    \label{fig:pareto_app}
    % \vspace{-5mm}
\end{figure}

% Tables~\ref{tab:codesign_results_all_atom} and~\ref{tab:aa_400_tri_results} illustrate how different noise scale parameters affect the trade-off between designability and diversity. 
Tables~\ref{tab:codesign_results_all_atom} and~\ref{tab:aa_400_tri_results} demonstrate the impact of varying noise scale parameters on the trade-off between designability and diversity (\textit{codes} vs. \textit{div} vs. \textit{opt} settings, see \tref{tab:model_hparams}).
% Our analysis also reveals that both local translation and frame-based side chain parameterizations are useful. The choice of which to use depends on behavior one is trying to target. The local frame yields better diversity with comparable co-designability for the high co-designability setups. In contrast, the local translation yields better co-designability and diveristy in the high diveristy setups.

Moreover, we find that both local translation and frame-based side chain parameterizations are useful (\textit{local frame} vs. \textit{local trans}), but their relative effectiveness depends on the specific goals of the task. In particular, the local frame is advantageous in high co-designability settings, where it achieves better diversity with comparable co-designability. In contrast, local translation is more effective in high diversity settings, where it yields better co-designability and diversity. See ~\sref{app:sec:multimodal_flow_matching} for definitions of local translation and frame-based parameterizations.

Tables~\ref{tab:codesign_results_all_atom} and~\ref{tab:aa_400_tri_results} also illustrate the effect of incorporating triangle updates, which demonstrate improved performance up to a length of 400, despite being trained only up to 256. This is notable, especially when compared to the other \ourmodel variants, which were finetuned to a length of 400. Further details on the triangle update layers can be found in Appendix~\ref{app:tri_details}.

\begin{table}[t]
\centering
\caption{Impact of consistent synthetic data on \laprot. All models generate 100 proteins for lengths $\in$ [100, 500] with step size 100. Baselines taken directly from \cite{geffner2025laproteina}.}
\begin{tabular}{lcc}
\toprule
Model & CODES-AA (\%)$\,\uparrow$ & DIV-AA$\,\uparrow$ \\
\midrule
P(all-atom) & 36.7 & 134 \\
\laprot (\geniedata)  & 68.4 & 206 \\
\midrule
\laprot (\ourdata) & 86.8 & 318 \\
\bottomrule
\end{tabular}
\label{tab:la_proteina_comparison}
\end{table}
Table~\ref{tab:la_proteina_comparison} further demonstrates the impact of our introduced consistent synthetic data on even longer lengths that the original \laprot was trained and evaluated on.

\subsubsection{Pareto Frontier}
We include ~\fref{fig:pareto_app}, an updated pareto frontier to include \ourmodel trained with additional triangle multiplicative updates.
Adding 4M worth of triangle multiplicative updates to our 222M \ourmodel further pushes the Pareto frontier. We emphasize that \textit{\ourmodel-tri} is only trained up to length 256 but shows the ability to generalize to longer proteins. Given the increased time and memory costs, we leave further improvements of the \basemodel and \ourmodel transformers to future work.
These triangle operations are typically seen as required for protein modeling success. In contrast, we are able to take advantage of scaling our data and simpler transformer architectures to yield strong performance.

\subsubsection{Atomistic Side Chain Initialization}
\label{app:sec:side_chain_init}

\begin{table}[t]
\centering
\caption{\textbf{Ablation of the side chain initialization.} All results here use the same model weights and sampling hyperparameters for a \ourmodel model with ``local trans'' non-$C_\alpha$ coordinates.}
\scalebox{1.0}{
\begin{tabular}{lcc}
\toprule
Method & CODES-AA (\%)$\,\uparrow$ & DIV-AA$\,\uparrow$ \\
\midrule
Gaussian Initialization & 56.8 & 177 \\
Zero Initialization & 60.8 & 196 \\
Learned clean data objective& 38.2 & 76 \\
Learned vector field (default) & 81.4 & 262 \\
\bottomrule
\end{tabular}
}

\label{tab:init_aa_abl}
\end{table}

The side chain structures of proteins generated by \ourmodel are of variable atom sequence length as a function of generation time, because the residue types may change during the generation process (through a series of remasking and unmasking operations). As a result \ourmodel must be able to handle the resetting and regeneration of accurate side chain structures subject to the discrete sampling process of the discrete flow matcher (see~\sref{app:side_chain_init}). This dynamic coupling requires careful handling of the initialization point as seen in \tref{tab:init_aa_abl}, demonstrating the importance of learning a meaningful side chain initialization, instead of using naive zero or random Gaussian initialization. 
Furthermore, we see that if using a ``clean data prediction objective'', where we try to predict the side chain structure from the mask token directly rather than using our introduced vector field-like augmentation (c.f.~\sref{app:side_chain_init}), the model struggles to generate accurate side chains.
\subsection{Atomistic Side Chain Evaluation}
\begin{figure}[t!]
\centering
  % \vspace{-4mm}
    \includegraphics[width=0.70\linewidth]{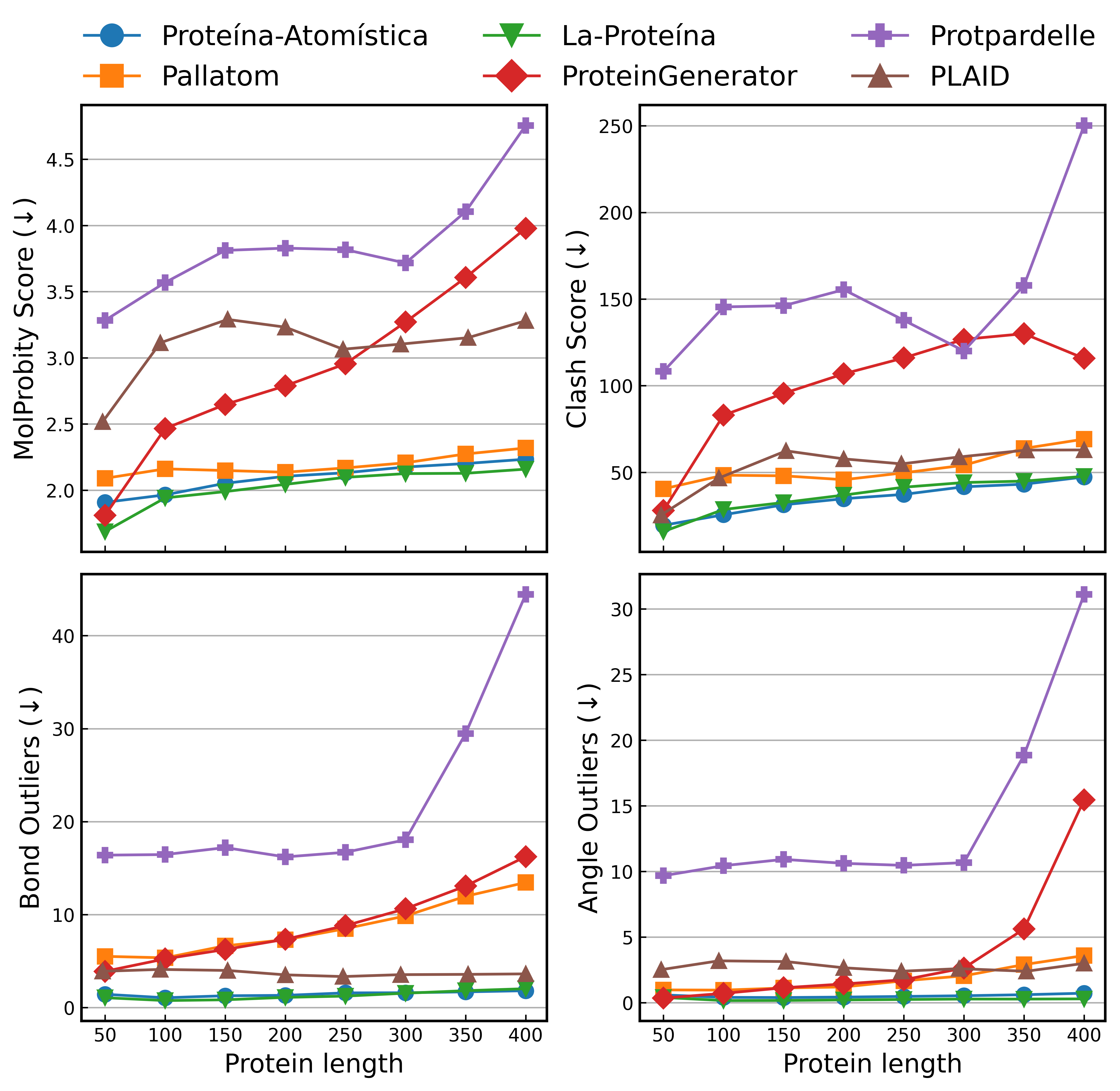}
  \vspace{-1mm}
    \caption{\small \textbf{Side chain structure evaluations.} Lower scores indicate higher side chain quality.}
    \label{fig:sc_eval}
  % \vspace{-2mm}
\end{figure}

To evaluate the generated atomistic protein structures, we compute: (1) MolProbity score~\citep{davis2007molprobity}, (2) clash scores, (3) bond length outliers, and (4) angle outliers, as shown in~\fref{fig:sc_eval}. MolProbity (MP) score is a composite evaluation metric of macromolecular structures. It measures geometric and stereochemical quality, including steric clashes, backbone dihedral angles, and side-chain conformations. Lower MP score indicates higher structure quality. The clash, bond and angle metrics focus on measuring the physical correctness of the atomistic details of the generated side-chains.
% \begin{wrapfigure}{r}{6.5cm}
%   \vspace{-4mm}
%     \includegraphics[width=\linewidth]{figures/sc_eval.png}
%   \vspace{-6mm}
%     \caption{\small \textbf{Side chain structure evaluations.} Lower scores indicate higher side chain quality.}
%     \label{fig:sc_eval}
%   \vspace{-2mm}
% \end{wrapfigure}%

% \begin{figure}[htb]
%     \centering
%     \includegraphics[width=\linewidth]{figures/sc_eval_horitontal.png}
%     \caption{\small \textbf{\ourmodel generates the most accurate side chains across all lengths}. \kk{Can you please turn this into a 2x2 large side figure? The design is too close to the plots in the other submission, let's be different at least stylistically.}}
%     \label{fig:sc_eval}
% \end{figure}
% \input{appendix/G_motif_scaffolding}
\section{Metric Definitions and Baselines}

% \subsection{Metric Definitions}
% \dr{@danny}

\subsection{De Novo Design Metrics}
To assess the performance of our models, we employ standardized metrics~\citep{geffner2025proteina, watson2023rfdiffusion, qu2024pallatom} for de novo protein design, adapting them for backbone-sequence co-design and all-atom contexts. The metrics used include:

\begin{enumerate}
    \item \textbf{Designability (DES)}: This measures the ability to inverse fold a generated protein backbone using ProteinMPNN~\citep{dauparas2022robust} and refold the generated sequences. We report two variants: \textbf{DES-M1} (single shot) and \textbf{DES-M8} (best of 8 sequences), where DES-M1 evaluates the designability of a single sequence generated by ProteinMPNN, and DES-M8 evaluates the designability of the best sequence out of 8 generated sequences.
    \item \textbf{Co-designability (CODES)}: Similar to DES-M1, but using the model's output sequence instead of ProteinMPNN-generated sequences. We also report \textbf{All-Atom Co-designability (CODES-AA)}, an extension of CODES that uses all-atom scRMSD.
    \item \textbf{Diversity}: We evaluate the structural diversity of samples by counting the number of Foldseek~\citep{vankempen2024foldseek} clusters formed by the filtered subset of backbones, using a TM-score threshold of $0.5$. Higher cluster counts indicate greater diversity.
      \begin{itemize}
            \item \textbf{DIV-AA}: Diversity metric filtered for All-Atom Co-designable samples (CODES-AA)
            \item \textbf{DIV-CA}: Diversity metric filtered for Co-designable samples (CODES)
            \item \textbf{DIV-M8}: Diversity metric filtered for Designable samples (DES-M8)
            \item \textbf{DIV-M1}: Diversity metric filtered for Designable samples (DES-M1)
      \end{itemize}
 \item \textbf{Novelty}: This metric evaluates a model's ability to generate structures that are distinct from those in predefined reference sets (PDB and $\mathcal{D}_{\textrm{Genie2}}$). We report the average maximum TM-score between designable structures and the reference sets, with lower scores indicating greater novelty. Specifically, we report PDB novelty for co-designable samples (\textbf{NOV-PDB}) and all-atom co-designable samples (\textbf{NOV-PDB-AA}), as well as their counterparts with respect to Genie2 (\textbf{NOV-AFDB} and \textbf{NOV-AFDB-AA}).
\end{enumerate}

Our evaluation protocol involves generating samples across a range of lengths, from $50$ to $250$ to align with prior work such as P(all-atom) as well as $50$ to $400$ to evaluate a more difficult spectrum. We then compute the aforementioned metrics across these samples. For designability, we use ProteinMPNN to generate sequences for each backbone and ESMFold~\citep{lin2023esm2} to predict structures, calculating the self-consistency RMSD (scRMSD) between predicted and original structures. A sample is considered designable if its scRMSD is under $2$\AA.

% \subsubsection{Side Chain Accuracy Metrics}
% To evaluate the atomistic protein structures generated by \textit{\ourmodel}, we compute: (1) MolProbity score~\citep{davis2007molprobity}, (2) clash scores, (3) bond length outliers, and (4) angle outliers. MolProbity (MP) score is a composite evaluation metric of macromolecular structures. It measures geometric and stereochemical quality, including steric clashes, backbone dihedral angles, and side-chain conformations. Lower MP score indicates higher structure quality. The clash, bond and angle metrics focus on measuring the physical correctness of the atomistic details of the generated side-chains. Across all lengths \textit{\ourmodel} generates more accurate side chains compared to prior methods (\fref{fig:sc_eval}). \ourmodel achieves a length-averaged MP score of 2.097 compared to 4.307 of P(all-atom) (the next closest performing model from~\tref{tab:aa_400_results}). The next closest average MP score is ProteinGenerator with 2.940, but it has the lowest all atom co-designability.
\subsection{Side Chain Accuracy Metrics}

To evaluate the atomistic protein structures generated by \textit{\ourmodel}, we compute several metrics that assess the accuracy and physical correctness of the generated side chains. These metrics include:

\begin{enumerate}
    \item \textbf{MolProbity Score}: The MolProbity (MP) score is a composite evaluation metric that assesses the geometric and stereochemical quality of macromolecular structures~\citep{davis2007molprobity}. It is a combination of several individual metrics, including:
        \begin{itemize}
            \item \textbf{Clashscore}: Measures the number of steric clashes between atoms in the protein structure.
            \item \textbf{Ramachandran outliers}: Refers to the percentage of residues with dihedral angles ($\phi$ and $\psi$) that fall outside the allowed regions of the Ramachandran plot.
            \item \textbf{Rotamer outliers}: Refers to the percentage of residues with side-chain conformations that are inconsistent with the expected rotameric states.
        \end{itemize}
        A lower MolProbity (MP) score indicates higher structure quality. Notably, a score of $\geq3$ indicates significant stereochemical issues, highlighting potential problems with the structure's accuracy. The MP score is a widely used and reliable metric for evaluating protein structure quality.
    \item \textbf{Clash Scores}: Clash scores measure the number of steric clashes between atoms in the protein structure. Steric clashes occur when two or more atoms are too close to each other, resulting in unfavorable interactions. A lower clash score indicates fewer steric clashes and a more physically realistic structure.
    \item \textbf{Bond Length Outliers}: Bond length outliers refer to the percentage of bonds in the protein structure that deviate significantly from their expected lengths. A lower percentage of bond length outliers indicates a more accurate structure.
    \item \textbf{Angle Outliers}: Angle outliers refer to the percentage of bond angles in the protein structure that deviate significantly from their expected values. A lower percentage of angle outliers indicates a more accurate structure.
\end{enumerate}

These metrics provide a comprehensive evaluation of the accuracy and physical correctness of the generated side chains. By assessing the MolProbity score, clash scores, bond length outliers, and angle outliers, we can better understand the strengths and weaknesses of \textit{\ourmodel} and prior atomistic models in generating accurate atomistic protein structures.

Across all lengths, \textit{\ourmodel} generates more accurate side chains compared to prior methods (\fref{fig:sc_eval}). \ourmodel achieves a length-averaged MP score of 2.097 compared to 4.307 of P(all-atom) (the next closest performing model from~\tref{tab:aa_400_results}). The next closest is ProteinGenerator, which has an average MP score of 2.940 but has the lowest all-atom co-designability.

\subsection{Baselines}
In this section, we discuss the sampling configurations of the baselines we compared to in this paper. For backbone design methods not included below nor introduced by us, the results were taken from Geffner et al.~\citep{geffner2025proteina}.

\textbf{Pallatom:}
We used the code and checkpoint from the \href{https://github.com/levinthal/Pallatom}{public Pallatom repository}. We used the default configuration suggested: \texttt{t\_min=0.01}, \texttt{t\_max=1}, \texttt{gamma=0.2}, \texttt{step\_scale=2.25}, and \texttt{T=200}. No training code is provided at this time.

\textbf{Protpardelle:}
We used the code and checkpoint from the \href{https://github.com/ProteinDesignLab/protpardelle}{public Protpardelle repository}. We used the default \texttt{uncond\_sampling.yml} file provided in the repository for unconditional sampling. For motif scaffolding, we prepared the \texttt{.pdb} files of each task based on the corresponding contigs and then used the provided \texttt{cond\_sampling.yml} configuration for sampling.

\textbf{ProteinGenerator:}
We used the code from the \href{https://github.com/RosettaCommons/protein_generator}{public ProteinGenerator repository}. We used the base checkpoint set in the repository. We followed the default configuration for unconditional sampling except for the number of sampling steps. Since we sampled proteins with length up to 400 residues, we increased the number of sampling steps from the default 25 to 100 for better generation quality, as recommended in the repository. 

\textbf{PLAID:}
We used the code from the \href{https://github.com/amyxlu/plaid}{public PLAID repository}. We used the 100M parameter checkpoint hosted on the \href{https://huggingface.co/amyxlu/plaid/tree/main}{PLAID HuggingFace repo} as it is the only loadable option. Since PLAID only supports sampling proteins with length divisible by 4, the actual length we sampled are $[48, 96, 152, 200, 248, 296, 352, 400]$. We used the default unconditional sampling configuration in the repository.

\textbf{DPLM-2:} We use the code from the \href{https://github.com/bytedance/dplm/tree/main}{DPLM Repository} specifically the pull request from \href{https://github.com/bytedance/dplm/tree/dplm2}{DPLM-2 branch}. We follow the instructions in the README.md to generate proteins from their 650M co-generation model using the indicated inference configuration and settings.

\textbf{MultiFlow:} We use the code from the \href{https://github.com/jasonkyuyim/multiflow}{MultiFlow Repository}. We use the provided \texttt{inference\_unconditional} config provided adjusted for the appropriate length intervals.

\textbf{CarbonNovo:} We use the code from the \href{https://github.com/CarbonMatrixLab/carbonnovo}{CarbonNovo Repository}. We use the provided predict.py to generate proteins of the desired lengths.

\textbf{FoldFlow-2:} We use the code from the \href{https://github.com/DreamFold/FoldFlow}{FoldFlow Repository}. We use the \texttt{runner/inference.py} script with both provided FoldFlow-2 weights with \texttt{model=ff2}. We use the default sampling parameters provided in inference.yaml.

% \textbf{\laprot:} We use the code from the \href{https://github.com/NVIDIA-Digital-Bio/la-proteina}{\laprot Repository}. We the training and inference code provided as well as the inference settings taken from~\cite{geffner2025laproteina}.

\section{Limitations}
\label{app:sec:limitation}
While \ourmodel performs well, it faces challenges in balancing natural sequence distribution learning with generated sample diversity. Key limitations include: increased computational cost and decreased speed associated with full-atom modeling compared to backbone-only approaches; the inability to capture protein dynamics; and the lack of guarantees for desired function or binding affinity. These limitations highlight exciting directions for future research. Future work can additionally explore similar techniques for conditional tasks such as motif scaffolding and binder design, as well as the generation of even longer protein sequences as done in \laprot~\citep{geffner2025laproteina}.
\section{Broader Impact}
\label{app:sec:impact}
Our method advances the field of de novo protein design by enabling joint generation of sequences and all-atom structures, with potential applications in drug discovery, enzyme engineering, and biomaterials. While this capability could accelerate the development of novel therapeutics and sustainable biocatalysts, it raises ethical considerations, such as the risk of misuse for harmful purposes. Additionally, the model’s performance depends on the quality of training data, which may inherit biases from structure prediction tools like AlphaFold2 and ESMFold. We emphasize responsible use and encourage further research into safety measures and bias mitigation to ensure positive societal impact.
% \input{appendix/L_license}

% \newpage
% \input{checklist}

\end{document}